\documentclass[american,preprint, review, sort&compress]{elsarticle}
\pdfoutput=1 
\usepackage[T1]{fontenc}
\usepackage[latin9]{inputenc}
\usepackage{geometry}
\usepackage{array}
\usepackage{float}
\usepackage{units}
\usepackage{multirow}
\usepackage{amsmath}
\usepackage{stackrel}
\usepackage{graphicx}
\usepackage{tablefootnote}
\usepackage[dvipsnames]{xcolor}

\makeatletter

\providecommand{\tabularnewline}{\\}
\floatstyle{ruled}
\newfloat{algorithm}{tbp}{loa}
\providecommand{\algorithmname}{Algorithm}
\floatname{algorithm}{\protect\algorithmname}

\usepackage{graphicx}
\usepackage[margin=10pt,font=footnotesize,labelsep=period]{caption}
\usepackage{geometry}
\usepackage{algorithm,algpseudocode}
\usepackage[section]{placeins}

\@ifundefined{showcaptionsetup}{}{%
 \PassOptionsToPackage{caption=false}{subfig}}
\usepackage{subfig}
\makeatother

\usepackage{babel}
\begin{document}

\title{A massively parallel explicit solver for elasto-dynamic problems
exploiting octree meshes}

\begin{frontmatter}{}

\author[unsw]{Junqi Zhang\corref{cor1}}

\ead{junqi.zhang@unsw.edu.au}

\cortext[cor1]{Corresponding author}

\author[unsw]{Ankit Ankit}

\author[ude]{Hauke Gravenkamp}

\author[unsw]{Sascha Eisentr{\"a}ger}

\author[unsw]{Chongmin Song}

\address[unsw]{School of Civil and Environmental Engineering, University of New
South Wales, Sydney NSW 2052, Australia}

\address[ude]{Department of Civil Engineering, University of Duisburg-Essen, 45141
Essen, Germany}
\begin{abstract}
Typical areas of application of explicit dynamics are impact, crash
test, and most importantly, wave propagation simulations. Due to the
numerically highly demanding nature of these problems, efficient automatic
mesh generators and transient solvers are required. To this end, a
parallel explicit solver exploiting the advantages of balanced octree
meshes is introduced. To avoid the hanging nodes problem encountered
in standard finite element analysis (FEA), the scaled boundary finite
element method (SBFEM) is deployed as a spatial discretization scheme.
Consequently, arbitrarily shaped star-convex polyhedral elements are
straightforwardly generated. Considering the scaling and transformation
of octree cells, the stiffness and mass matrices of a limited number
of unique cell patterns are pre-computed. A recently proposed mass
lumping technique is extended to 3D yielding
a well-conditioned diagonal mass matrix. This enables us to leverage
the advantages of explicit time integrator, i.e., it is possible to
efficiently compute the nodal displacements without the need for solving
a system of linear equations. We implement the proposed scheme together
with a central difference method (CDM) in a distributed computing
environment. The performance of our parallel explicit solver is evaluated
by means of several numerical benchmark examples, including complex
geometries and various practical applications. A significant speedup
is observed for these examples with up to one billion of degrees of
freedom and running on {\normalsize{}up to 16,384} computing cores.
\end{abstract}

\end{frontmatter}{}

\section{Introduction\label{sec:Introduction}}

Numerical methods for the solution of partial differential equations
(PDEs) have been widely used in engineering due to their feasibility
and reliability in handling problems with complex geometries and boundary
conditions. The finite element method (FEM) is one of the most popular
numerical methods, in which a computational domain is spatially discretized
into small subdomains of simple shapes, referred to as elements. As
a result, the governing PDEs are transformed into a system of linear
algebraic equations, which can be easily solved using modern computers.

Dynamic loading is ubiquitous in engineering practice and can be caused
by various physical processes, e.g., earthquakes~\citep{Dupros2010},
blasts~\citep{Brun2012}, impacts~\citep{Bettinotti2017} and many
others~\citep{Zhang2019b,Meister2020}. The FEM has been
used in structural dynamics to obtain the response history of structures,
which can be determined based on several methods, such as the modal
superposition technique, the Ritz-vector analysis or direct time integration
schemes~\citep{Cook1989}. Direct time integration schemes are often
based on finite difference approximations to achieve a temporal discretization.
Equilibrium equations are solved at each time step to compute the
unknown variables (displacements). Such integration schemes can be
classified further into two categories, namely, explicit and implicit
methods~\citep{Brun2012}. In an explicit method, the results of
the current time step are calculated based on one or more previous
time steps only, while in implicit methods the result depends not
only on previous time steps but also on the nodal velocity and acceleration
of the current step~\citep{Patzak2001}. In explicit methods, if
the mass and damping matrices are diagonal, the new nodal displacements
can be obtained efficiently without solving a system of linear equations~\citep{Talebi2012}.
Furthermore, the internal force vector can be assembled independently,
i.e., element-by-element (EBE). Therefore, it is not necessary to
assemble the global stiffness matrix, which yields advantages with
respect to the required memory. Due to the possibility of employing
EBE techniques, we are in a position to exploit parallel processing
strategies, which can significantly improve the numerical efficiency
by making use of the capabilities of multi-processor computers. In
recent years, the parallel implementation of explicit dynamics has
also been investigated in other numerical approaches such as meshfree
methods~\citep{Li2014a}, the virtual element
method~\citep{Park2019} and isogeometric analysis~\citep{Anitescu2019}.

Parallel processing is an attractive technique in high-performance
computing to speed up the algorithm by partitioning a serial job into
several smaller jobs and distributing them to individual processors
for execution~\citep{Rek2016}. A commonly used parallel processing
paradigm is single program, multiple data (SPMD)\footnote{Note that in Flynn's taxonomy (classification of computer architectures)
the same approach is referred to as single instruction, multiple data
(SIMD) which exploits data level parallelism.}, i.e., different processors execute the same command on different
data sets simultaneously~\citep{Altman2014}. In explicit dynamics,
the element set can be partitioned and assigned to several processors
so that several local nodal force vectors can be assembled at the
same time, which will then be synchronized across all processors to
form a global nodal force vector~\citep{Ma2020}. To this end, the
workload should be distributed equally to the processors, and the
amount of data communication should be as small as possible, because
load-imbalance will hinder the data synchronization, which constitutes
the bottleneck in current parallel processing strategies and limits
the attainable speedup. It is worth mentioning that GPU-accelerated
parallelization also attracted a lot of research interest with the
development of modern GPU architectures such as CUDA~\citep{Tang2020}.
In the current contribution, our focus is on parallelization based
on the use of multiple CPUs, and therefore, GPU-based schemes are
only mentioned for the sake of completeness but certainly constitute
a very promising direction for future applications.

Here, an explicit dynamic analysis methodology based on the use of
balanced octree meshes is presented. Due to the ease of creating such
meshes in the scaled boundary finite element method (SBFEM), our approach
utilizes this particular numerical method for the spatial discretization.
The SBFEM was first proposed by~\citet{Song1997} and initially developed
for dynamic analysis in unbounded domains and later extended to a
broad range of applications~\citep{Deeks2002,Song2004a,Bazyar2008}.
It has since been developed into a general-purpose numerical method
for the solution of PDEs~\citep{Ooi2020}.
The basic idea is to divide the problem domain into subdomains satisfying
the so-called scaling requirement (Section~\ref{subsec:Polyhedron}).
In the SBFEM, a semi-analytical approach is adopted to construct an
approximate solution in a subdomain. To this end, only the boundary
of the subdomain is discretized in a finite element sense, and the
solution in the radial direction must be obtained analytically. This
feature of the SBFEM enables a straightforward derivation of polytopal
(polygonal~\citep{Chiong2014,Gravenkamp2018}/polyhedral~\citep{Talebi2016})
elements with an arbitrary number of nodes, edges and faces to be
used in the analysis. Due to the versatility in element shapes, the
difficulty in mesh generation can be greatly reduced. In three-dimensional
analyses, the surface discretization typically consists of quadrilateral
and triangular elements~\citep{Saputra2017a}. However, it is well-known
that quadrilateral elements are more accurate, and therefore, in an
attempt to employ only quadrilateral elements in the surface discretization,
a novel transition element approach has been developed in Refs.~\citep{Saputra2020,Gravenkamp2020}.
Based on this methodology, arbitrary element types can be coupled~\citep{Duczek2020}.
In recent years, researchers have endeavored to apply this method
to address image-based analysis~\citep{Saputra2017a,Saputra2017,Gravenkamp2017},
acoustics~\citep{Liu2018,Liu2019a}, contact~\citep{Xing2018,Xing2019},
domain decomposition~\citep{Zhang2019a}, wave propagation~\citep{Gravenkamp2012,Gravenkamp2012a,Gravenkamp2017a}
and many other problems~\citep{Zou2019,Zhang2020,Liu2020,Natarajan2020,Eisentrager2020}.
The first application of SBFEM in the context of explicit dynamics
has been reported in Ref.~\citep{Qu2020}.

In the proposed explicit method, a balanced octree mesh (a recursive
partition of the three-dimensional space into smaller octants where
adjacent elements can only differ by one octree level) is generated
to approximate the geometry. The octree mesh generation is highly
efficient. An octree based algorithm has been developed for automatic
mesh generation of digital images~\citep{Saputra2017a} and stereolithography
(STL) models~\citep{Liu2017}. The application of octree meshes in
the standard finite element analysis is hindered by the existence
of hanging nodes. On the other hand, octree meshes are highly complementary
with the SBFEM~\citep{Song2018,Zhang2020a}. An advantage of the
octree meshes benefiting parallel computing is that only a limited
number of octree patterns are present, and therefore, the element
stiffness and mass matrices can be pre-computed. The nodal force vector
is calculated element-wise without the need for assembling the global
stiffness matrix, while the global mass matrix is lumped into a well-conditioned
diagonal matrix based on a recently developed technique~\citep{Gravenkamp2020a}.
All elements of the same shape (i.e., same nodal pattern) are grouped,
and their nodal displacement vectors are assembled as a matrix. In
the next step, the nodal forces are calculated by a simple matrix
multiplication. A parallel version of the solver is developed, in
which the mesh is first partitioned into several parts of almost equal
size with minimum connection~\citep{Chan1994}, and each
processor assembles the nodal force vector of one individual part.
Data exchange between individual processors is only needed at the
interfaces of the connected parts. The local nodal displacement vector
for the next time step is efficiently computed on each processor due
to the diagonal structure of the mass matrix before a global update
is conducted.

The remainder of this article is organized as follows: In Section~\ref{sec:SBFEM},
a brief description of polyhedral elements constructed using the SBFEM
is provided. The central difference method, a widely used explicit
time integration scheme which forms the basis of our parallel solver,
and its implementation using octree mesh is presented in Section~\ref{sec:Dynamics}.
Our parallel processing strategy is discussed in Section~\ref{sec:Parallel}.
In Section~\ref{sec:Examples}, the performance including the attainable
speedup is demonstrated by means of five numerical examples of increasing
complexity of the geometric models, number of degrees of freedom and
computing cores. General remarks and a summary are given in Section~\ref{sec:Summary}.

\section{The scaled boundary finite element method\label{sec:SBFEM}}

In this section, the construction of polyhedral elements using the
SBFEM is briefly discussed. For the sake of brevity, only the key
equations that are necessary for the implementation are presented.
Readers interested in a detailed derivation and additional explanations
are referred to the monograph by~\citet{Song2018}.

\subsection{Polyhedral elements based on scaled boundary finite element method\label{subsec:Polyhedron}}

\begin{figure}[tb]
\hfill{}\includegraphics{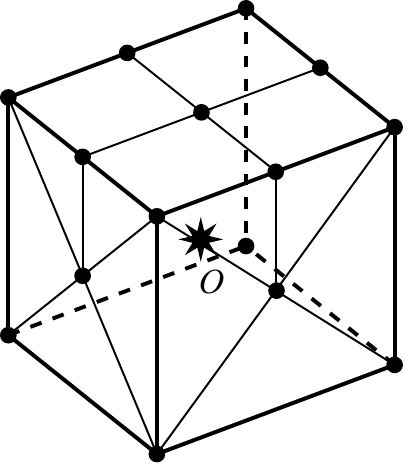}\hfill{}
\caption{A polyhedral element obtained by scaling the surface mesh toward the
scaling center $O$\label{fig:Subdomain}}
\end{figure}

The derivation of a polyhedron as a volume element in the SBFEM is
illustrated in Fig.~\ref{fig:Subdomain}. A so-called scaling center
(denoted as ``$O$'' in Fig.~\ref{fig:Subdomain}) is defined,
which should be directly visible from the entire boundary of the volume
element (condition of star-convexity). This requirement on the geometry
can always be satisfied by subdividing a concave polyhedron. One distinct
advantage of the SBFEM is that only the boundary of the volume element
needs to be discretized by means of surface elements. The volume of
the element is then obtained by scaling the discretized boundary towards
the scaling center. In this work, only four-node quadrilateral and
three-node triangular elements are used as surface elements.

\begin{figure}[tb]
\hfill{}%
\begin{minipage}[b]{0.2\textwidth}%
\subfloat[A triangular element\label{fig:Shape_fun_a}]{\hspace{3em}\includegraphics{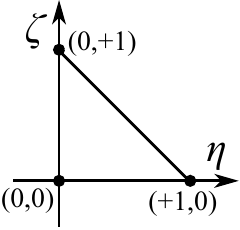}\hspace{3em}}

\subfloat[A quadrilateral element\label{fig:Shape_fun_b}]{\hspace{3em}\includegraphics{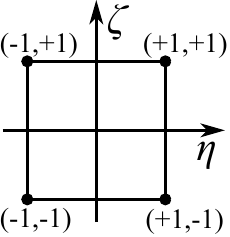}\hspace{3em}}%
\end{minipage}\hfill{}\subfloat[Radial coordinate $\xi$ emanating from the scaling center $O$ to
the boundary of the volume element\label{fig:Shape_fun_c}]{\includegraphics{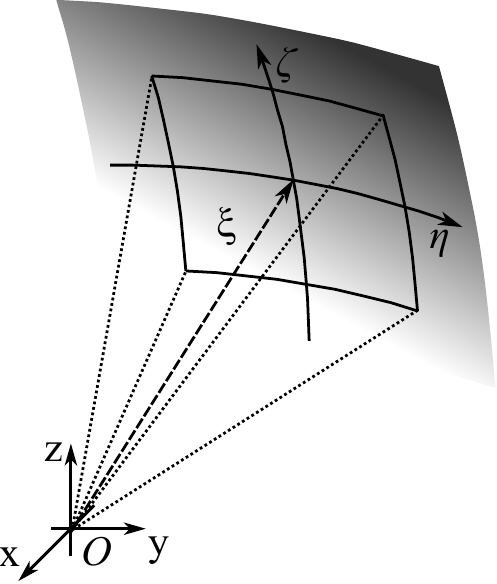}}\hfill{}

\caption{Scaled boundary coordinates $\xi$, $\eta$ and $\zeta$\label{fig:Shape_function}}
\end{figure}

The surface elements that are required to discretize the boundary
of the SBFEM volume element are conventional finite elements. Triangular
and quadrilateral elements in their local coordinate systems are shown
in Fig.~\ref{fig:Shape_fun_a} and Fig.~\ref{fig:Shape_fun_b},
respectively. The shape functions of a surface element are assembled
in the matrix $\mathrm{\mathbf{N}}\!\left(\eta,\zeta\right)$, where
$\eta$ and $\zeta$ are the local coordinates of the element. As
illustrated in Fig.~\ref{fig:Shape_fun_c}, a dimensionless radial
coordinate $\xi$ spanning $\left[0,1\right]$ from the scaling center
$O$ to the boundary is introduced for the purpose of scaling.

In the scaled boundary coordinates $\left(\xi,\eta,\zeta\right)$,
a point $\left(\hat{x},\hat{y},\hat{z}\right)$ inside the domain
can be defined as\begin{subequations}\label{subeq:xyz_hat} 
\begin{align}
\hat{x}\!\left(\xi,\eta,\zeta\right) & =\xi x\!\left(\eta,\zeta\right)=\xi\mathrm{\mathbf{N}}\!\left(\eta,\zeta\right)\mathrm{\mathbf{x}},\\
\hat{y}\!\left(\xi,\eta,\zeta\right) & =\xi y\!\left(\eta,\zeta\right)=\xi\mathrm{\mathbf{N}}\!\left(\eta,\zeta\right)\mathrm{\mathbf{y}},\\
\hat{z}\!\left(\xi,\eta,\zeta\right) & =\xi z\!\left(\eta,\zeta\right)=\xi\mathrm{\mathbf{N}}\!\left(\eta,\zeta\right)\mathrm{\mathbf{z}},
\end{align}
\end{subequations} where $\mathbf{x},\mathbf{y},\mathbf{z}$ are
the nodal coordinate vectors of the surface element in Cartesian coordinates.
The corresponding Jacobian matrix on the boundary $\left(\xi=1\right)$
is 
\begin{equation}
\mathrm{\mathbf{J}}\!\left(\eta,\zeta\right)=\left[\begin{array}{ccc}
x\!\left(\eta,\zeta\right) & y\!\left(\eta,\zeta\right) & z\!\left(\eta,\zeta\right)\\
x\!\left(\eta,\zeta\right)_{,\eta} & y\!\left(\eta,\zeta\right)_{,\eta} & z\!\left(\eta,\zeta\right)_{,\eta}\\
x\!\left(\eta,\zeta\right)_{,\zeta} & y\!\left(\eta,\zeta\right)_{,\zeta} & z\!\left(\eta,\zeta\right)_{,\zeta}
\end{array}\right].\label{eq:J_bd}
\end{equation}
The unknown displacement functions $\mathrm{\mathbf{u}}\!\left(\xi\right)$
are introduced on the radial lines connecting the scaling center and
the boundary nodes 
\begin{equation}
\mathrm{\mathbf{u}}\!\left(\xi\right)=\underset{e}{\sum}\mathrm{\mathbf{u}}^{e}\!\left(\xi\right),\label{eq:u_ue}
\end{equation}
where $\underset{e}{\sum}$ indicates a standard finite element assembly
process. Within the volume element, the displacements at a point inside
a sector, formed by a surface element $e$, are interpolated as 
\begin{equation}
\mathrm{\mathbf{u}}\!\left(\xi,\eta,\zeta\right)=\mathrm{\mathbf{N}}^{\mathrm{u}}\!\left(\eta,\zeta\right)\mathrm{\mathbf{u}}^{e}\!\left(\xi\right),\label{eq:u_Nuue}
\end{equation}
where $\mathrm{\mathbf{N}}^{\mathrm{u}}$ is the shape functions written in matrix form.  
In scaled boundary coordinates, the linear differential operator $\mathrm{\mathbf{L}}$
for the strain-displacement relationship 
is expressed as~\citep{Song1997} 
\begin{equation}
\mathrm{\mathbf{L}}=\mathrm{\mathbf{b}}_{\mathrm{1}}\!\left(\eta,\zeta\right)\frac{\partial}{\partial\xi}+\frac{1}{\xi}\!\left(\mathrm{\mathbf{b}}_{\mathrm{2}}\!\left(\eta,\zeta\right)\frac{\partial}{\partial\eta}+\mathrm{\mathbf{b}}_{\mathrm{3}}\!\left(\eta,\zeta\right)\frac{\partial}{\partial\zeta}\right),\label{eq:L}
\end{equation}
where $\mathrm{\mathbf{b}}_{1}\!\left(\eta,\zeta\right)$, $\mathrm{\mathbf{b}}_{2}\!\left(\eta,\zeta\right)$
and $\mathrm{\mathbf{b}}_{3}\!\left(\eta,\zeta\right)$ are defined
in~\citep{Song2018}.

Using Eqs.~(\ref{eq:u_Nuue})
and~(\ref{eq:L}), the strains are expressed
as
\begin{equation}
\mathrm{\boldsymbol{\varepsilon}}\!\left(\xi,\eta,\zeta\right)=\mathrm{\mathbf{B}}_{\mathrm{1}}\!\left(\eta,\zeta\right)\mathrm{\mathbf{u}}^{e}\!\left(\xi\right),_{\xi}+\frac{1}{\xi}\mathrm{\mathbf{B}}_{2}\!\left(\eta,\zeta\right)\mathrm{\mathbf{u}}^{e}\!\left(\xi\right),\label{eq:eps_B1B2ue}
\end{equation}
with\begin{subequations}\label{subeq:B1B2} 
\begin{align}
\mathrm{\mathbf{B}}_{\mathrm{1}}\!\left(\eta,\zeta\right)= & \mathrm{\mathbf{b}}_{\mathrm{1}}\!\left(\eta,\zeta\right)\mathrm{\mathbf{N}}^{\mathrm{u}}\!\left(\eta,\zeta\right),\label{eq:B1}\\
\mathrm{\mathbf{B}}_{\mathrm{2}}\!\left(\eta,\zeta\right)= & \mathrm{\mathbf{b}}_{\mathrm{2}}\!\left(\eta,\zeta\right)\mathrm{\mathbf{N}}^{\mathrm{u}}\!\left(\eta,\zeta\right),_{\eta}+\mathrm{\mathbf{b}}_{\mathrm{3}}\!\left(\eta,\zeta\right)\mathrm{\mathbf{N}}^{\mathrm{u}}\!\left(\eta,\zeta\right),_{\zeta}.\label{eq:B2}
\end{align}
\end{subequations} Assuming linear elasticity, the stresses are equal
to 
\begin{equation}
\mathrm{\boldsymbol{\sigma}}\!\left(\xi,\eta,\zeta\right)=\mathrm{\mathbf{D}}\!\left(\mathrm{\mathbf{B}}_{\mathrm{1}}\!\left(\eta,\zeta\right)\mathrm{\mathbf{u}}^{e}\!\left(\xi\right),_{\xi}+\frac{1}{\xi}\mathrm{\mathbf{B}}_{2}\!\left(\eta,\zeta\right)\mathrm{\mathbf{u}}^{e}\!\left(\xi\right)\right),\label{eq:sig_B1B2ue}
\end{equation}
where $\mathrm{\mathbf{D}}$ is the elasticity matrix.

\subsection{Solution of the scaled boundary finite element equation for elasto-statics\label{subsec:Solution}}

The scaled boundary finite element equation in displacements can be
derived based on Galerkin\textquoteright s method or the virtual work
principle, which were reported in Refs.~\citep{Song1997} and~\citep{Deeks2002},
respectively. For each volume element, the equation can be written
as 
\begin{equation}
\mathrm{\mathbf{E}_{0}}\xi^{2}\mathrm{\mathbf{u}}\!\left(\xi\right),_{\xi\xi}+\left(2\mathbf{E}_{0}-\mathbf{E}_{1}+\mathbf{E}_{1}^{\mathrm{T}}\right)\xi\mathrm{\mathbf{u}}\!\left(\xi\right),_{\xi}+\left(\mathbf{E}_{1}^{\mathrm{T}}-\mathrm{\mathbf{E}_{2}}\right)\mathrm{\mathbf{u}}\!\left(\xi\right)=\boldsymbol{0}.\label{eq:SBFE}
\end{equation}
The coefficient matrices $\mathbf{E}_{0}^{e}$, $\mathbf{E}_{1}^{e}$
and $\mathbf{E}_{2}^{e}$ of a surface element $e$ are obtained by
\begin{subequations}\label{subeq:E0E1E2} 
\begin{align}
\mathbf{E}_{0}^{e}= & \int_{e}\mathrm{\mathbf{B}_{1}^{\mathrm{T}}\mathrm{\mathbf{D}}\mathbf{B}_{1}}\mathrm{|\mathrm{\mathbf{J}}|d\eta d\zeta},\\
\mathbf{E}_{1}^{e}= & \int_{e}\mathrm{\mathbf{B}_{2}^{\mathrm{T}}\mathrm{\mathbf{D}}\mathbf{B}_{1}}\mathrm{|\mathrm{\mathbf{J}}|d\eta d\zeta},\\
\mathbf{E}_{2}^{e}= & \int_{e}\mathrm{\mathbf{B}_{2}^{\mathrm{T}}\mathrm{\mathbf{D}}\mathbf{B}_{2}}\mathrm{|\mathrm{\mathbf{J}}|d\eta d\zeta}.
\end{align}
\end{subequations} They are assembled to form the coefficient matrices
$\mathbf{E}_{0}$, $\mathbf{E}_{1}$ and $\mathbf{E}_{2}$ of a volume
element in Eq.~(\ref{eq:SBFE}) according to the element connectivity
data. This procedure is identical to that known from the FEM. According
to~\citet{Song2004}, the internal nodal forces $\mathrm{\mathbf{q}}\!\left(\xi\right)$
on a surface with a constant $\xi$ can be formulated as 
\begin{equation}
\mathrm{\mathbf{q}}\!\left(\xi\right)=\xi\!\left(\mathbf{E}_{0}\xi\mathrm{\mathbf{u}}\!\left(\xi\right),_{\xi}+\mathbf{E}_{1}^{\mathrm{T}}\mathrm{\mathbf{u}}\!\left(\xi\right)\right).\label{eq:q_E0E1}
\end{equation}
The order of the differential equations in Eq.~(\ref{eq:SBFE}) can
be reduced by one. 
The solution obtained
from eigenvalue decomposition is given in
the following form\begin{subequations}\label{subeq:u_q_phixic} 
\begin{alignat}{1}
\mathrm{\mathbf{u}}\!\left(\xi\right) & =\mathbf{\mathrm{\Phi}}_{\mathrm{u1}}\xi^{\mathbf{-\boldsymbol{\lambda}}-0.5\mathrm{\mathbf{I}}}\mathbf{c_{\mathrm{1}}},\\
\mathrm{\mathbf{q}}\!\left(\xi\right) & =\mathbf{\mathrm{\Phi}}_{\mathrm{q1}}\xi^{-\mathbf{\boldsymbol{\lambda}}+0.5\mathrm{\mathbf{I}}}\mathbf{c_{\mathrm{1}}},
\end{alignat}
\end{subequations} where $-\boldsymbol{\lambda}$ is the
eigenvalues, $\mathbf{\mathbf{\mathrm{\Phi}}_{\mathrm{u1}}}$ and 
$\mathbf{\mathbf{\mathrm{\Phi}}_{\mathrm{q1}}}$ are the sub-matrices
of the eigenvector matrix, while $\mathbf{c}_{1}$ and $\mathbf{c}_{2}$
denote the integration constants. On the boundary $\left(\xi=1\right)$, the stiffness
matrix of the volume element ($\mathrm{\mathbf{q}}\!\left(\xi=1\right)=\mathbf{K}\mathrm{\mathbf{u}}\!\left(\xi=1\right)$)
is obtained by eliminating the integration constants $\mathbf{c}_{1}$
as 
\begin{equation}
\mathbf{K}=\mathbf{\mathbf{\mathrm{\Phi}}_{\mathrm{q1}}}\mathbf{\mathbf{\mathrm{\Phi}}}_{\mathbf{\mathrm{u1}}}^{-1}.\label{eq:K}
\end{equation}
In order to obtain the element mass matrix, a coefficient matrix $\mathbf{M}_{0}$
is introduced
\begin{align}
\mathbf{M}_{0} & =\underset{e}{\sum}\mathbf{M}_{0}^{e},
\end{align}
where the element coefficient matrix $\mathbf{M}_{0}^{e}$ is defined
as
\begin{align}
\mathbf{M}_{0}^{e} & =\int_{e}\mathrm{\mathbf{N}}^{\mathrm{u}}\!\left(\eta,\zeta\right)^{\mathrm{T}}\rho\mathrm{\mathbf{N}}^{\mathrm{u}}\!\left(\eta,\zeta\right)|\mathrm{\mathbf{J}}|\mathrm{d}\eta\mathrm{d}\zeta.
\end{align}
An abbreviation $\mathbf{m}_{0}$ is introduced as
\begin{align}
\mathbf{m}_{0} & =\mathbf{\mathrm{\Phi}}_{\mathrm{u1}}^{\mathrm{T}}\mathbf{M}_{0}\mathbf{\mathrm{\Phi}}_{\mathrm{u1}}.
\end{align}
To integrate analytically along the radial direction, a matrix $\mathbf{m}$
is constructed, the components of which can be computed analytically
based on the matrix $\mathbf{m}_{0}$
\begin{align}
\mathrm{m}_{ij} & =\int_{0}^{1}\xi^{\lambda_{i}}\mathrm{m}_{0ij}\xi^{\lambda_{j}}\xi^{2}\mathrm{d}\xi\nonumber \\
 & =\frac{\mathrm{m}_{0ij}}{\lambda_{i}+\lambda_{j}+3},
\end{align}
and the consistent mass matrix $\mathbf{M}$ is written as
\begin{align}
\mathbf{M} & =\mathbf{\mathrm{\Phi}}_{\mathrm{u1}}^{-\mathrm{T}}\mathbf{m}\mathbf{\mathrm{\Phi}}_{\mathrm{u1}}^{-1}.
\end{align}

\section{Explicit dynamics\label{sec:Dynamics}}

In this section, the central difference equations for explicit dynamics
are presented, including the discussion of an SBFEM-specific mass
lumping technique and the application of octree cell patterns. The
explicit solver is first presented in a serial version, which lays
the foundation for the parallel explicit solver introduced in Section~\ref{sec:Parallel}.

\subsection{Explicit time integration\label{subsec:Time}}

In our implementation of the central difference method (CDM), a fixed
time step size is assumed for the sake of simplicity of the algorithm.
However, it should be kept in mind that adaptive time-stepping techniques
would also be possible. The displacement vector of all degrees of
freedom (DOFs) at the $n$-th time step is referred to as $\mathbf{U}_{n}$.
The velocity and acceleration vectors are the derivatives of the displacement
vector with respect to time and are denoted by $\dot{\mathbf{U}}_{n}$
and $\ddot{\mathbf{U}}_{n}$, respectively. The equilibrium equation
at the $n$-th time step using Newton's second law can be written
as
\begin{equation}
\mathbf{M}\ddot{\mathbf{U}}_{n}+\mathbf{C}\dot{\mathbf{U}}_{n}+\mathbf{R}_{n}^{\mathrm{int}}=\mathbf{R}_{n}^{\mathrm{ext}},\label{eq:Equ}
\end{equation}
 where $\mathbf{M}$ is the mass matrix, $\mathbf{C}$ denotes the
damping matrix, $\mathbf{R}_{n}^{\mathrm{ext}}$ represents the external
load vector, and the internal force vector for linear analyses is
given as
\begin{equation}
\mathbf{R}_{n}^{\mathrm{int}}=\mathbf{K}\mathbf{U}_{n}.\label{eq:Rint}
\end{equation}
 Using the CDM, the derivatives of $\mathbf{U}_{n}$ can be approximated
as\begin{subequations}\label{subeq:Dndot} 
\begin{align}
\dot{\mathbf{U}}_{n} & =\frac{1}{2\Delta t}\!\left(\mathbf{U}_{n+1}-\mathbf{U}_{n-1}\right),\\
\ddot{\mathbf{U}}_{n} & =\frac{1}{\Delta t^{2}}\!\left(\mathbf{U}_{n+1}-2\mathbf{U}_{n}+\mathbf{U}_{n-1}\right),
\end{align}
\end{subequations}where $\Delta t$ denotes the time step. In the
next step, we substitute Eqs.~(\ref{subeq:Dndot}) into Eq.~(\ref{eq:Equ}),
\begin{equation}
\frac{1}{\Delta t^{2}}\mathbf{M}\!\left(\mathbf{U}_{n+1}-2\mathbf{U}_{n}+\mathbf{U}_{n-1}\right)+\frac{1}{2\Delta t}\mathbf{C}\!\left(\mathbf{U}_{n+1}-\mathbf{U}_{n-1}\right)+\mathbf{R}_{n}^{\mathrm{int}}=\mathbf{R}_{n}^{\mathrm{ext}},\label{eq:EquDn}
\end{equation}
which can be solved for the displacement at the next time step, i.e.,
$\mathbf{U}_{n+1}$, expressed as
\begin{align}
\left(\frac{1}{\Delta t^{2}}\mathbf{M}+\frac{1}{2\Delta t}\mathbf{C}\right)\mathbf{U}_{n+1} & =\mathbf{R}_{n}^{\mathrm{ext}}-\mathbf{R}_{n}^{\mathrm{int}}+\frac{2}{\Delta t^{2}}\mathbf{M}\mathbf{U}_{n}-\!\left(\frac{1}{\Delta t^{2}}\mathbf{M}-\frac{1}{2\Delta t}\mathbf{C}\right)\mathbf{U}_{n-1}.\label{eq:Dn1}
\end{align}
If mass-proportional damping is considered, 
\begin{equation}
\mathbf{C}=\alpha\mathbf{M},\label{eq:C}
\end{equation}
where $\alpha$ is a constant, Eq.~(\ref{eq:Dn1}) can be written
as
\begin{align}
\mathbf{U}_{n+1} & =\frac{\Delta t^{2}\mathbf{M}^{-1}\!\left(\mathbf{R}_{n}^{\mathrm{ext}}-\mathbf{R}_{n}^{\mathrm{int}}\right)+2\mathbf{U}_{n}-\!\left(1-\frac{\alpha}{2}\Delta t\right)\mathbf{U}_{n-1}}{1+\frac{\alpha}{2}\Delta t}.\label{eq:DnM}
\end{align}
We observe that the displacement at the next step is only related
to the displacement of the current and previous time steps. To start
the procedure, a fictitious point $\mathbf{U}_{-1}$ is added, which
is calculated as
\begin{equation}
\mathbf{U}_{-1}=\mathbf{U}_{0}-\Delta t\dot{\mathbf{U}}_{0}+\frac{\Delta t^{2}}{2}\ddot{\mathbf{U}}_{0}.\label{eq:D0}
\end{equation}
One important drawback of explicit methods is that they are generally
only conditionally stable, which means there exists a critical time
step $\Delta t_{\mathrm{cr}}$ which should not be exceeded, otherwise
the results will diverge. Thus, they are often only used in high-frequency
problems where a small time step size is already demanded by the physics
of the problem. The exact value of $\Delta t_{\mathrm{cr}}$ involves
the calculation of the maximum natural frequency of the whole structure,
which is time and memory consuming. In our implementation, the stability
limit is estimated by the maximum frequency on element level, i.e.,
the critical time step (in the CDM) is
\begin{equation}
\Delta t_{\mathrm{cr}}=\frac{2}{\omega_{\max}},\label{eq:tcr}
\end{equation}
where $\omega_{\max}$ is the maximum frequency of all elements~\citep{Cook1989,De2016}.

\subsection{Implementation using octree mesh\label{subsec:Octree}}

In this section, we introduce the mesh generation technique based
on octree algorithm, which is highly complementary to the SBFEM. Unique
octree cell patterns are discussed, which enables an efficient calculation
of nodal force vector without the assembly of global stiffness matrix.

\subsubsection{Balanced octree mesh for SBFEM}

\begin{figure}
\hfill{}\subfloat[A balanced octree mesh\label{fig:Octree_mesh}]{\qquad{}\includegraphics[width=0.2\textwidth]{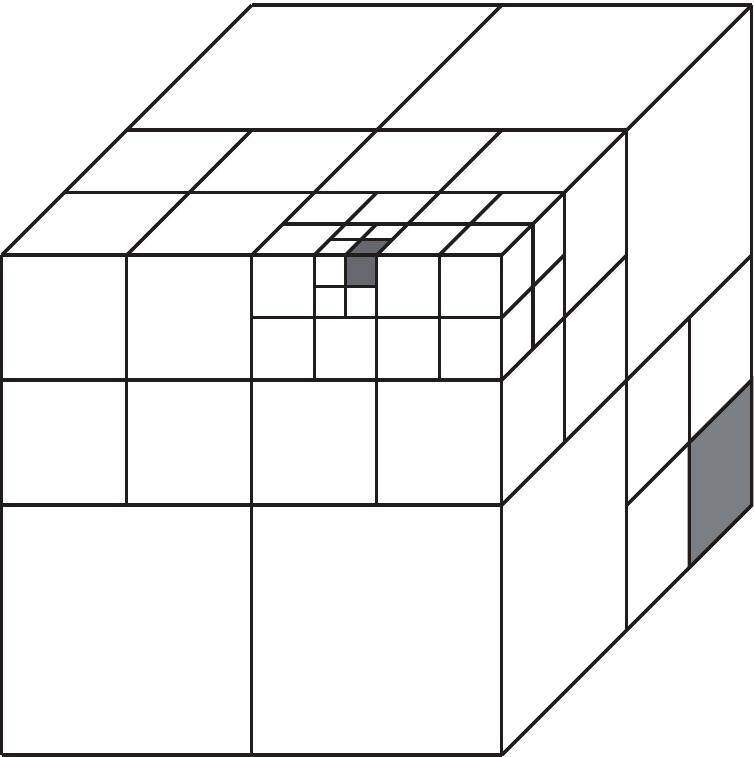}\qquad{}

}\hfill{}\subfloat[A balanced octree cell indicating all possible corner ($\bullet$),
mid-edge ($\circ$) and center ($+$) nodes\label{fig:Octree_cell}]{\qquad{}\includegraphics[width=0.2\textwidth]{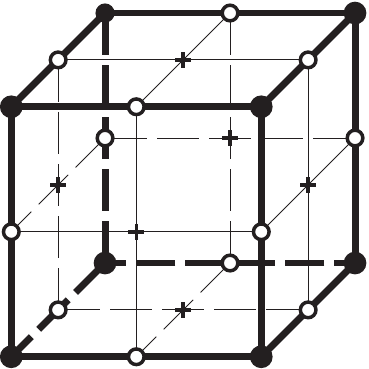}\qquad{}

}\hfill{}

\caption{Example of balanced octree mesh and octree cell patterns~\citep{Song2018}\label{fig:Octree_mesh_cell}}
\end{figure}

The octree algorithm is a highly-efficient hierarchical-tree based
technique for mesh generation. It discretizes a problem domain into
cells by recursively bisecting the cell edges until some specified
stopping criteria are met~\citep{Saputra2017a}. An example of octree
mesh in shown in Fig.~\ref{fig:Octree_mesh}, in which each octree
cell is modeled as a scaled boundary finite element. An octree mesh
is usually balanced by limiting the length ratio between two adjacent
cells to $2$ ($2:1$ rule). All the possible node locations of a
balanced octree cell is indicated in Fig.~\ref{fig:Octree_cell}.
There are $2^{12}=4096$ patterns in total~\citep[Chapter 9,][]{Song2018}.

\begin{figure}[tb]
\hfill{}\subfloat[$n_{\mathrm{h}}=0$]{\includegraphics[width=0.15\textwidth]{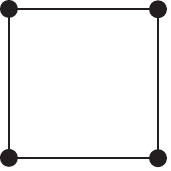}}\hfill{}\subfloat[$n_{\mathrm{h}}=1$]{\includegraphics[width=0.15\textwidth]{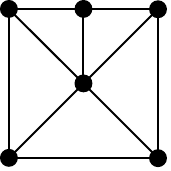}}\hfill{}\subfloat[$n_{\mathrm{h}}=2$ adjacent edges]{\includegraphics[width=0.15\textwidth]{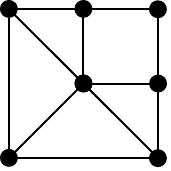}}\hfill{}

\hfill{}\subfloat[$n_{\mathrm{h}}=2$ opposite edges]{\includegraphics[width=0.15\textwidth]{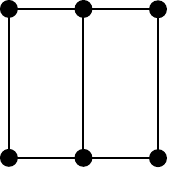}}\hfill{}\subfloat[$n_{\mathrm{h}}=3$]{\includegraphics[width=0.15\textwidth]{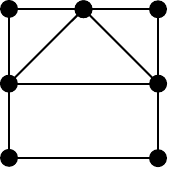}}\hfill{}\subfloat[$n_{\mathrm{h}}=4$]{\includegraphics[width=0.15\textwidth]{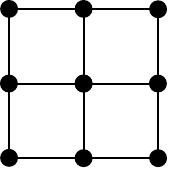}}\hfill{}

\caption{Discretization of the polygonal faces on the boundary of the cubic
octree volume elements. The number of 'hanging nodes' on the edges
for each discretization pattern is indicated by $n_{\mathrm{h}}$.\label{fig:OctreeDist}}
\end{figure}

When considering rotations, the faces of the cubic octree cells have
only six unique node arrangements as shown in Fig.~\ref{fig:OctreeDist},
depending on the number of mid-edge nodes $n_{\mathrm{h}}$ and their
locations. The faces are treated as polygons in the SBFEM and discretized
using rectangular and isosceles triangular elements, which are considered
as high-quality elements.

\subsubsection{Unique octree cell patterns}

The 4096 patterns can be further transformed into 144 unique cases
using rotation and mirroring operators. The 90 degree rotation matrices
about the $x$, $y$ and $z$ axes are
\begin{equation}
\mathbf{T}_{\mathrm{x}}=\left[\begin{array}{ccc}
1 & 0 & 0\\
0 & 0 & 1\\
0 & -1 & 0
\end{array}\right],\enskip\mathbf{T}_{\mathrm{y}}=\left[\begin{array}{ccc}
0 & 0 & 1\\
0 & 1 & 0\\
-1 & 0 & 0
\end{array}\right],\enskip\mathbf{T}_{\mathrm{z}}=\left[\begin{array}{ccc}
0 & 1 & 0\\
-1 & 0 & 0\\
0 & 0 & 1
\end{array}\right],\label{eq:Rotate}
\end{equation}
and the mirror matrices are
\begin{equation}
\mathbf{R}_{\mathrm{x}}=\left[\begin{array}{ccc}
-1 & 0 & 0\\
0 & 1 & 0\\
0 & 0 & 1
\end{array}\right],\enskip\mathbf{R}_{\mathrm{y}}=\left[\begin{array}{ccc}
1 & 0 & 0\\
0 & -1 & 0\\
0 & 0 & 1
\end{array}\right],\enskip\mathbf{R}_{\mathrm{z}}=\left[\begin{array}{ccc}
1 & 0 & 0\\
0 & 1 & 0\\
0 & 0 & -1
\end{array}\right].\label{eq:Reflect}
\end{equation}
These transformation matrices can be arranged and combined to form
48 unique transformations of an octree cell in 3D. Alternatively,
the 48 transformations can be constructed directly by filling $-1$
or $1$ to each row and column of a $3\times3$ empty matrix, i.e.,
there is exactly one nonzero value ($1$ or $-1$) in each row and
each column. The number of transformations is equal to $3\times2\times1\times2^{3}=48$.
{The 48  transformations involve only permutations of nodal numbers in an element, which can be easily implemented using an indexing vector.}
Each of the 4096 patterns can be transformed into one of the 144 unique
patterns using one of the 48 transformations. All the unique octree
patterns are presented in~\ref{sec:Octree_pattern}.

{The 144 patterns are determined solely by the presence or absence of mid-edge nodes, which is independent of the polyhedral element formulation selected. In our implementation, the octree meshes with triangulated boundaries are handled by the SBFEM, however it is important to note that the same approach can also be applied to other polyhedral elements with well-conditioned lumped mass matrices. The use of element patterns significantly improves computational efficiency, in terms of time and memory requirements, especially on a modern high-performance computing system.}

\begin{figure}[tb]
\hfill{}\subfloat[Octree master cell: $\mathrm{c}_{1}$\label{fig:OctTrans_a}]{\includegraphics[width=0.05\textwidth]{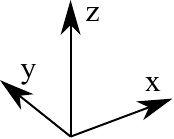}\enskip{}\includegraphics[width=0.2\textwidth]{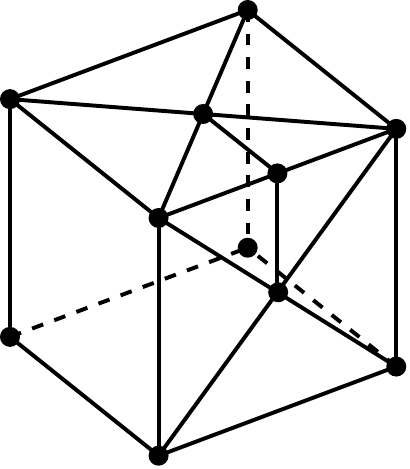}}\hfill{}\subfloat[Octree cell with a pattern identical to $\mathrm{c}_{1}$\label{fig:OctTrans_b}]{\includegraphics[width=0.2\textwidth]{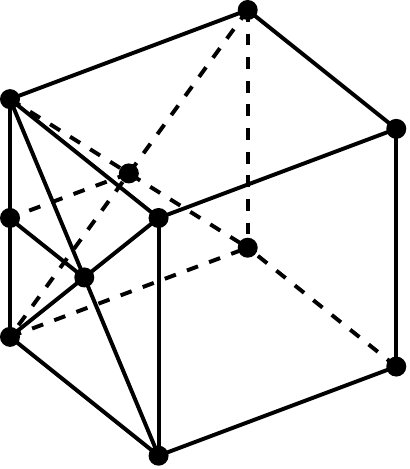}}\hfill{}\subfloat[Transformation matrix $\mathbf{Q}_{1}=\mathbf{T}_{\mathrm{x}}\mathbf{T}_{\mathrm{z}}$\label{fig:OctTrans_c}]{%
\begin{minipage}[b][1\totalheight][t]{0.25\columnwidth}%
$\mathbf{Q}_{1}=\left[\begin{array}{ccc}
0 & 1 & 0\\
0 & 0 & 1\\
1 & 0 & 0
\end{array}\right]$%
\end{minipage}}\hfill{}

\hfill{}\subfloat[Octree master cell: $\mathrm{c}_{2}$\label{fig:OctTrans_d}]{\includegraphics[width=0.05\textwidth]{Transform_0}\enskip{}\includegraphics[width=0.2\textwidth]{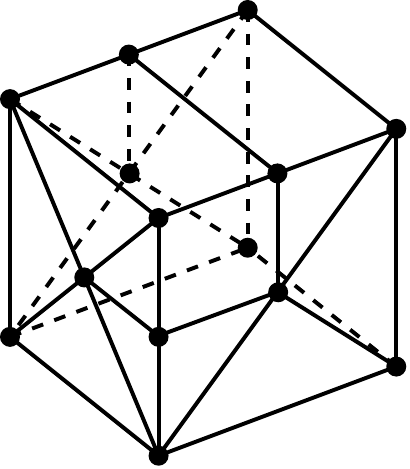}}\hfill{}\subfloat[Octree cell with a pattern identical to $\mathrm{c}_{2}$\label{fig:OctTrans_e}]{\includegraphics[width=0.2\textwidth]{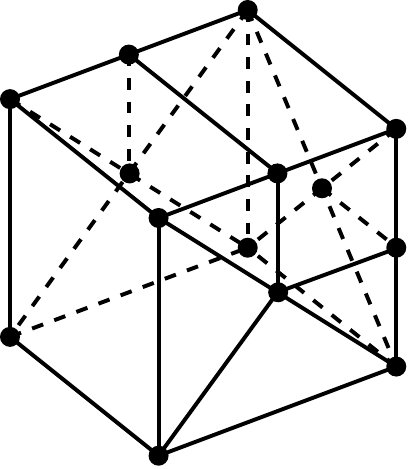}}\hfill{}\subfloat[Transformation matrix $\mathbf{Q}_{2}=\mathbf{R}_{\mathrm{x}}$\label{fig:OctTrans_f}]{%
\begin{minipage}[b][1\totalheight][t]{0.25\columnwidth}%
$\mathbf{Q}_{2}=\left[\begin{array}{ccc}
-1 & 0 & 0\\
0 & 1 & 0\\
0 & 0 & 1
\end{array}\right]$%
\end{minipage}}\hfill{}

\caption{Transformation of octree cells\label{fig:OctTrans}}
\end{figure}

Two sample transformations are illustrated in Fig.~\ref{fig:OctTrans}.
The octree cell in Fig.~\ref{fig:OctTrans_b} can be transformed
into the corresponding master cell $\mathrm{c}_{1}$ in Fig.~\ref{fig:OctTrans_a}
by rotating 90 degrees about the $x$ axis and then rotating 90 degrees
about the $z$ axis (following the right hand rule as positive rotation
direction). The rotation matrix can be constructed as $\mathbf{Q}_{1}=\mathbf{T}_{\mathrm{x}}\mathbf{T}_{\mathrm{z}}$.
Similarly, the octree cell in Fig.~\ref{fig:OctTrans_e} can be transformed
into $\mathrm{c}_{2}$ in Fig.~\ref{fig:OctTrans_d} by mirroring
about the $y$-$z$ plane using $\mathbf{Q}_{2}=\mathbf{R}_{\mathrm{x}}$.

When calculating the nodal force vector of an octree cell, the nodal
displacement vector is transformed using the same transformation matrix
$\mathbf{Q}$, which is then multiplied by the stiffness matrix of
the corresponding master cell. The obtained nodal force vector must
be transformed back using the inverse of transformation matrix, $\mathbf{Q}^{-1}$.

\subsubsection{Implementation exploiting octree cell patterns\label{subsec:Pattern}}

As an alternative to Eq.~(\ref{eq:Rint}), $\mathbf{R}_{n}^{\mathrm{int}}$
can be calculated EBE so that the assembly of the global stiffness
matrix $\mathbf{K}$ is not required.

\begin{equation}
\mathbf{R}_{n}^{\mathrm{int}}=\stackrel[e=1]{n_{\mathrm{e}}}{\sum}\mathbf{K}_{e}\mathbf{U}_{e,n},\label{eq:Rinte}
\end{equation}
where $\sum$ represents the assembly process according to the element
connectivity, $n_{\mathrm{e}}$ is the total number of elements, $\mathbf{K}_{e}$
is the element stiffness matrix of volume element $e$, and $\mathbf{U}_{e,n}$
is the nodal displacement vector of the element at the $n$-th time
step.

In an octree mesh, elements of the same shape (see~\ref{sec:Octree_pattern})
can be grouped together, and therefore, their nodal force vectors
can be calculated in one step using a single matrix-multiplication.
Eq.~(\ref{eq:Rinte}) can be rewritten as
\begin{equation}
\mathbf{R}_{n}^{\mathrm{int}}=\stackrel[p=1]{n_{\mathrm{p}}}{\sum}\mathbf{K}_{p}\!\left[\begin{array}{cccc}
\mathbf{U}_{p1} & \mathbf{U}_{p2} & \cdots & \mathbf{U}_{pk}\end{array}\right]_{n}\mathrm{diag}\!\left(\mathbf{S}\right),\label{eq:Rintep}
\end{equation}
where $n_{\mathrm{p}}$ is the number of unique patterns in the octree
mesh, $\mathbf{K}_{p}$ is the stiffness matrix of a master element
of pattern $p$ with unit size, $\mathbf{U}_{pi}$ is the nodal displacement
vector of the $i$-th element of pattern $p$, and $\mathrm{diag}\!\left(\mathbf{S}\right)$
is a diagonal matrix containing the edge lengths of each element of
pattern $p$. Each column obtained from the right-hand side will be
assembled into $\mathbf{R}_{n}^{\mathrm{int}}$ using the element
connectivity data. As mentioned before, the spatial discretization
is based on a balanced octree mesh. This holds two important advantages:
(i) all surface elements are well-shaped and (ii) for a cubic volume
element only a limited number of surface patterns exist. These are
the prerequisites to exploit a pre-computation technique. To this
end, the stiffness and mass matrices of all master elements corresponding
to the different patterns $p$ can be computed beforehand. For the
pre-computation strategy, a unit size of the master element is assumed.
Due to the limited surface element shapes, the stiffness and mass
matrices for the master element only need to be scaled by suitable
factors including information on the element size and material properties.

\begin{figure}[tb]
\hfill{}\subfloat[Master element\label{fig:Scale_1}]{\includegraphics{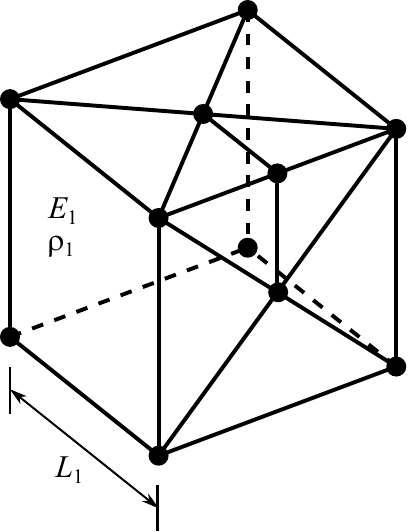}}\hfill{}\subfloat[Scaled element\label{fig:Scale_2}]{\includegraphics{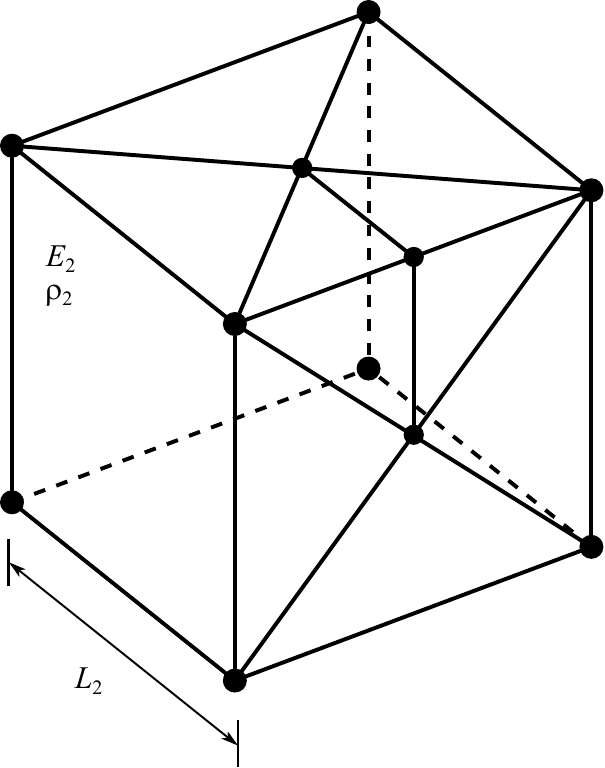}}\hfill{}

\caption{Scaling of an octree element\label{fig:Scale}}
\end{figure}

The stiffness matrix of an element is proportional to its Young's
modulus and edge length. On the other hand, the mass matrix of an
element is proportional to the mass density and the cube of edge length.
Here, we define the ratios between Young's modulus, mass density and
edge length as $\eta_{E}=\nicefrac{E_{2}}{E_{1}}$, $\eta_{\rho}=\nicefrac{\rho_{2}}{\rho_{1}}$
and $\eta_{L}=\nicefrac{L_{2}}{L_{1}}$ (see Fig.~\ref{fig:Scale}).
Therefore, the stiffness and mass matrices, $\mathbf{K}_{2}$ and
$\mathbf{M}_{2}$, of an element can be derived easily from a master
element of same pattern and Poisson's ratio as\begin{subequations}\label{subeq:KM_scale}
\begin{equation}
\mathbf{K}_{2}=\eta_{E}\eta_{L}\mathbf{K}_{1},
\end{equation}
\begin{equation}
\mathbf{M}_{2}=\eta_{\rho}\eta_{L}^{3}\mathbf{M}_{1},
\end{equation}
\end{subequations}where $\mathbf{K}_{1}$ and $\mathbf{M}_{1}$ are
stiffness and mass matrices of the master element, respectively. In
cases where materials with different Poisson's ratios are present,
the master element matrices have be computed for each material separately.
The critical time step is estimated based on the maximum eigenfrequency
of the element in Eq.~(\ref{eq:tcr}), therefore, considering Eq.~(\ref{subeq:KM_scale}),
$\Delta t_{\mathrm{cr}}{}_{2}$ of an element can be calculated from
the known time step of the master element $\Delta t_{\mathrm{cr}}{}_{1}$,
\begin{equation}
\Delta t_{\mathrm{cr}}{}_{2}=\sqrt{\frac{\eta_{\rho}}{\eta_{E}}}\eta_{L}\Delta t_{\mathrm{cr}}{}_{1}.\label{eq:tcr_scale}
\end{equation}

\subsection{Mass lumping in the SBFEM\label{subsec:Mass}}

It can be inferred from Eq.~(\ref{eq:DnM}) that if the mass matrix
$\mathbf{M}$ is given in diagonal form, it is possible to advance
in time by simple matrix-vector products without the need for solving
a system of linear equations. The reason for this increased efficiency
lies in the trivial inversion of a diagonal matrix. In the context
of the finite and spectral element methods (SEM), different mass lumping
schemes have been reported and thoroughly assessed in Refs.~\citep{Duczek2019,Duczek2019a}.

\begin{figure}[tb]
\noindent\begin{minipage}[b][1\totalheight][t]{1\columnwidth}%
\hfill{}\subfloat[Fully populated mass matrix $\mathbf{M}$ \label{fig:MassLump_1}]{\quad{}\includegraphics{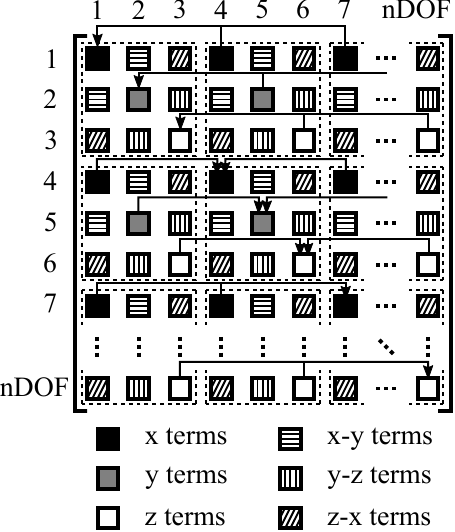}\quad{}}\hfill{}\subfloat[Block-diagonal mass matrix\label{fig:MassLump_2}]{%
\begin{minipage}[b][1\totalheight][t]{0.3\textwidth}%
\includegraphics{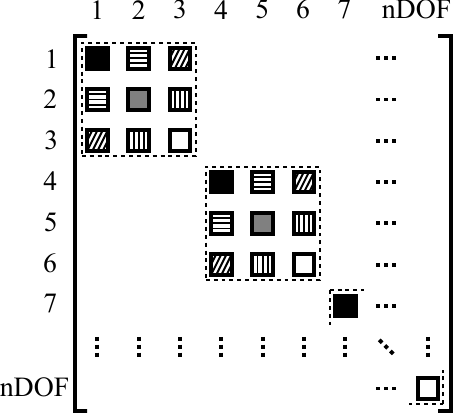}\vspace{12mm}
\end{minipage}}\hfill{}\subfloat[Diagonal mass matrix $\mathbf{M}_{\mathrm{diag}}$\label{fig:MassLump_3}]{%
\begin{minipage}[b][1\totalheight][t]{0.3\textwidth}%
\includegraphics{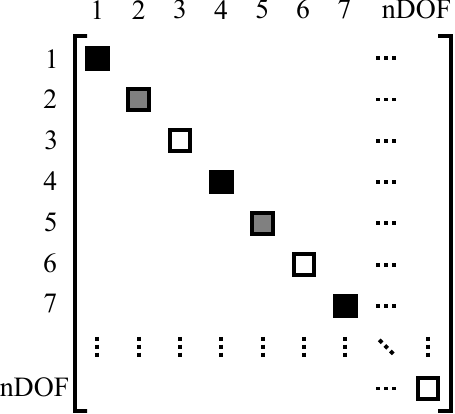}\vspace{12mm}
\end{minipage}}\hfill{}%
\end{minipage}

\caption{Lumping of the mass matrix\label{fig:MassLump}}
\end{figure}

Different from the FEM and SEM, the displacement interpolations inside
the scaled boundary finite elements are obtained semi-analytically,
and the $x$, $y$ and $z$ displacement components are coupled. In
our proposed mass lumping scheme, only those components of the mass
matrix that correspond to DOFs in the same coordinate direction are
summed up. The basic approach is illustrated in Fig.~\ref{fig:MassLump_1}
for the DOFs related to $x$ direction. The procedure is repeated
in the same fashion for the components corresponding to $y$ and $z$
directions. This method results in a block-diagonal matrix as shown
in Fig.~\ref{fig:MassLump_2}, from which a diagonal mass matrix
$\mathbf{M}_{\mathrm{diag}}$ is then extracted (Fig.~\ref{fig:MassLump_3}).
The off-diagonal terms, which sum up to zero, are ignored. The total
mass of the system is conserved in the formulation of the SBFEM. A
detailed explanation and verification of this mass lumping approach
has been reported by~\citet{Gravenkamp2020a}.

\begin{figure}
\hfill{}\subfloat[Element with 8 nodes]{\includegraphics[width=0.45\textwidth]{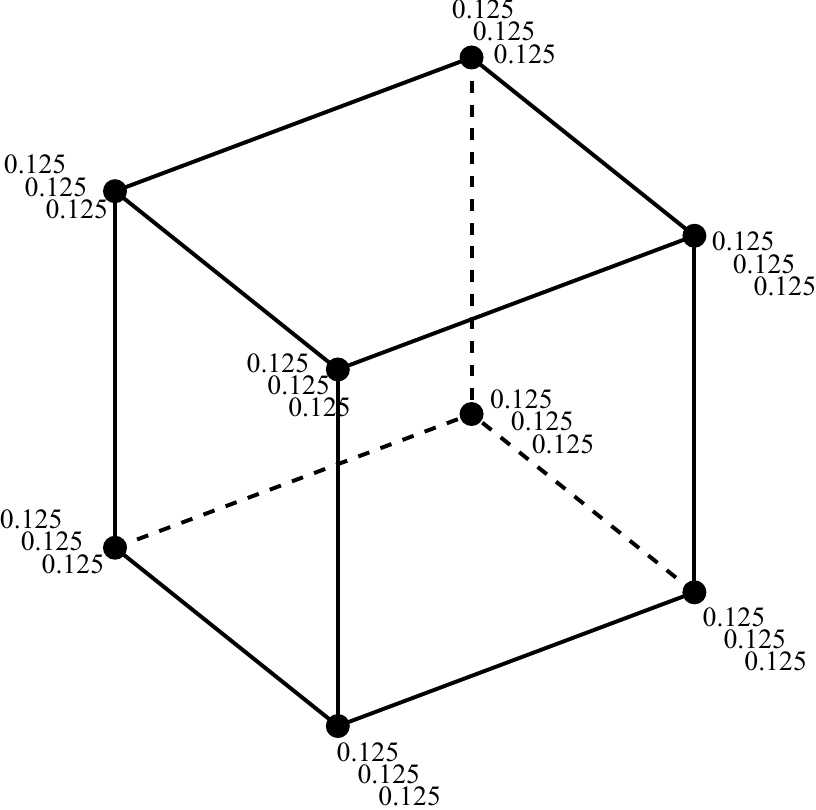}}\hfill{}\subfloat[Element with 11 nodes]{\includegraphics[width=0.45\textwidth]{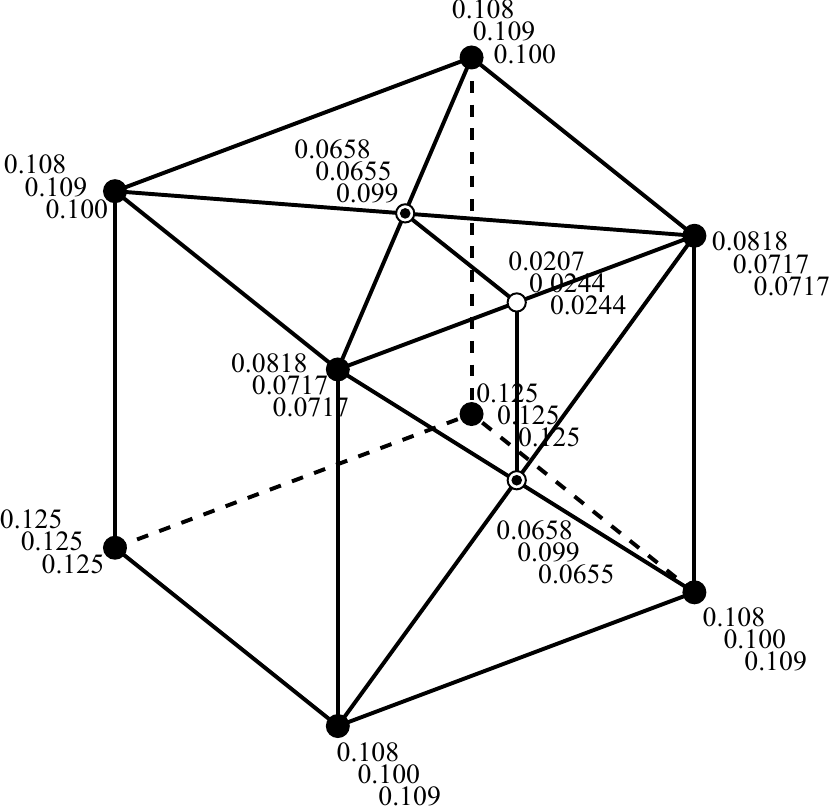}}\hfill{}

\hfill{}\subfloat[Element with 13 nodes]{\includegraphics[width=0.45\textwidth]{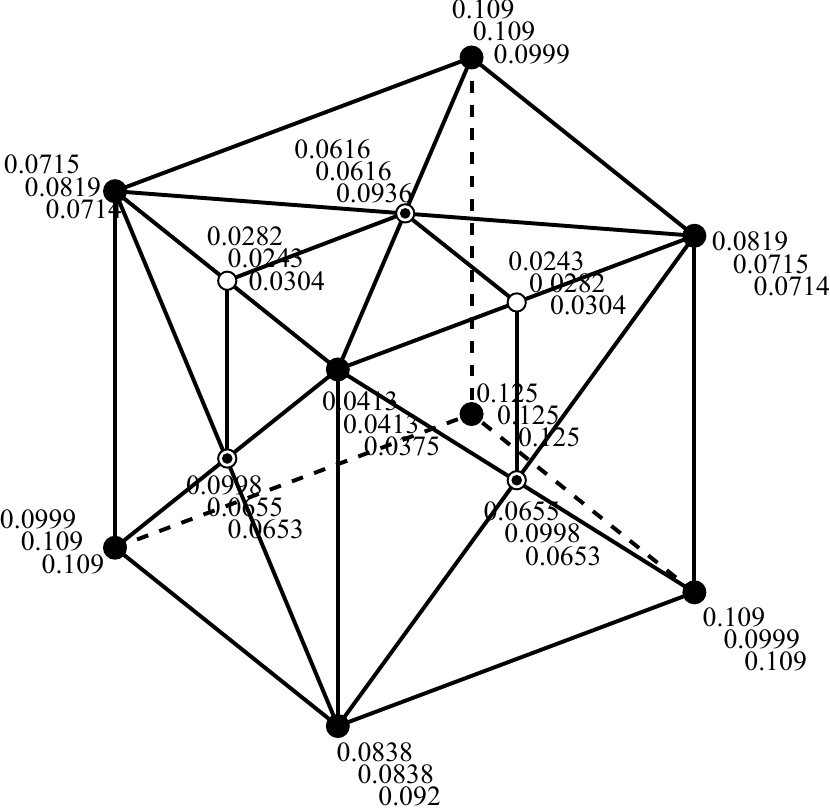}}\hfill{}\subfloat[Element with 26 nodes]{\includegraphics[width=0.45\textwidth]{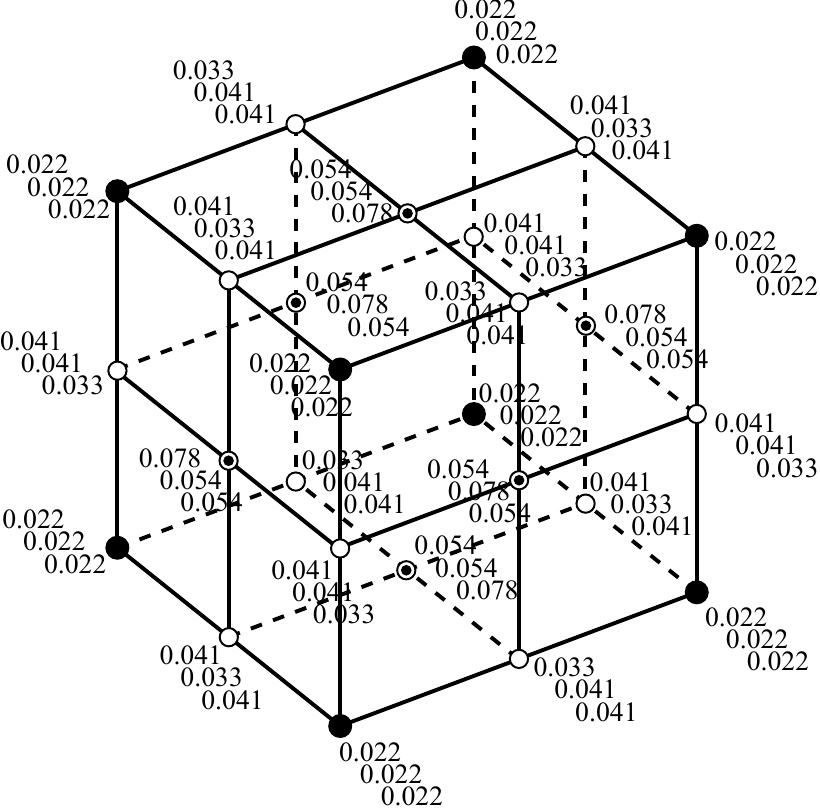}}\hfill{}

\caption{Lumped mass matrices with Poisson's ratio $\nu=0$\label{fig:LumpMass1}}
\end{figure}

\begin{figure}
\hfill{}\subfloat[Element with 8 nodes]{\includegraphics[width=0.45\textwidth]{LumpMass_01}}\hfill{}\subfloat[Element with 11 nodes]{\includegraphics[width=0.45\textwidth]{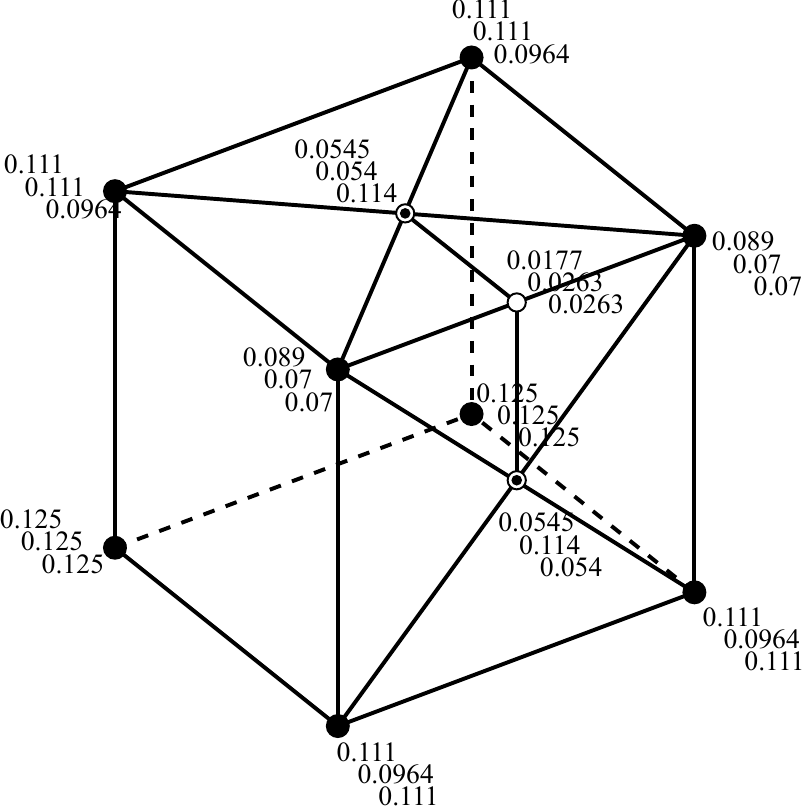}}\hfill{}

\hfill{}\subfloat[Element with 13 nodes]{\includegraphics[width=0.45\textwidth]{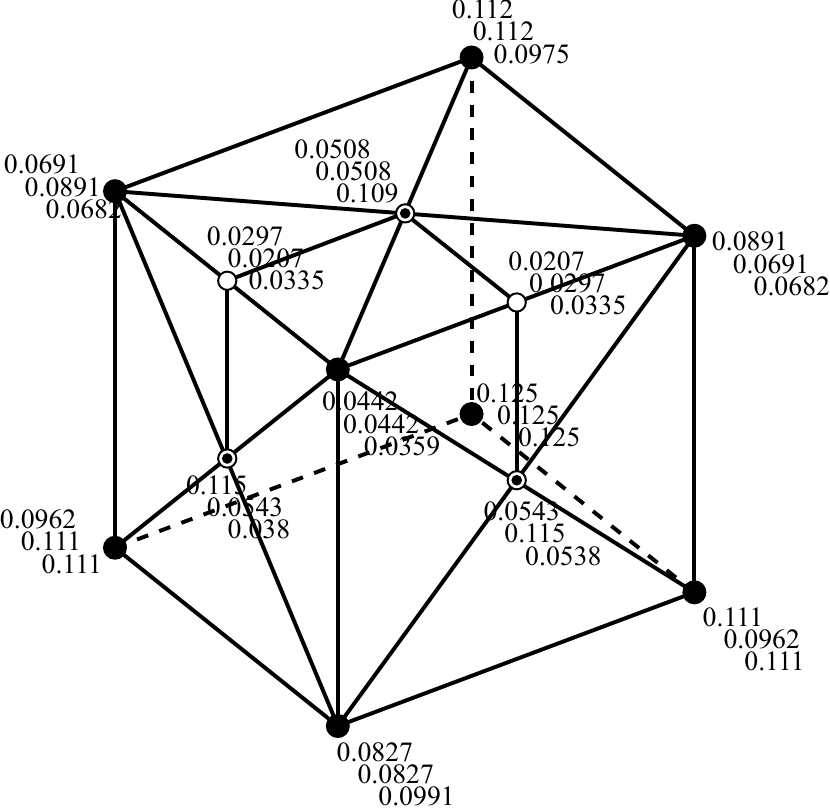}}\hfill{}\subfloat[Element with 26 nodes]{\includegraphics[width=0.45\textwidth]{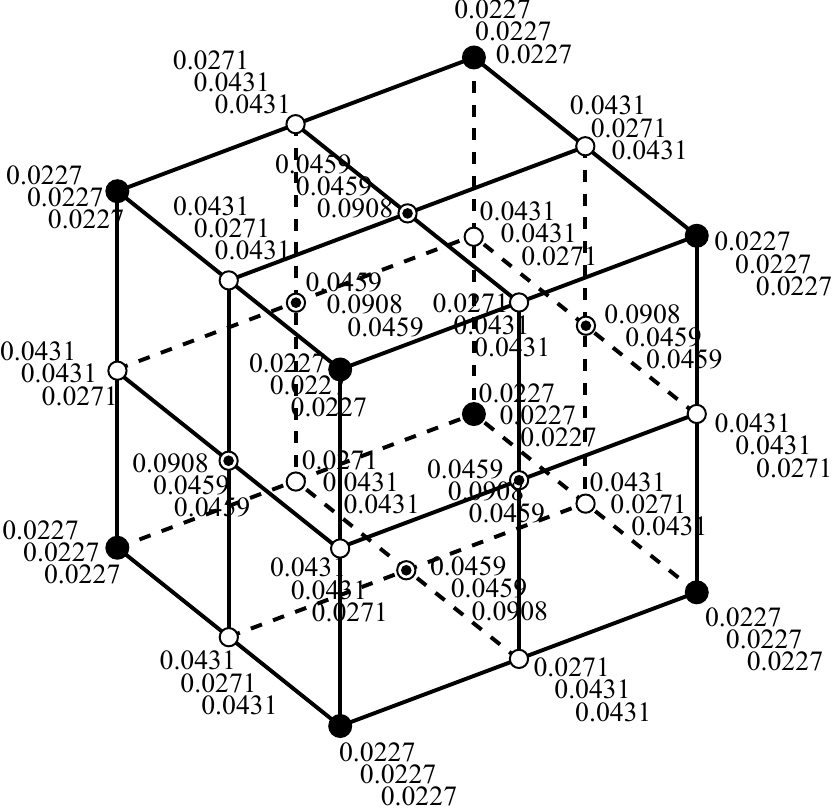}}\hfill{}

\caption{Lumped mass matrices with Poisson's ratio $\nu=0.3$\label{fig:LumpMass2}}
\end{figure}

Due to the mass matrix $\mathbf{M}$ being available in diagonal format
$\mathbf{M}_{\mathrm{diag}}$, we can re-write the expression for
the displacement vector in the next time step as
\begin{equation}
\mathbf{U}_{n+1}=\frac{\Delta t^{2}\mathbf{M}_{\mathrm{diag}}^{-1}\!\left(\mathbf{R}_{n}^{\mathrm{ext}}-\mathbf{R}_{n}^{\mathrm{int}}\right)+2\mathbf{U}_{n}-\!\left(1-\frac{\alpha}{2}\Delta t\right)\mathbf{U}_{n-1}}{1+\frac{\alpha}{2}\Delta t}.\label{eq:DnMdiag}
\end{equation}
Lumping the mass matrix avoids solving a system of linear equations
and as a result, each component in $\mathbf{U}_{n+1}$ can be calculated
independently of the other entries. Lumped mass matrices of several
octree patterns mentioned in Section~\ref{subsec:Pattern} are presented
below as ratios to the total mass of the octree cells. Poisson's ratios
are 0 and 0.3 in Fig.~\ref{fig:LumpMass1} and Fig.~\ref{fig:LumpMass2},
respectively.

\clearpage{}

\subsection{Explicit solver in serial\label{subsec:Solver_serial}}

Based on the previous sections, an explicit solver can be developed
making use of lumped mass matrices and the unique octree patterns.
The data structure of the solver is presented in Table~\ref{tab:Data_structure}.
Element stiffness and mass matrices of the 144 unique octree patterns
can be pre-computed and stored as $\mathbf{K}_{e}$ and $\mathbf{M}_{e}$.
It is noted that if multiple materials with different Poisson's ratios
are present in the model, each material will be assigned its own $\mathbf{K}_{e}$
and $\mathbf{M}_{e}$ matrices. For each specific problem, a matrix
$\mathbf{V}$ is constructed by concatenating the DOF vector of each
element, which is already arranged according to the transformation
matrix to the master cells. Furthermore, $n_{\mathrm{e}}$ is the
number of elements, and $n_{\mathrm{dof}}$ denotes the total number
of DOFs. The vector $\mathrm{\boldsymbol{\varGamma}_{D}}$ stores
the DOFs which are fixed (Dirichlet boundary conditions), and $\mathrm{\boldsymbol{\varGamma}_{N}}$
is a sparse vector containing the location and magnitude of the external
nodal forces (Neumann boundary conditions). $\mathrm{\mathbf{T}}$
is an integer vector indicating the element type, while $\mathrm{\mathbf{S}}$
is a vector containing the edge length of each element. The variables
mentioned above are constant for each problem during the calculation.
In each time step, the displacement vectors corresponding to the previous
step, current step and the next step, $\mathbf{U}_{\mathrm{p}}$,
$\mathbf{U}_{\mathrm{c}}$ and $\mathbf{U}_{\mathrm{n}}$, are updated
accordingly. $\mathbf{R}^{\mathrm{ele}}$ is a matrix containing the
nodal force vector of each element, which are grouped by element type.
$\mathbf{R}^{\mathrm{int}}$ is the internal force vector which is
assembled from $\mathbf{R}^{\mathrm{ele}}$, while $\mathbf{R}^{\mathrm{ext}}$
is the external force vector calculated at the time $t$. The pseudo-code
of the explicit solver is presented in Algorithm~\ref{alg:Solver_serial}.

\begin{table}
\caption{Data structure of explicit solver\label{tab:Data_structure}}

\begin{tabular}{|>{\centering}p{0.1\textwidth}|c|c|c|c|>{\centering}p{0.4\textwidth}|}
\hline 
 & Variable & Data type & Variable type & Size & Description\tabularnewline
\hline 
\hline 
\multirow{2}{0.1\textwidth}{Constant} & $\mathbf{K}_{e}$ & float & hyper-matrix\tablefootnote{Hyper-matrix: This term denotes that each component of the matrix
is a matrix itself. In MATLAB this can be for example realized as
a cell-array.} & 144 & Element stiffness matrices of the 144 unique octree cell patterns\tabularnewline
\cline{2-6} \cline{3-6} \cline{4-6} \cline{5-6} \cline{6-6} 
 & $\mathbf{M}_{e}$ & float & hyper-matrix & 144 & Diagonal mass matrices of the 144 unique octree cell patterns\tabularnewline
\hline 
\multirow{6}{0.1\textwidth}{Pre-compute} & $\mathbf{V}$ & integer & hyper-vector & $n_{\mathrm{e}}$ & DOF vector of each element\tabularnewline
\cline{2-6} \cline{3-6} \cline{4-6} \cline{5-6} \cline{6-6} 
 & $\boldsymbol{\varGamma}_{\mathrm{D}}$ & integer & vector & < $n_{\mathrm{dof}}$ & Dirichlet boundary condition (the DOFs which are fixed)\tabularnewline
\cline{2-6} \cline{3-6} \cline{4-6} \cline{5-6} \cline{6-6} 
 & $\boldsymbol{\varGamma}_{\mathrm{N}}$ & float & vector & $n_{\mathrm{dof}}$ & Neumann boundary condition (the location and magnitude of external
forces)\tabularnewline
\cline{2-6} \cline{3-6} \cline{4-6} \cline{5-6} \cline{6-6} 
 & $\mathbf{T}$ & integer & vector & $n_{\mathrm{e}}$ & Pattern of each element, from 1 to 144\tabularnewline
\cline{2-6} \cline{3-6} \cline{4-6} \cline{5-6} \cline{6-6} 
 & $\mathbf{S}$ & float & vector & $n_{\mathrm{e}}$ & Edge length of each element\tabularnewline
\cline{2-6} \cline{3-6} \cline{4-6} \cline{5-6} \cline{6-6} 
 & $\mathbf{M}_{\mathrm{diag}}$ & float & matrix & $n_{\mathrm{dof}}$ & Global diagonal mass matrix\tabularnewline
\hline 
\multirow{3}{0.1\textwidth}{Time stepping} & $\mathbf{U}_{\mathrm{p}}$, $\mathbf{U}_{\mathrm{c}}$, $\mathbf{U}_{\mathrm{n}}$ & float & vector & $n_{\mathrm{dof}}$ & Displacement vector of previous step, current step and next step\tabularnewline
\cline{2-6} \cline{3-6} \cline{4-6} \cline{5-6} \cline{6-6} 
 & $\mathbf{R}^{\mathrm{ele}}$ & float & hyper-vector & $n_{\mathrm{e}}$ & Internal nodal force vector of each element\tabularnewline
\cline{2-6} \cline{3-6} \cline{4-6} \cline{5-6} \cline{6-6} 
 & $\mathbf{R}^{\mathrm{ext}}$, $\mathbf{R}^{\mathrm{int}}$ & float & vector & $n_{\mathrm{dof}}$ & External and internal force vectors\tabularnewline
\hline 
\end{tabular}
\end{table}

\begin{algorithm}
\begin{algorithmic}[1]

\State{Input mesh}

\State{Load pre-computed master element stiffness and mass matrices
$\mathbf{K}_{e}$ and $\mathbf{M}_{e}$ }

\State{Identify element patterns $\mathrm{\mathbf{T}}$ and edge
lengths $\mathbf{S}$, and link to corresponding master element}

\State{Assemble global $\mathbf{V}$ matrix and group based on element
type}

\State{Assemble global diagonal mass matrix $\mathbf{M}_{\mathrm{diag}}$}

\State{Apply boundary conditions}

\State{Initialize nodal displacement vectors $\mathbf{U}_{\mathrm{p}}$,
$\mathbf{U}_{\mathrm{c}}$ and $\mathbf{U}_{\mathrm{n}}$}

\For{$t$ in $\left[0,T\right]$}

\State{Initialize global nodal force vector: $\mathbf{R}^{\mathrm{int}}=\left\{ \mathbf{0}\right\} $}

\For{$p=1:n_{\mathrm{p}}$}

\State{Assemble $\mathbf{U}_{\mathrm{c}p}$ of $p$-th pattern using
Eq.~(\ref{eq:Rintep})}

\State{Calculate $\mathbf{R}_{p}^{\mathrm{ele}}$ of $p$-th pattern:
$\mathbf{R}_{p}^{\mathrm{ele}}=\mathbf{K}_{p}\mathbf{U}_{\mathrm{c}p}\mathrm{diag}\!\left(\mathbf{S}_{p}\right)$}

\For{$e=1:n_{\mathrm{e}}$}

\State{Assemble the $e$-th column of $\mathbf{R}_{p}^{\mathrm{ele}}$
into $\mathbf{R}^{\mathrm{int}}$}

\EndFor

\EndFor

\State{Calculate external force $\mathbf{R}_{t}^{\mathrm{ext}}$
at current time}

\State{Calculate next step displacement $\mathbf{U}_{\mathrm{n}}$
using Eq.~(\ref{eq:DnM})}

\State{$\mathbf{U}_{\mathrm{p}}=\mathbf{U}_{\mathrm{c}}$; $\mathbf{U}_{\mathrm{c}}=\mathbf{U}_{\mathrm{n}}$}

\EndFor

\end{algorithmic}

\caption{Explicit solver in serial\label{alg:Solver_serial}}
\end{algorithm}

\FloatBarrier

\section{Parallel processing\label{sec:Parallel}}

Due to the fact that in explicit dynamics, the nodal force vector
can be assembled in an element-wise fashion, it is natural to exploit
parallel processing techniques to improve the efficiency of the solver.
The basic idea is to partition the mesh into different parts and assign
these parts to multiple processors. On each processor the nodal force
vector is calculated, and only the interface DOFs of each part need
to be synchronized. In this section, details regarding the partitioning
strategy and the parallel solver are presented.

\subsection{Mesh partition\label{subsec:Partition}}

\begin{figure}[tb]
\hfill{}\subfloat[A sample quadtree mesh\label{fig:MeshDual_a}]{\includegraphics{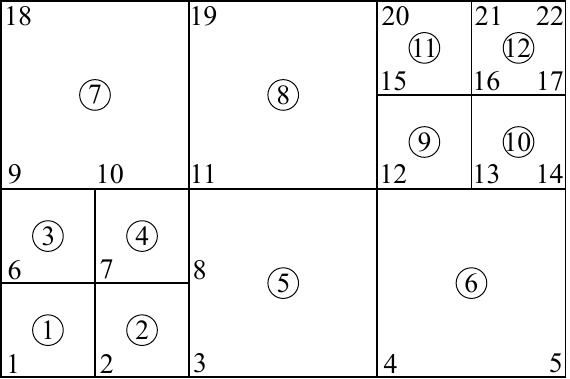}}\hfill{}\subfloat[Dual graph and its optimal partition (shaded line)\label{fig:MeshDual_b}]{\includegraphics{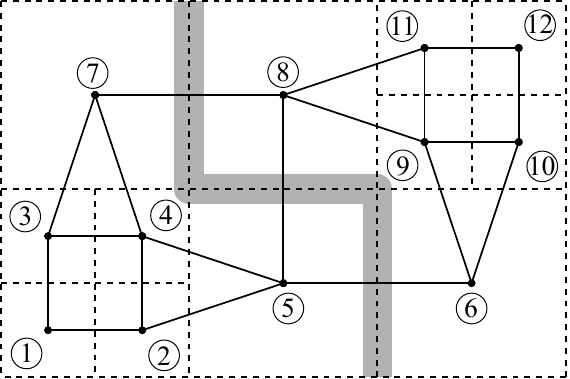}}\hfill{}

\caption{A quadtree mesh and its dual graph\label{fig:MeshDual}}
\end{figure}

Mesh partition serves a vital role in parallel processing as the quality
of partition directly affects the efficiency of the solver. There
are mainly two types of mesh partitioning approaches, namely node-
and element-cut strategies~\citep{Krysl2001}. In this work, as most
of the computational effort is spent on the element processing stage,
the node-cut strategy is adopted. There are two main considerations
that must be taken into account in order to achieve a good partitioning
of the mesh:
\begin{enumerate}
\item Size of the parts: An identical size of the individual parts ensures
that a balanced workload is achieved, i.e., all processors complete
their tasks at the same time. If that is not the case, some processors
are idle, and important resources are wasted as the speedup entirely
depends on the slowest processor. Although different patterns of elements
require a different amount of calculation, it is reasonable to estimate
the amount of work on each processor using the number of elements.
\item Interface between parts: The interface between individual parts should
be as small as possible. While data communication between processors
cannot be avoided in explicit dynamics because the nodal force vectors
need to be synchronized in each time step, the target is to minimize
the amount of data communication which is the main bottleneck in parallel
processing in terms of the attainable speedup. The amount of data
that needs to be transferred is measured by the number of nodes on
the interface.
\end{enumerate}
In order to partition a mesh, it is first converted into its dual
graph. Each element in the mesh is represented by a vertex $\mathrm{v}$
in the graph, while each interface shared by two elements is represented
by an edge $\mathrm{e}$ connecting two vertices. The set of vertices
is denoted as $V$ and the set of edges is referred to as $E$. A
typical quadtree mesh in 2D and its dual graph $G=\left\{ V,E\right\} $
are depicted in Fig.~\ref{fig:MeshDual}. An optimal partition can
be obtained by cutting through nodes $\left\{ \mathrm{n}_{4},\mathrm{n}_{12},\mathrm{n}_{11},\mathrm{n}_{19}\right\} $
and dividing the mesh into two parts with six elements each. Cutting
through $\left\{ \mathrm{n}_{3},\mathrm{n}_{8},\mathrm{n}_{11},\mathrm{n}_{12},\mathrm{n}_{15},\mathrm{n}_{20}\right\} $
is sub-optimal since more nodes are involved in the cutting, and thus
more results need to be communicated.

However, in general, it is difficult to find the optimal partition,
which is known as an NP-hard problem (non-deterministic polynomial-time
hardness~\citep{Knuth1974}). There are several techniques in practice
that can provide a reasonably good approximation of the optimal partition,
including spectral partitioners and geometric separators. The spectral
partition method is based on the topology of the graph~\citep{Cvetkovic1980},
while the geometric separator is based on the coordinates of the vertices~\citep{Gilbert1998},
which can be represented by the geometric center of the elements in
an octree mesh. It is noted that spectral methods are often significantly
slower than geometric separators if the mesh size is large, as it
involves an eigenvalue decomposition. The algorithms presented here
only partition the mesh into two parts, but they can be applied recursively
to obtain partitions with an integer power of two parts.

Many mesh partition packages have been developed in recent years based
on these techniques~\citep{Karypis1998,Ogawa2003,Lasalle2016}. The
mesh partition package \textit{meshpart} developed by~\citet{Meshpart2018}
is used in this work, in which a spectral partitioner is used for
examples with less than 5 million DOFs, while a geometric separator
is used for all other cases.

\subsection{Explicit solver in parallel\label{subsec:Solver_parallel}}

\begin{figure}[tb]
\hfill{}\includegraphics{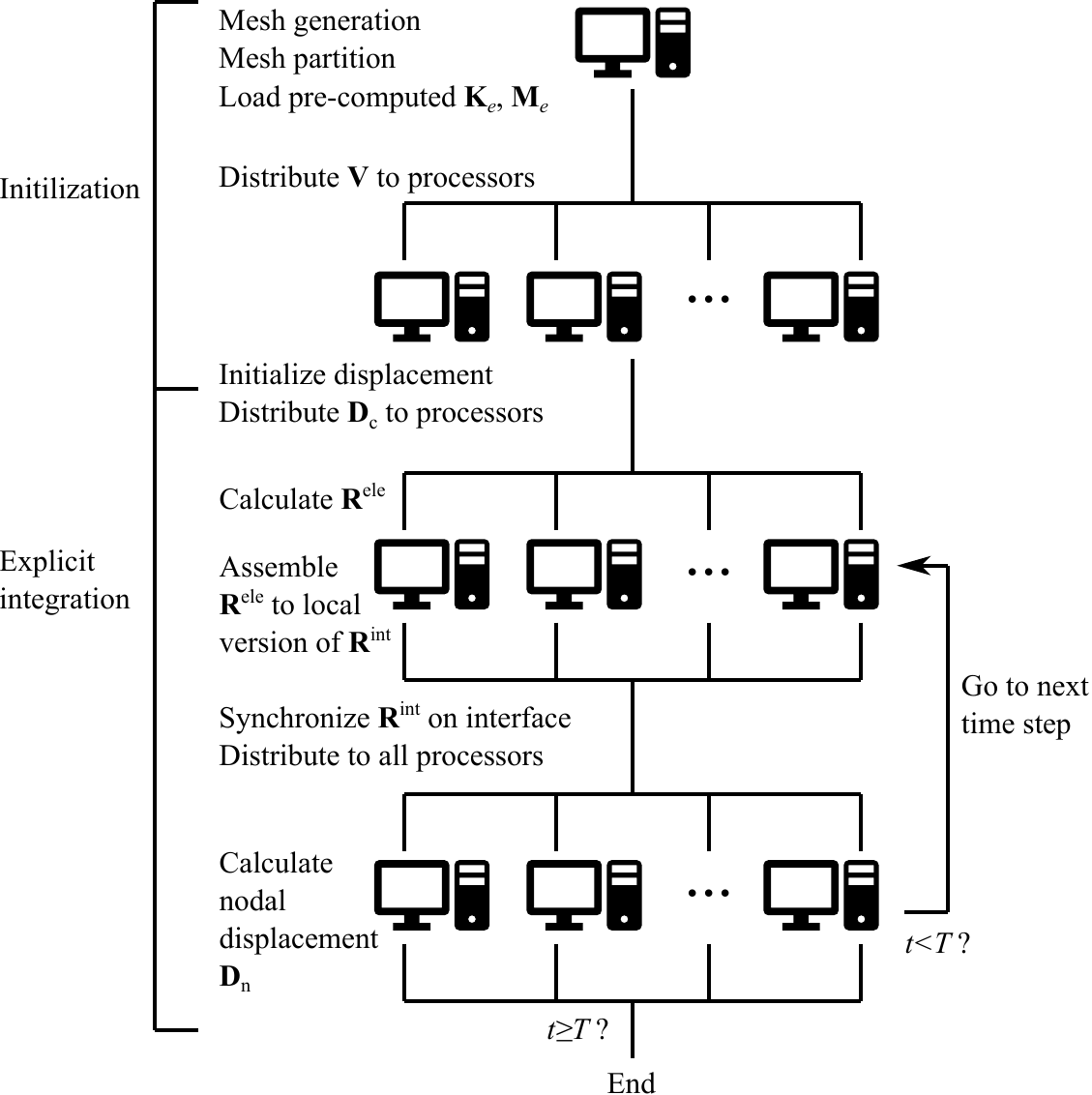}\hfill{}

\caption{Parallel code structure\label{fig:Parallel_code}}
\end{figure}

The explicit solver provided in Algorithm~\ref{alg:Solver_serial}
can be easily parallelized based on the structure shown in Fig.~\ref{fig:Parallel_code}.
The mesh generation and mesh partition are executed on the head node
(the main computer node which distributes jobs to individual processors),
and the element matrices of the 144 unique master cells are stored
as a static library, which is loaded every time the code runs. The
mesh partition package will generate an array of size $n_{\mathrm{e}}$,
which stores a flag for each element, indicating the part to which
it belongs. The element DOF vectors are grouped based on this array,
and they are assigned to the different processors. The displacement
vectors are initialized and distributed to the processors, where each
processor only stores the chunk associated with its own nodes. At
each time step, the nodal force vector is calculated using Eq.~(\ref{eq:Rintep})
on each processor locally. This local version of the nodal force vector
can be divided into two parts, the one corresponding to the nodes
inside each part (internal DOFs) and those shared with other parts
(external DOFs). The values of the latter part must be updated (corrected)
by the contributions of the adjacent part(s). The values of the external
nodes in the local nodal force vectors are summed and synchronized
for all the processors so that the contribution from all elements
connected to the interface is included. As the size of the interface
is relatively small compared to the total size, this operation can
be done in a relatively short time. After the updated nodal force
vector is distributed to each processor, the local portion of displacement
of the next step $\mathbf{U}_{n+1}$ is calculated.

At the beginning of the program, the octree mesh is generated and
partitioned. Pre-computed stiffness and mass matrices of the master
cells are imported and linked to the individual elements using corresponding
transformation matrices. The vector $\mathbf{V}$ is constructed and
distributed to individual processors based on the partitioning result.
The initial displacement vector $\mathbf{U}_{0}$ and fictitious $\mathbf{U}_{-1}$
are calculated and distributed to processors. These steps are considered
as an initialization stage, and are not included in the timing as
they take constant time independent of the number of time steps. The
speedup of the parallel solver is measured based on the time spent
after the initialization stage, including the calculation of the nodal
force vector, the synchronization between processors and the calculation
of the nodal displacement vector.

\begin{figure}[tb]
\hfill{}\includegraphics{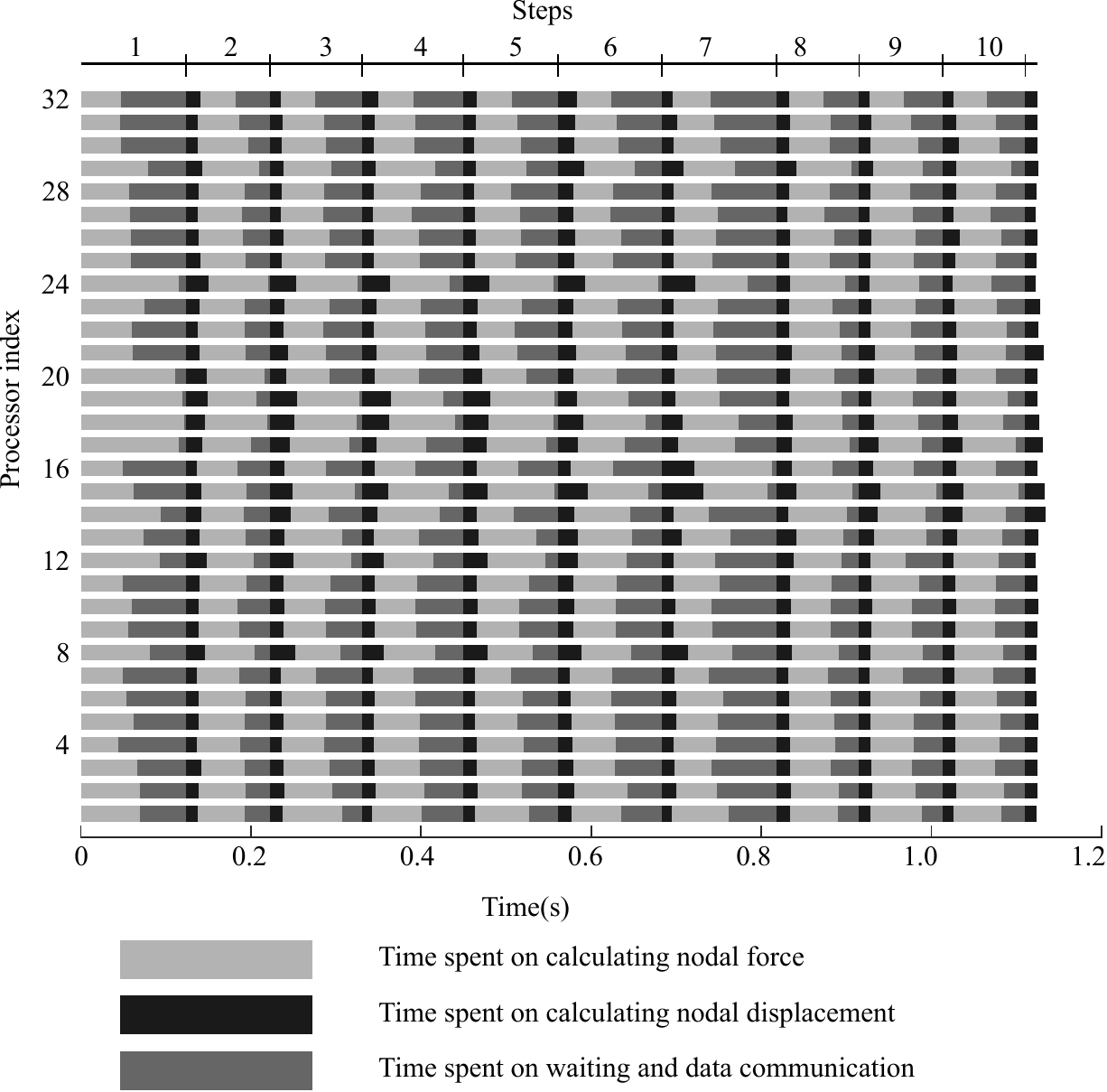}\hfill{}

\caption{A sample of time distribution of parallel processing\label{fig:Time_distribution}}
\end{figure}

A sample timing of a parallel computation using 32 processors is shown
in Fig.~\ref{fig:Time_distribution}. It can be observed that the
speed of these processors is usually different, as well as the amount
of work distributed to each processor. The light gray bar represents
the time spent by each processor to calculate the nodal force vector
locally, and the dark gray bar indicates the time for each processor
to wait for the slower ones to finish their job and communicate the
results. The black bar represents the time required to calculate the
displacement vector for the next time step once the overall nodal
force is obtained.

In parallel computing, the accurate measurement of computational time
is difficult since different processors usually start working at different
time. Here we use the reasonable assumption that all processors commence
the calculation when the first (earliest) processor starts, and all
of them finish when the slowest processor completes the assigned task.
In the numerical examples, the average calculation time for all processors
(including light gray and black bars in Fig.~\ref{fig:Time_distribution})
and communication and waiting time (dark gray bars) are used to indicate
the performance of the proposed method.\clearpage{}

\section{Numerical examples\label{sec:Examples}}

\begin{figure}
\hfill{}\subfloat[Load history of a Ricker wavelet\label{fig:Ricker}]{\includegraphics{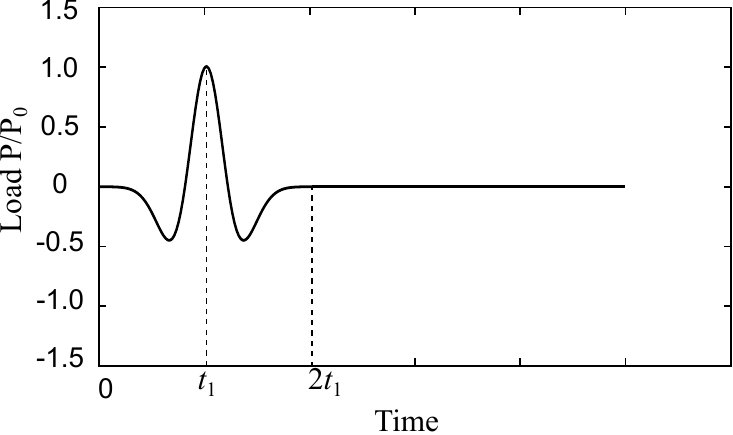}}\hfill{}\subfloat[Spectrum of a Ricker wavelet\label{fig:Ricker_spectrum}]{\includegraphics{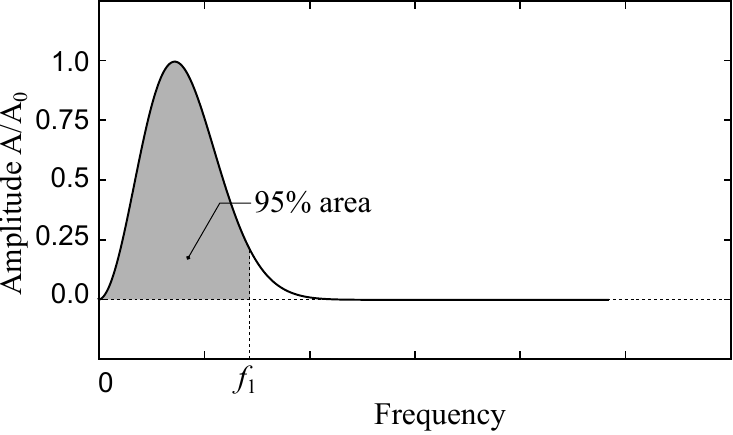}}\hfill{}

\hfill{}\subfloat[Load history of a triangular pulsed\label{fig:Impact}]{\includegraphics{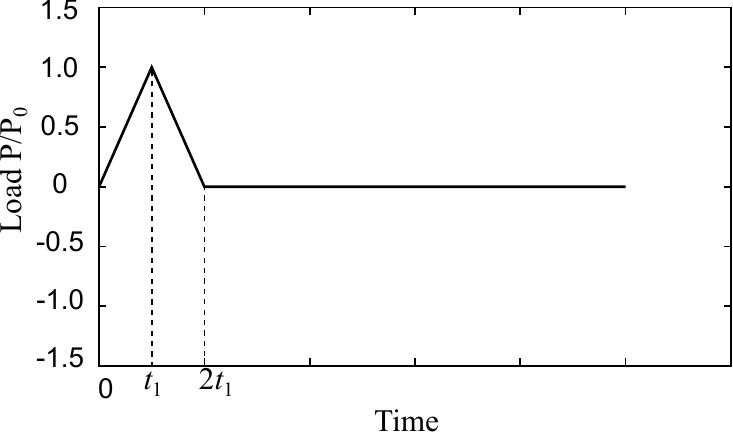}}\hfill{}\subfloat[Spectrum of a triangular pulse\label{fig:Impact_spectrum}]{\includegraphics{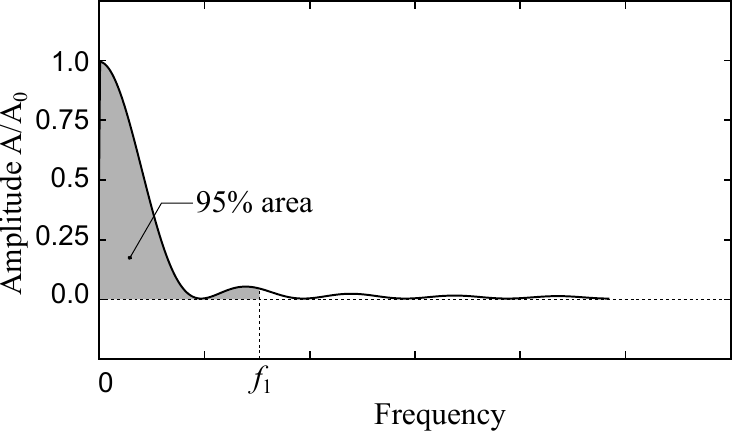}}\hfill{}

\hfill{}\subfloat[Load history of a sine-burst signal\label{fig:Lamb}]{\includegraphics{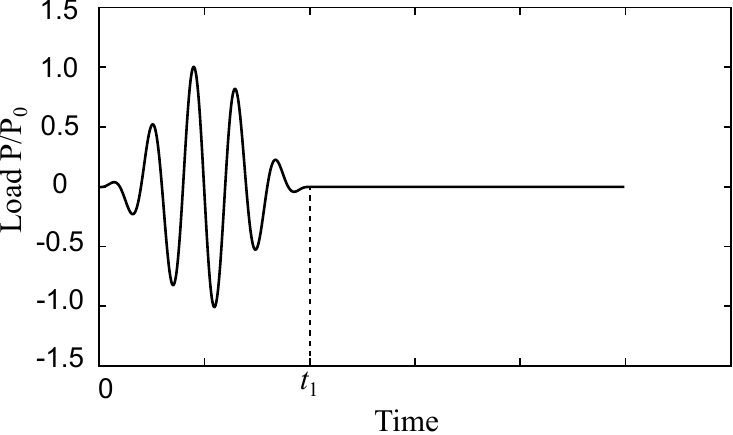}}\hfill{}\subfloat[Spectrum of a sine-burst signal\label{fig:Lamb_spectrum}]{\includegraphics{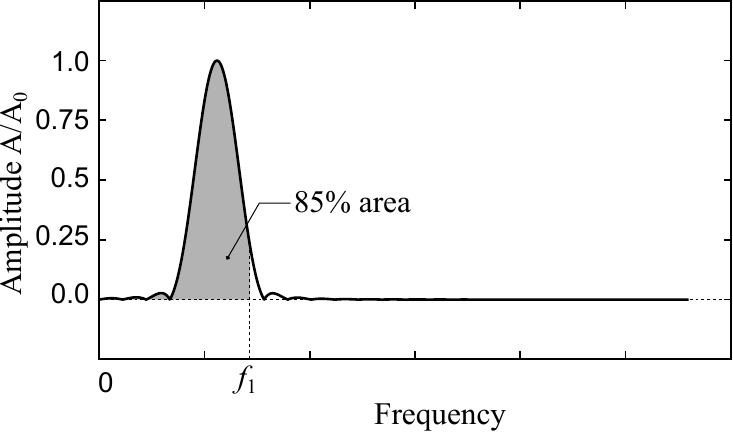}}\hfill{}

\caption{Dynamic loading for the examples\label{fig:Load}}
\end{figure}

In this section, five numerical examples are presented to verify the
proposed procedure and to demonstrate the attainable speedup of the
novel parallel explicit solver with increasing number of computing
cores. The efficiency $\eta$ of the parallel computation is measured
by the ratio of the speedup and the number of cores (strong scaling).
The different transient amplitude functions that are used for our
numerical examples to describe the time-dependent excitations are
depicted in Fig.~\ref{fig:Load}. Depending on the problem, we choose
either a Ricker wavelet (see Fig.~\ref{fig:Ricker}), a triangular
pulse (see Fig.~\ref{fig:Impact}), or a sine-burst signal (see Fig.~\ref{fig:Lamb}).
The analytical form of the Ricker wavelet in the time domain is
\begin{equation}
P\!\left(t\right)=\!\left(1-\!\left(\frac{t-t_{1}}{t_{1}/5}\right)^{2}\right)\exp\!\left(-\frac{1}{2}\!\left(\frac{t-t_{1}}{t_{1}/5}\right)^{2}\right)P_{0},\label{eq:Ricker}
\end{equation}
where $P_{0}$ is the maximum load. The triangular impact load is
expressed as
\begin{equation}
P\!\left(t\right)=\!\begin{cases}
\!\left(1-\!\left|\frac{t}{t_{1}}-1\right|\right)P_{0} & \textrm{if }0\leq t\leq2t_{1}\\
0 & \textrm{if }t>2t_{1}
\end{cases},\label{eq:Triangle}
\end{equation}
while the sine-burst (sine excitation modulated by a Hann window)
is expressed as
\begin{equation}
P\!\left(t\right)=\!\begin{cases}
\sin\!\left(\frac{2\pi nt}{t_{1}}\right)\sin^{2}\!\left(\frac{\pi t}{t_{1}}\right)P_{0} & \textrm{if }0\leq t\leq t_{1}\\
0 & \textrm{if }t>t_{1}
\end{cases},\label{eq:Lamb}
\end{equation}
where $n$ is the number of cycles. The central frequency $f_{\mathrm{m}}$
is calculated as 
\begin{equation}
f_{\mathrm{m}}=\frac{n}{t_{1}}.
\end{equation}
To obtain the frequency spectrum of the given excitation functions
we apply the Fourier transform. The frequency spectrum of the Ricker
wavelet is depicted in Fig.~\ref{fig:Ricker_spectrum} and can be
expressed as
\begin{equation}
A\!\left(f\right)=\!\left(\dfrac{2}{\sqrt{\pi}}\right)\!\left(\dfrac{f}{f_{m}}\right)^{2}\exp\!\left(-\!\left(\dfrac{f}{f_{m}}\right)^{2}\right),\label{eq:Ricker_A}
\end{equation}
where the central frequency $f_{\mathrm{m}}$ is
\begin{equation}
f_{\mathrm{m}}=\frac{5}{\sqrt{2}\pi t_{1}}.\label{eq:Ricker_fm}
\end{equation}
Similarly, the amplitude spectrum of the Fourier transform of the
triangular pulse is shown in Fig~\ref{fig:Impact_spectrum} and can
be written as
\begin{equation}
A\!\left(f\right)=\!\left(\frac{\sin\!\left(\pi ft_{1}\right)}{\pi ft_{1}}\right)^{2},\label{eq:Triangle_A}
\end{equation}
A critical frequency $f_{1}$ is identified such that the area under
the curve on the left of $f_{1}$ is equal to a certain percentage
of the total area, e.g., $95\%$. This approach ensures that most
of the energy content of the signal is accounted for. For a Ricker
wavelet, $f_{1}$ is approximately $2.2/t_{1}$, while in the case
of triangular impact $f_{1}$ is $1.8/t_{1}$. For the sine-burst
excitation, $f_{1}$ is identified corresponding to $85\%$ of the
area, which is $5.8/t_{1}$ when $n=5$. The determined critical frequency
is an important parameter that is closely related to the required
spatial discretization. Depending on the value of $f_{1}$ the wave
velocities of the propagating modes are calculated from which the
number of nodes per wavelength is derived.

In infinite elastic media the only propagating waves are bulk waves
which can be distinguished in two types, i.e., dilatational/pressure/primary
and shear/secondary waves. The dilatational wave velocity $v_{\mathrm{P}}$
and shear wave velocity $v_{\mathrm{S}}$ for an isotropic, homogeneous,
linear-elastic medium are calculated as\begin{subequations}\label{subeq:wavespeed}
\begin{align}
v_{\mathrm{P}} & =\sqrt{\frac{E\!\left(1-\nu\right)}{\rho\!\left(1+\nu\right)\!\left(1-2\nu\right)}},\\
v_{\mathrm{S}} & =\sqrt{\frac{E}{2\rho\!\left(1+\nu\right)}},
\end{align}
\end{subequations}where $E$ is Young's modulus, $\nu$ denotes Poisson's
ratio, and $\rho$ stands for the mass density of the material. Therefore,
the minimum dilatational wavelength $L_{\mathrm{P}}$ and shear wavelength
$L_{\mathrm{S}}$ considered in the structure are\begin{subequations}\label{subeq:wavelength}
\begin{align}
L_{\mathrm{P}} & =\frac{v_{\mathrm{P}}}{f_{1}},\\
L_{\mathrm{S}} & =\frac{v_{\mathrm{S}}}{f_{1}}.
\end{align}
\end{subequations}A common assumption on the spatial discretization
is that the maximum element size in the mesh should be limited to
have at least 10 (linear) finite elements per wavelength. This is
a bare minimum which will result in deviations of more than 1\% compared
to a reference solution. Therefore, \citet{Willberg2012} and \citet{PhDGravenkamp2014}
recommend roughly 30 nodes per wavelength when using linear shape
function in numerical methods for wave propagation analysis.

The in-house solver is written in Python 3.7, while the inter-processor
communication is implemented using MPI4Py~\citep{Dalcin2005,Dalcin2008,Dalcin2011},
which provides bindings for the Message Passing Interface (MPI) standard
with OpenMPI (4.0.2) as the back end. The distributed computing environment
used in this research is the supercomputer Gadi maintained by the
national computing infrastructure (NCI) in Canberra (Australia). A
typical compute node on Gadi has two sockets, each equipped with a
24-core Intel Xeon Scalable \textquoteleft Cascade Lake\textquoteright{}
processor and connected to $190\,\unit{GB}$ RAM. The compute nodes
and the storage systems are connected through a high-speed network
with a data transfer rate of up to $200\,\unit{Gbps}$. In order to
maximize the parallel performance, the MPI processes are mapped by
sockets within each compute node, which provides an optimal achievable
memory bandwidth to each MPI process on both sockets~\citep{Chang2018}.
For multi-node computations, the MPI processes are evenly distributed
among the compute nodes to fully exploit the computational capability
of Gadi.

\FloatBarrier

\subsection{Eigenfrequencies of a cube}

\begin{figure}
\hfill{}\includegraphics{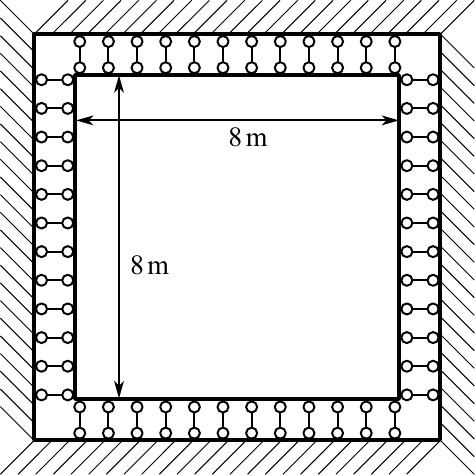}\hfill{}

\caption{Geometry and boundary conditions of the cube\label{fig:Cube_geo}}
\end{figure}

In the first example, we verify the lumped mass matrix by investigating
the eigenfrequencies of a cube with a Young's modulus of $E=1000\,\unit{Pa}$
and a Poisson's ratio of $\nu=0.3$, while the mass density of the
cube is chosen as $\rho=1\,\unitfrac{kg}{m^{3}}$. The edge length
of the cube is $L=8\,\unit{m}$. The displacements perpendicular to
the surfaces are constrained as shown in 2D in Fig.~\ref{fig:Cube_geo}.

\begin{figure}
\hfill{}\includegraphics{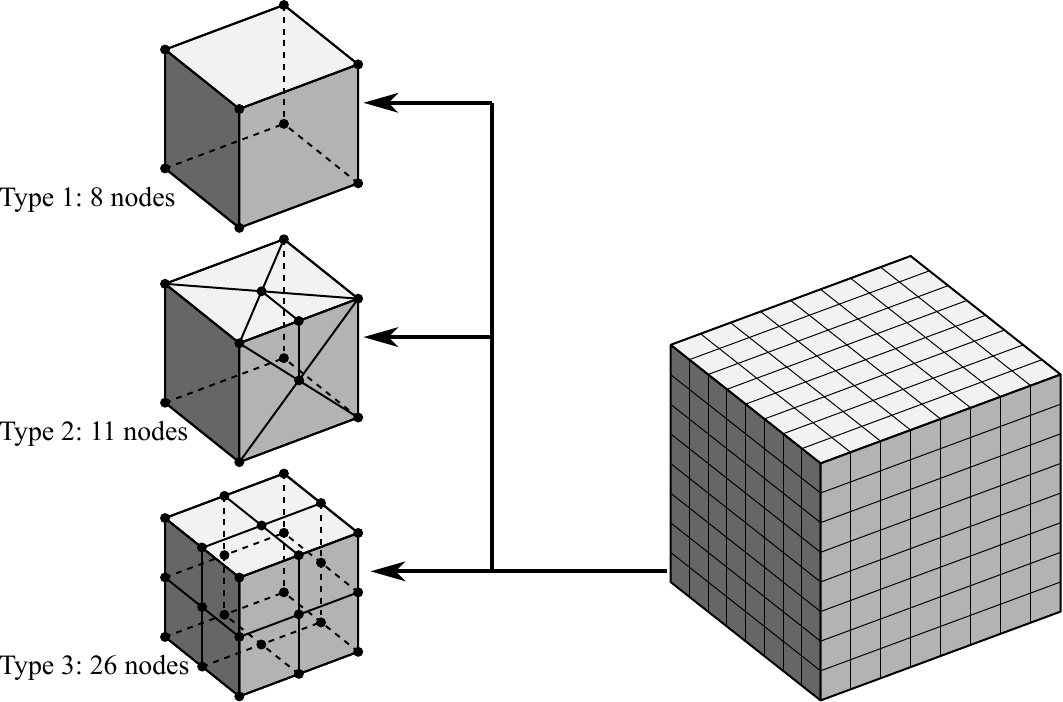}\hfill{}

\caption{Three types of meshes of the cube\label{fig:Cube_mesh}}
\end{figure}

\begin{figure}
\hfill{}\includegraphics[scale=0.9]{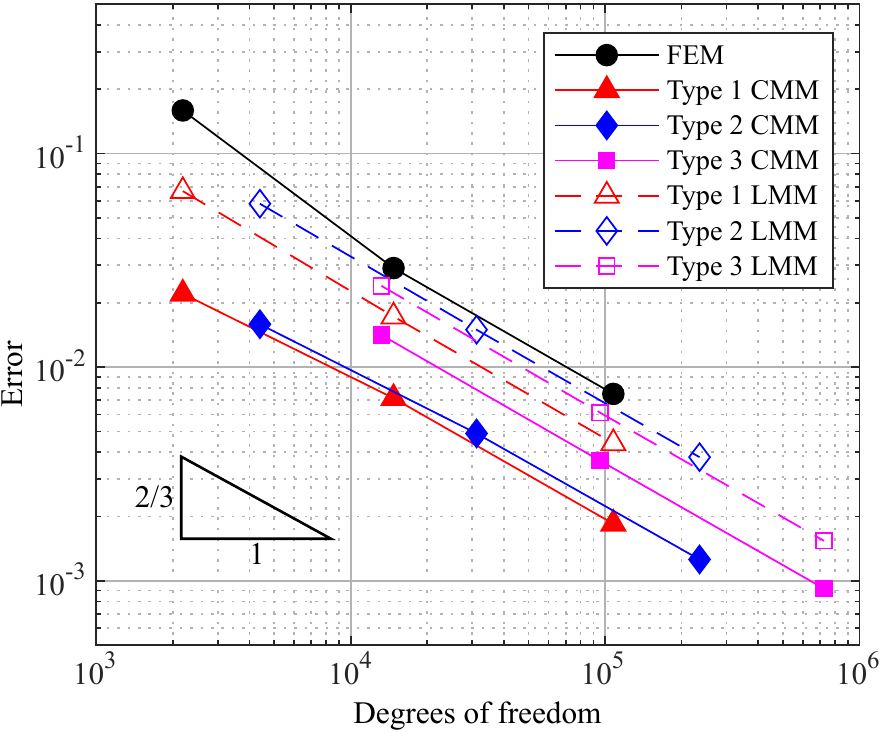}\hfill{}

\caption{Relative error of first 100 eigenfrequencies calculated from different
types of meshes using different methods\label{fig:Cube_curve}}
\end{figure}

Three types of meshes are generated based on three typical octree
patterns depicted in Fig.~\ref{fig:Cube_mesh}. Note that in the
coarsest mesh the edge length of an element is $1\,\unit{m}$. The
eigenfrequencies based on these meshes using either a consistent mass
matrix (CMM) or a lumped mass matrix (LMM) are calculated, as well
as the FEM result calculated using commercial software ABAQUS with
type 1 mesh. More detailed investigations regarding the use of LMMs
in the framework of the SBFEM are discussed in Ref.~\citep{Gravenkamp2020a}.
A convergence study of all element types is performed, and the $L_{2}$
norm of the relative error $e$ in the first 100 frequencies is presented
in Fig.~\ref{fig:Cube_curve}. The error $e$ is calculated as
\begin{equation}
e=\sqrt{\frac{\stackrel[i=1]{100}{\sum}\left(\omega_{i}^{\mathrm{num}}-\omega_{i}^{\mathrm{ref}}\right)^{2}}{\stackrel[i=1]{100}{\sum}\left(\omega_{i}^{\mathrm{ref}}\right)^{2}},}
\end{equation}
where $\omega_{i}^{\mathrm{num}}$ is the numerical solution of the
$i$-th eigenfrequency, and $\omega_{i}^{\mathrm{ref}}$ is the analytical
(reference) solution. The analytical solution is given in the form
\begin{equation}
\omega^{\mathrm{ref}}=\sqrt{\frac{E\pi^{2}}{\rho\left(1-\nu^{2}\right)L^{2}}\!\left(l^{2}+m^{2}+n^{2}\right)\gamma},\label{eq:Cube_f}
\end{equation}
where $l$, $m$ and $n$ are non-negative integers, and the parameter
$\gamma$ is defined as
\begin{equation}
\gamma=\!\begin{cases}
1 & \textrm{if two of \ensuremath{\left\{  l,m,n\right\} } are 0 }\\
\!\left\{ 1,\frac{1-\nu}{2}\right\}  & \textrm{if one of \ensuremath{\left\{  l,m,n\right\} } is 0}\\
\!\left\{ 1,\frac{1-\nu}{2},\frac{1-\nu}{2}\right\}  & \textrm{if none of \ensuremath{\left\{  l,m,n\right\} } is 0}
\end{cases},\label{eq:Cube_gama}
\end{equation}
Note that if all of $\left\{ l,m,n\right\} $ are zeros, there is
no solution to the equation.

The numerical results depicted in Fig.~\ref{fig:Cube_curve} highlight
that with the proposed approach, an optimal convergence rate (2/3)
is achieved independent of using either an LMM- or CMM-based formulation
in the SBFEM or the FEM formulation. Furthermore, we observe that
the different octree pattern types have no influence on the slope
of the curves. Although the same rates of convergence are attainable,
we have to note that in general the CMM-based approach yields slightly
more accurate results for the same number of DOFs, and the FEM solution
is less accurate than both LMM- or CMM-based approach, which results
in an offset of the error curves. This difference is, however, of
no concern for the following discussions as optimal convergence is
still observed.

\FloatBarrier

\subsection{Wave propagation in a beam}

In the second example, a cantilever beam is investigated, whose dimensions
are $16\,\unit{m}\times1\,\unit{m}\times1\,\unit{m}$. Its material
properties are: Young's modulus $E=10\,\unit{kPa}$, Poisson's ratio
$\nu=0$, and mass density $\rho=1\,\unitfrac{kg}{m^{3}}$. The beam
is fixed on one end, and a uniform pressure load is applied on the
other end, which follows the Ricker wavelet depicted in Fig.~\ref{fig:Ricker}
with the parameters $t_{1}=15\,\unit{ms}$ and $P_{0}=1\,\unit{Pa}$.
Therefore, the maximum frequency of interest $f_{1}$ is $150\,\unit{Hz}$.
The dilatational wave speed, calculated using Eq.~(\ref{subeq:wavespeed}),
is $100\,\unitfrac{m}{s}$, and the wavelength in the beam is $0.67\,\unit{m}$
according to Eq.~(\ref{subeq:wavelength}). This problem is equivalent
to a 1D wave propagation problem; therefore, Duhamel's integral is
applied to obtain a reference solution for the displacement response,
the details of which can be found in Ref.~\citep{Gravenkamp2020a}.

\begin{figure}
\hfill{}\includegraphics{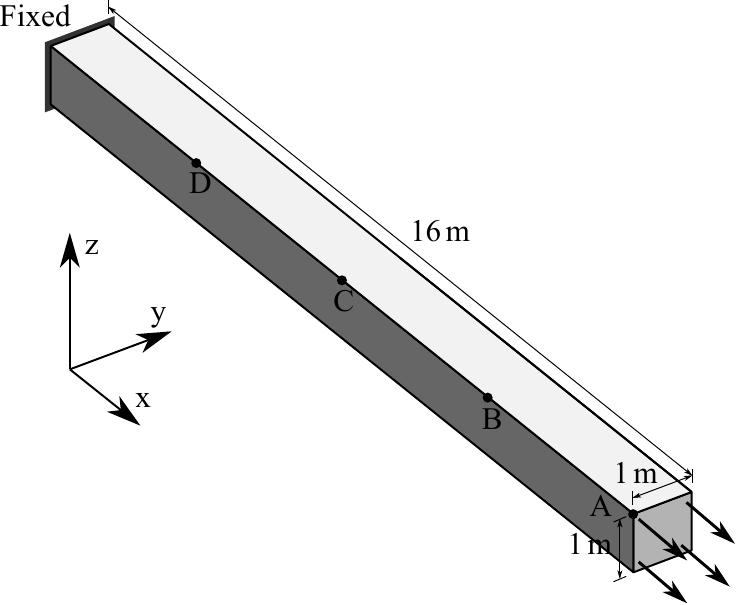}\hfill{}

\caption{Geometry and boundary conditions of the beam\label{fig:Beam_geo}}
\end{figure}

\begin{figure}
\hfill{}\includegraphics{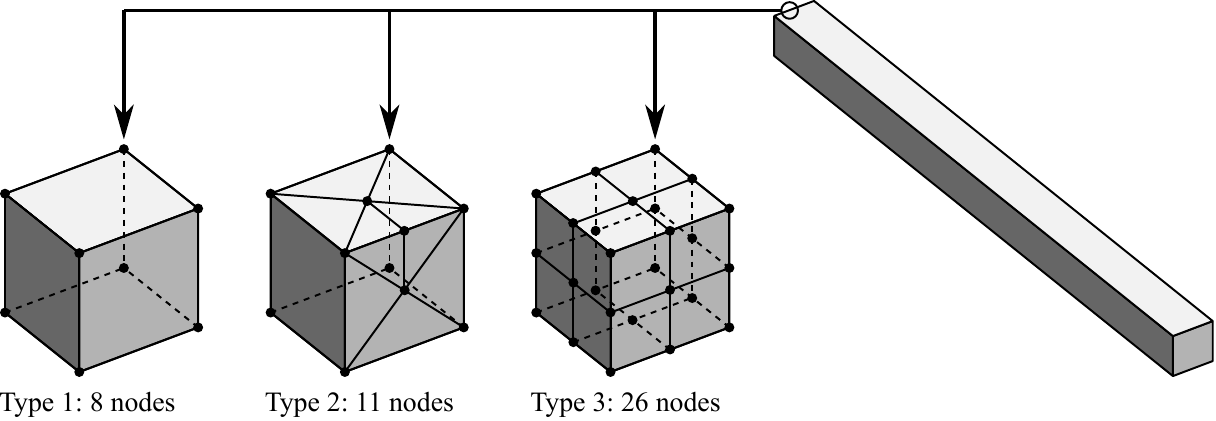}\hfill{}

\caption{Three types of meshes of the beam\label{fig:Beam_mesh}}
\end{figure}

\begin{figure}
\hfill{}\subfloat[Displacement history of four selected points\label{fig:Beam_disp_1}]{\includegraphics{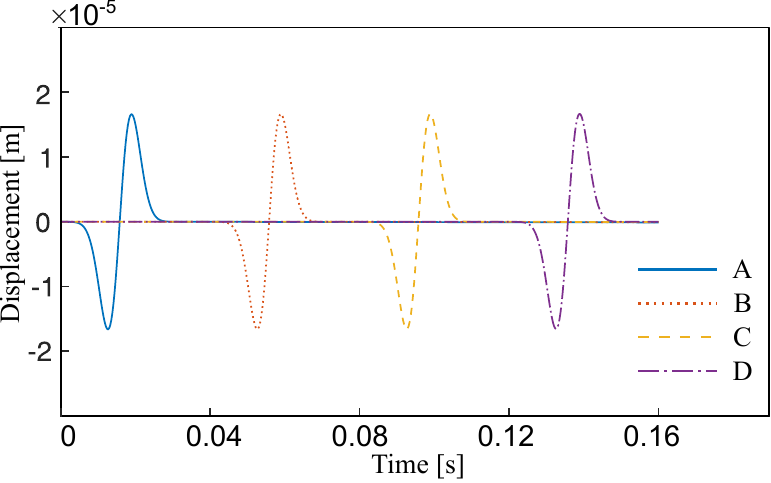}}\hfill{}\subfloat[Acceleration history of four selected points\label{fig:Beam_disp_2}]{\includegraphics{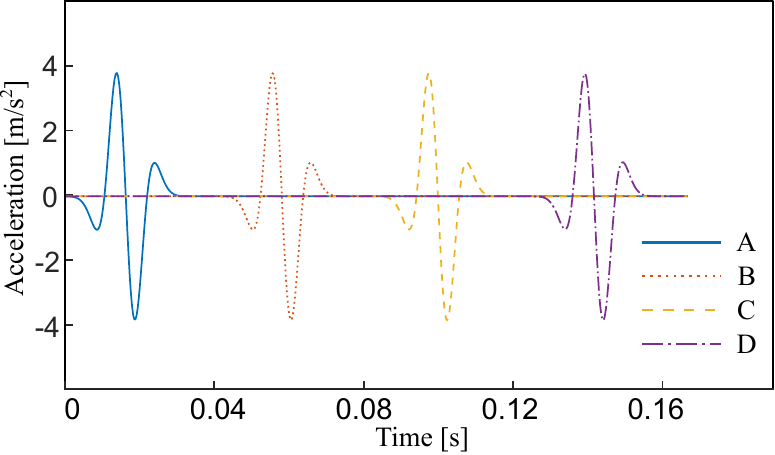}}\hfill{}

\caption{Displacement and acceleration histories of four selected points on
the beam\label{fig:Beam_disp}}
\end{figure}

\begin{figure}
\hfill{}\subfloat[Convergence curve of displacement\label{fig:Beam_curve_1}]{\includegraphics{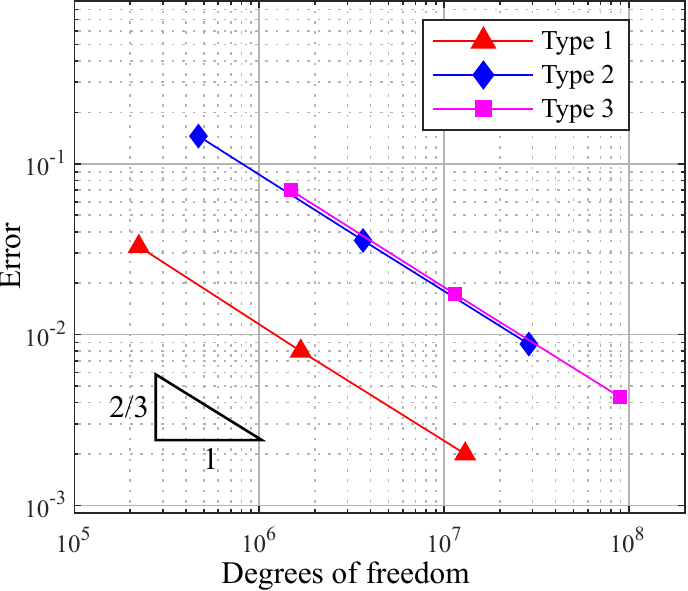}}\hfill{}\subfloat[Convergence curve of acceleration\label{fig:Beam_curve_2}]{\includegraphics{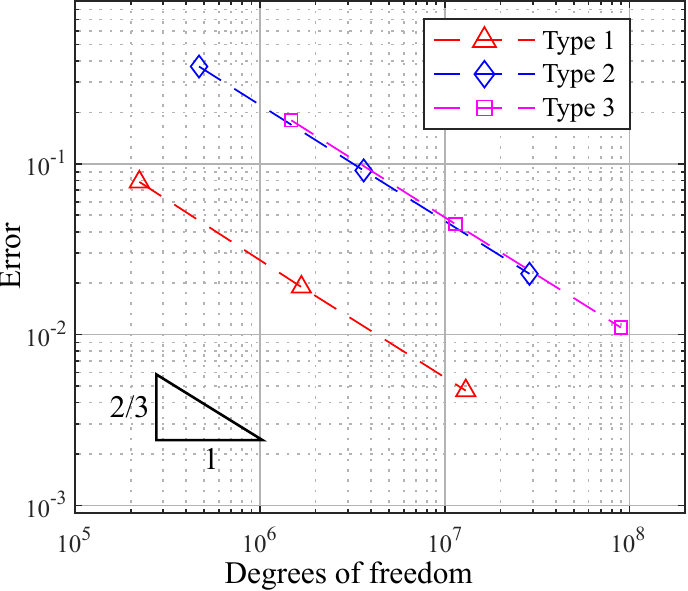}}\hfill{}

\caption{Convergence curves of different types of meshes\label{fig:Beam_curve}}
\end{figure}

Again, three types of meshes are generated based on three typical
octree patterns as shown in Fig.~\ref{fig:Beam_mesh}, and the convergence
behavior is examined. In the coarsest group of meshes, the element
size is set to $0.0625\,\unit{m}$; as a result, there are approximately
10 elements in a wavelength. The time steps in the three coarsest
meshes are $0.5897\,\unit{ms}$, $0.3141\,\unit{ms}$ and $0.3380\,\unit{ms}$,
which guarantees there are at least 10 time steps\footnote{The Shannon-Nyquist theorem establishes a sufficient condition for
the sampling rate of a signal. According to this theorem, at least
two sampling points are required for one period of the signal associated
with the highest frequency. This is, however, not sufficient in structural
dynamics, and therefore, it is often suggested to use 10 to 20 sampling
points. For highly accurate simulations, this should be increased
to more than 100 sampling points.} during a period of the highest frequency. The displacement and acceleration
history of four selected points are plotted in Fig.~\ref{fig:Beam_disp}.
The $L_{2}$ norm of the error of the selected points at all time
steps is calculated, and the convergence curves of these meshes are
presented in Fig.~\ref{fig:Beam_curve}.

As in the modal analysis example of the previous section, we observe
that the theoretical rates of convergence for both displacement and
acceleration are recovered by our approach, independent of the chosen
mesh type. These results confirm the correctness of the implementation
and provide confidence in the performance of our method. Thus, more
complex examples that are of practical interest are tackled in the
following sections.

\FloatBarrier

\subsection{Multi-story building induced ground vibration\label{subsec:Benchmark}}

\begin{figure}
\hfill{}\includegraphics{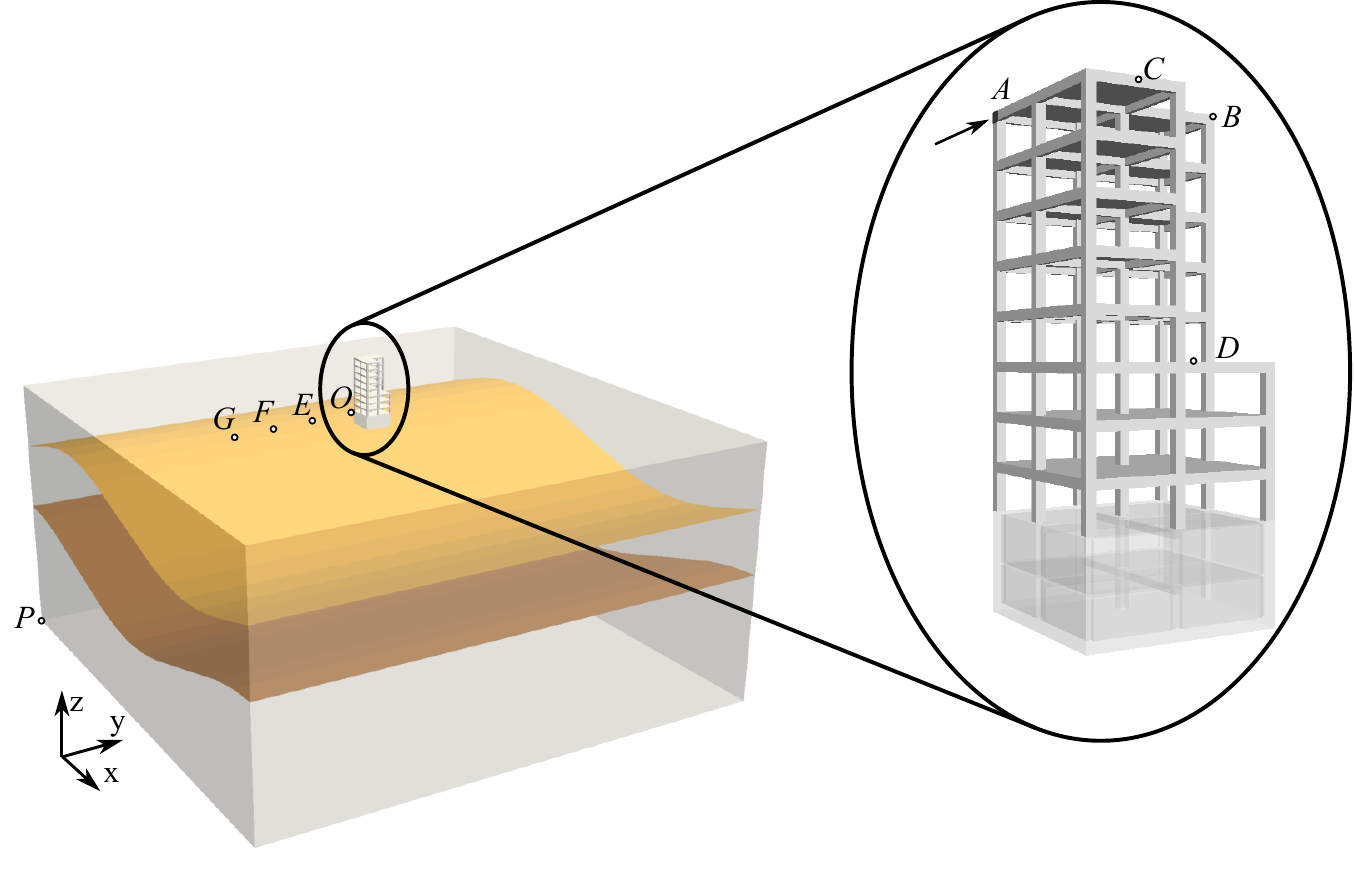}\hfill{}

\caption{Geometry of the frame and ground\label{fig:Frame_geo}}
\end{figure}

\begin{figure}
\hfill{}\includegraphics{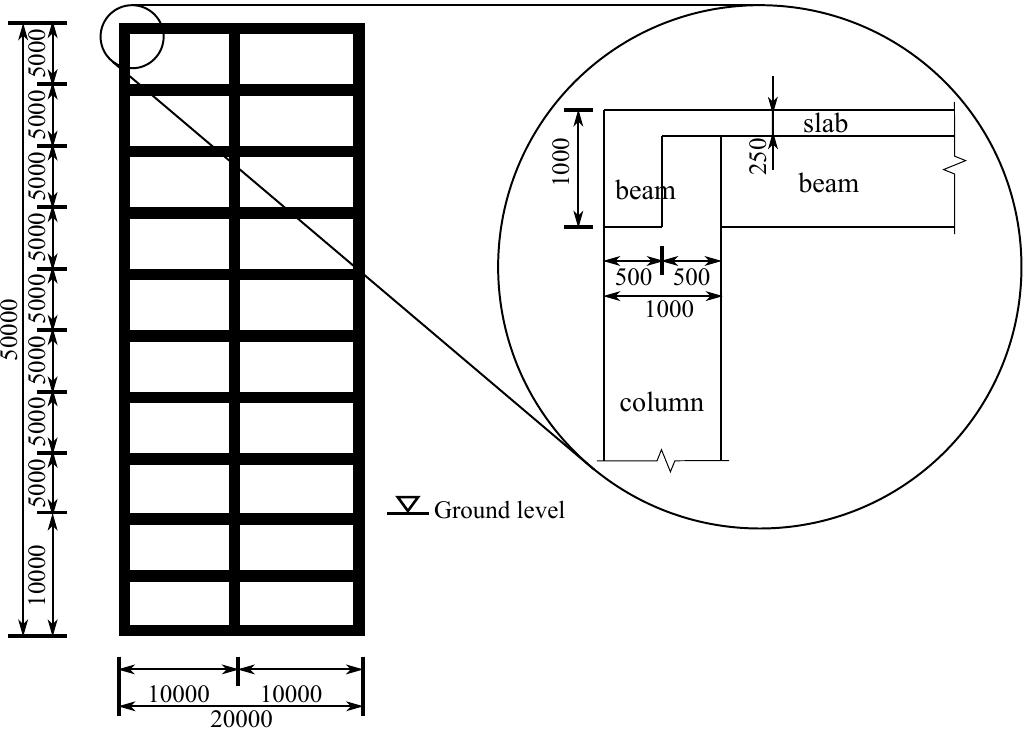}\hfill{}

\caption{Dimension of the frame (Unit: mm)\label{fig:Frame_dimen}}
\end{figure}

To further verify the proposed technique and evaluate its computational
performance, the ground vibration induced by a dynamic load applied
to a multi-story building is analyzed. The multi-story building and
the adjacent ground are depicted in Fig.~\ref{fig:Frame_geo}. The
building consists of columns, beams, slabs and a foundation with a
basement, and the dimensions are illustrated in Fig.~\ref{fig:Frame_dimen}.

The material properties of the frame are: Young's modulus $E=17\,\unit{GPa}$,
Poisson's ratio $\nu=0.2$ and mass density $\rho=2400\,\unitfrac{kg}{m^{3}}$.
The dilatational wave speed is equal to $2805\,\unitfrac{m}{s}$,
and the shear wave speed is $1718\,\unitfrac{m}{s}$. A region of
$400\,\unit{m}\times400\,\unit{m}\times200\,\unit{m}$ of the ground
adjacent to the building is included in the model. The coordinates
of point $P$ depicted in Fig.~\ref{fig:Frame_geo} are $\left(-183\,\unit{m},-187\,\unit{m},-200\,\unit{m}\right)$,
and the origin of the coordinate system is denoted as $O$. The ground
is made of three layers, and the shapes of these interfaces between
the individual layers are defined as (unit: m)\begin{subequations}
\begin{align}
z_{12} & =-60+20\times\sin\!\left(\frac{x+183}{200}\pi\right),\\
z_{23} & =-130+20\times\cos\!\left(\frac{x+183}{200}\pi\right).
\end{align}
\end{subequations}From the top to the bottom layers, the Young's
moduli are equal to $0.5\,\unit{GPa}$, $0.8\,\unit{GPa}$ and $1\,\unit{GPa}$,
respectively. All three layers have the same Poisson's ratio of $\nu=0.2$
and mass density of $\rho=2400\,\unitfrac{kg}{m^{3}}$. The corresponding
dilatational wave velocities are equal to $481\,\unitfrac{m}{s}$,
$609\,\unitfrac{m}{s}$ and $680\,\unitfrac{m}{s}$, and the shear
wave velocities are $295\,\unitfrac{m}{s}$, $373\,\unitfrac{m}{s}$
and $417\,\unitfrac{m}{s}$.

An area of $\unit[1]{m^{2}}$ at the side of the top of the building
($2.5\,\unit{m}\leq x\leq3.5\,\unit{m}$, $y=2.5\,\unit{m}$ and $39\,\unit{m}\leq z\leq40\,\unit{m}$),
indicated by region $A$ in Fig.~\ref{fig:Frame_geo}, is subjected
to an impact load described by Fig.~\ref{fig:Impact} with the peak
pressure of $P_{0}=10^{7}\,\unit{Pa}$. The half duration of the impact
is $t_{1}=250\,\unit{ms}$, and the maximum frequency of interest
is $8\,\unit{Hz}$.

A reference solution is obtained from a convergence study using the
commercial finite element package ABAQUS. To limit the size of the
problem, only the building including its foundation with basement
is modeled. The displacements perpendicular to the side faces below
the ground level and the bottom of the basement are constrained. The
displacement and acceleration responses at Points $B$, $C$ and $D$
($B\!\left(2.5\,\unit{m},22.5\,\unit{m},40\,\unit{m}\right)$, corner
of the building; $C\!\left(22.5\,\unit{m},7.5\,\unit{m},40\,\unit{m}\right)$,
middle of a beam; $D\!\left(17\,\unit{m},17\,\unit{m},15\,\unit{m}\right)$,
center of a slab) indicated in Fig.~\ref{fig:Frame_geo} are considered.
A series of analyses using uniform meshes of hexahedral elements (C3D8)
are performed. From the results of the convergence study we can infer
that an acceptable accuracy is obtained for meshes with an element
size of $0.0625\,\unit{m}$. Such a mesh contains 9,242,624 elements
and 32,186,598 DOFs.

\begin{figure}
\hfill{}\includegraphics[width=1\textwidth]{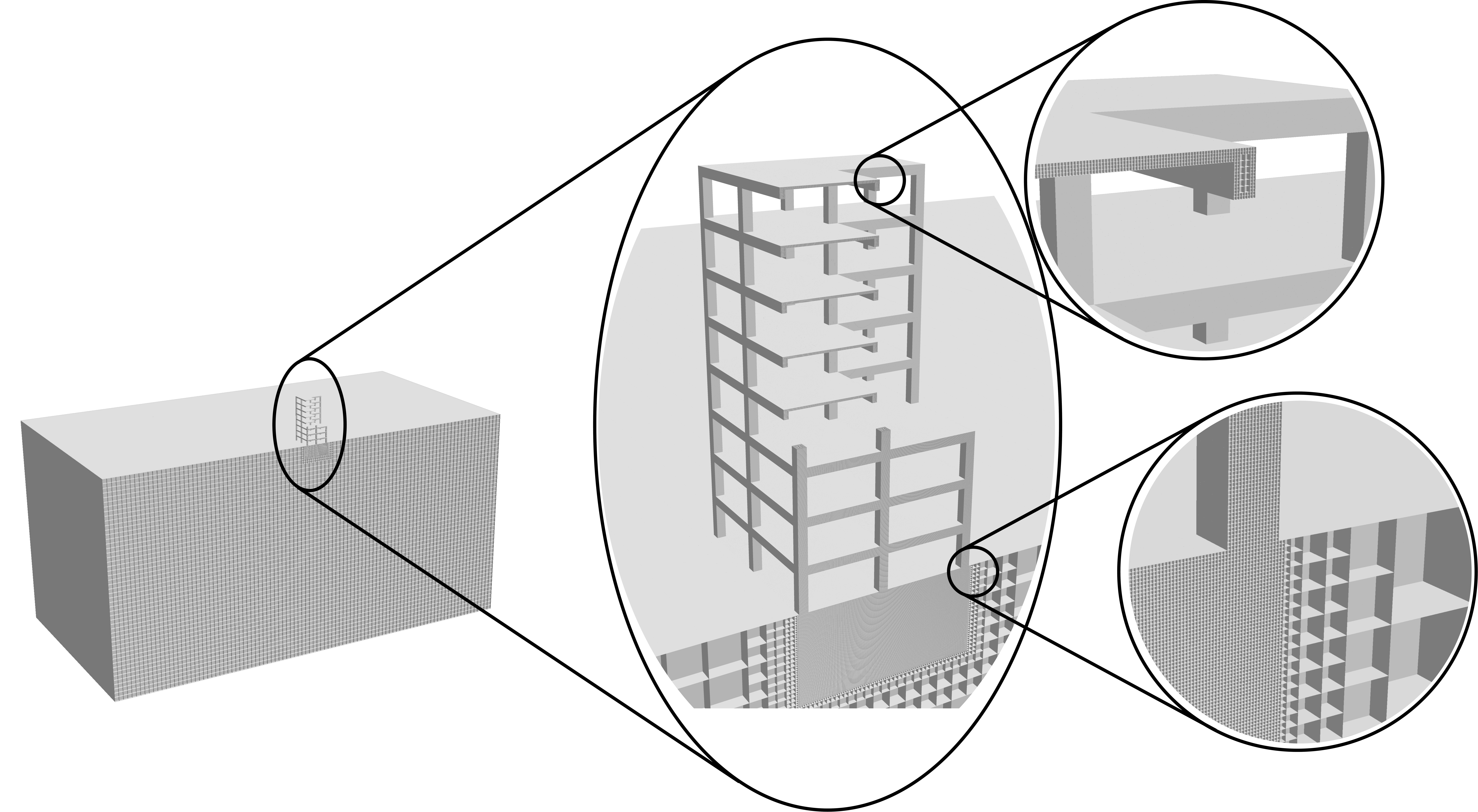}\hfill{}

\caption{Octree mesh of the frame and ground\label{fig:Frame_mesh}}
\end{figure}

\begin{figure}
\hfill{}\subfloat[Displacement of points $B$, $C$ and $D$]{\includegraphics{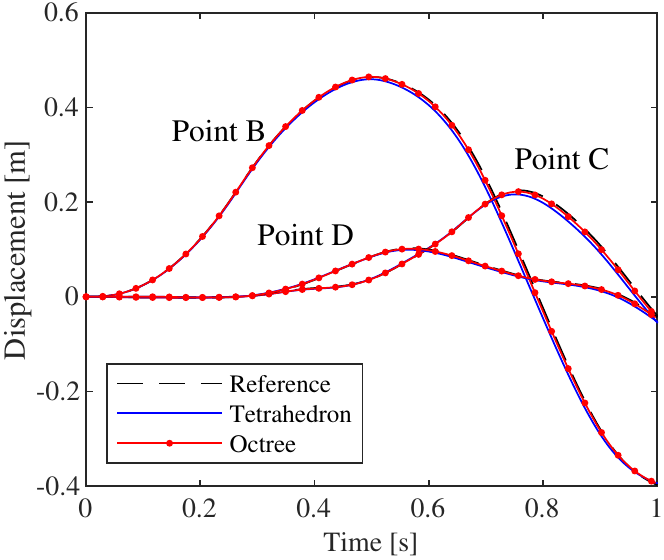}}\hfill{}\subfloat[Acceleration of points $B$]{\includegraphics{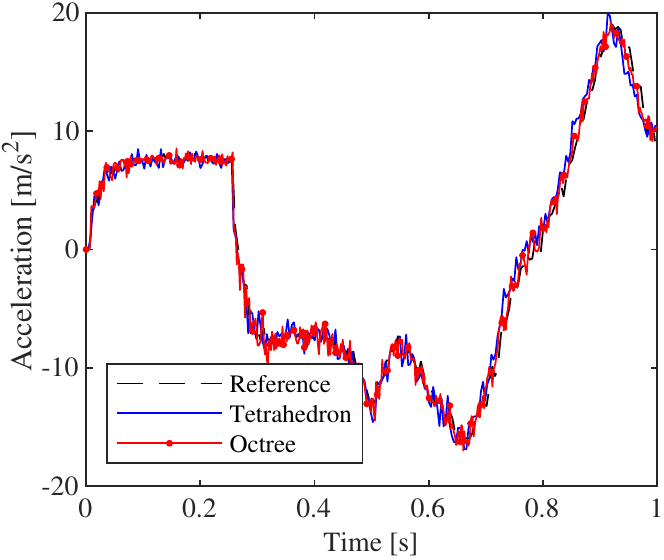}}\hfill{}

\hfill{}\subfloat[Acceleration of points $C$]{\includegraphics{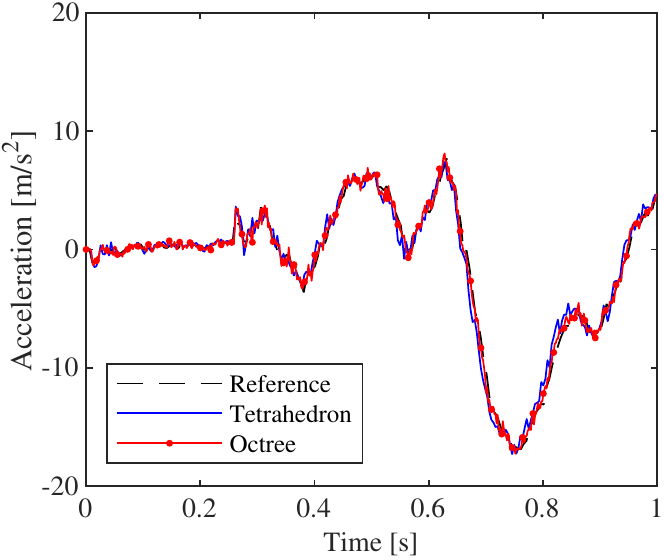}}\hfill{}\subfloat[Acceleration of points $D$]{\includegraphics{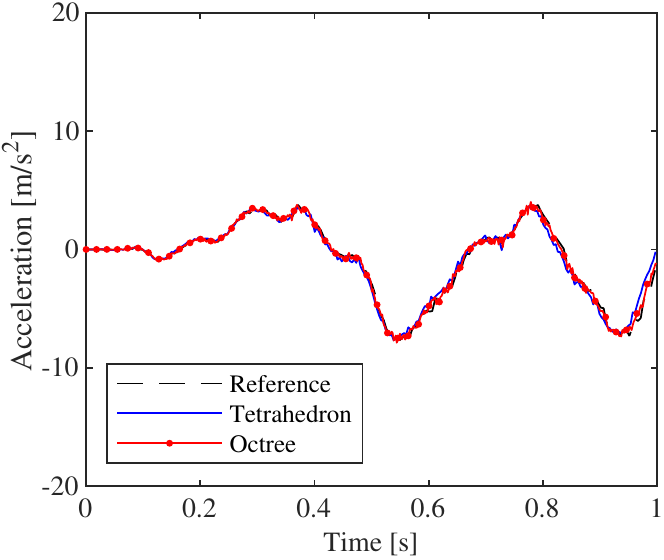}}\hfill{}

\caption{The displacement and acceleration history of the building using our
in-house code and ABAQUS\label{fig:Frame_history}}
\end{figure}

Generating a non-uniform hexahedral mesh for an efficient simulation
of wave propagation in the complete system including both building
and ground requires considerable human interventions. To compare with
the present technique using octree meshes that are generated fully
automatically (shown as a part of Fig.~\ref{fig:Frame_mesh}), non-structured
tetrahedral meshes are used in ABAQUS. Both types of meshes are generated
according to specified element sizes. Using the building model and
the reference solution, we identify a tetrahedral mesh and an octree
mesh that yield a similar level of accuracy in the displacement and
acceleration response histories as plotted in Fig.~\ref{fig:Frame_history}.
The minimum and maximum element sizes are $0.0625\,\unit{m}$ and
$0.25\,\unit{m}$, respectively. Using our SBFEM-based code in conjunction
with an octree mesh, the number of elements in the building model
is 6,018,480 and therefore, the number of DOFs is 25,789,146. On the
other hand, the tetrahedral mesh generated in ABAQUS has a minimum
element size of $0.05\,\unit{m}$ while the maximum element size is
$0.25\,\unit{m}$. Hence, there are 27,274,592 elements and 21,632,154
DOFs in the mesh.

To model the ground, both the tetrahedral and octree meshes are generated
with a maximum element size of $4\,\unit{m}$, which is about 1/10
of the shear wavelength of $36.9\,\unit{m}$ in the top layer at the
maximum frequency of interest. There are 7,339,939 elements and 30,482,412
DOFs in the octree mesh of the whole system (Fig.~\ref{fig:Frame_mesh}),
while there are 38,847,472 elements and 26,192,571 DOFs in the tetrahedral
mesh. The time step used in this example is $\Delta t=0.018\,\unit{ms}$
which was estimated using Eq.~(\ref{eq:tcr}). Damping coefficient
in Eq.~(\ref{eq:C}) is chosen as $\alpha=0.1$. A total duration
of $1\,\unit{s}$ is analyzed with 55,556 steps.

\begin{figure}
\hfill{}\subfloat[Displacement at points $E$, $F$ and $G$]{\includegraphics{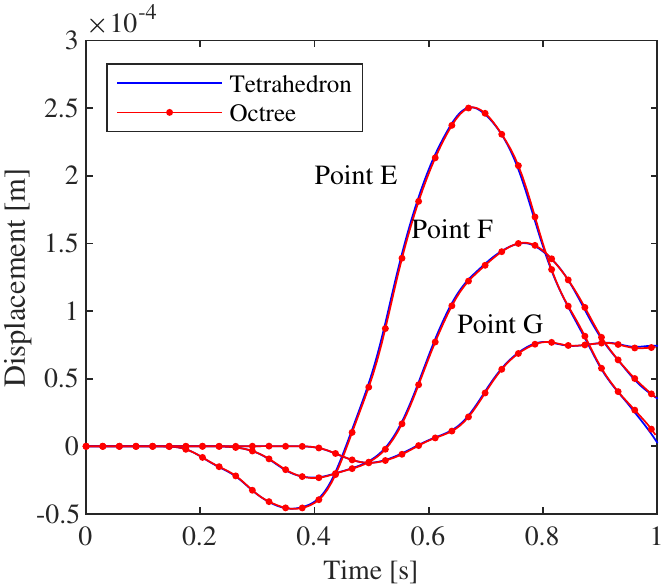}}\hfill{}\subfloat[Acceleration at point $E$]{\includegraphics{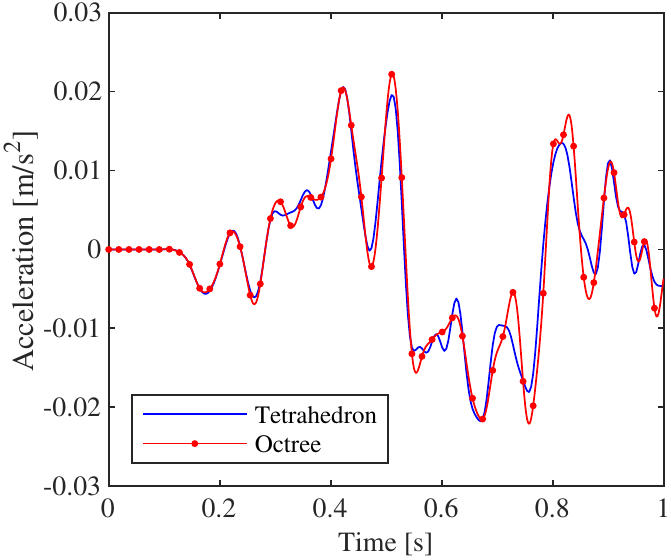}}\hfill{}

\hfill{}\subfloat[Acceleration at point $F$]{\includegraphics{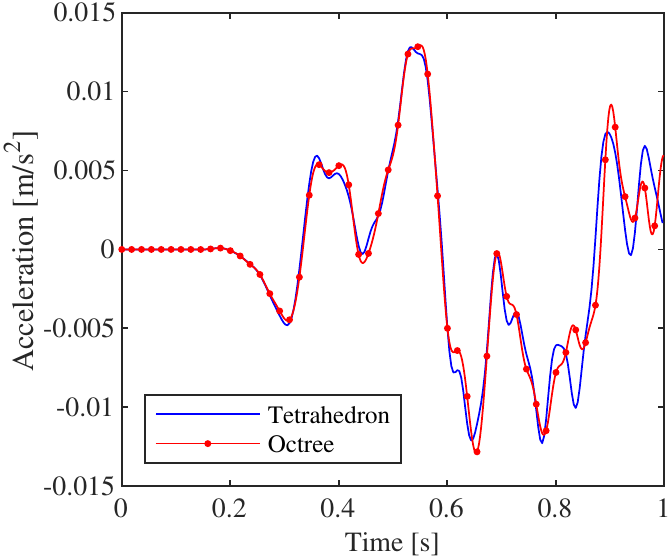}}\hfill{}\subfloat[Acceleration at point $G$]{\includegraphics{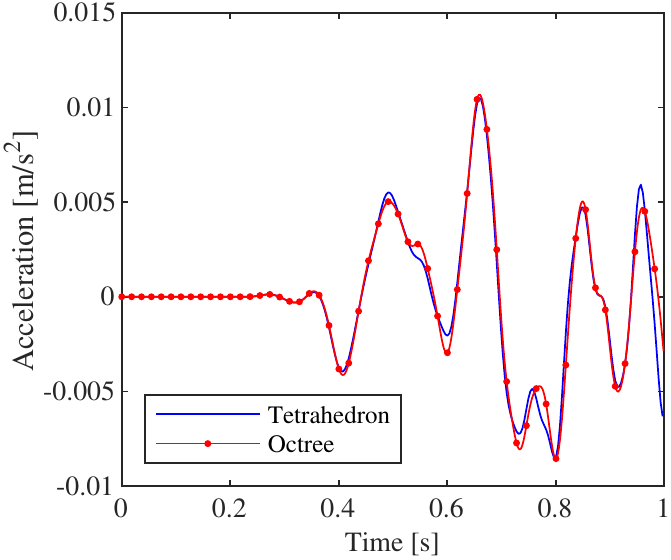}}\hfill{}

\caption{Displacement and acceleration history at selected points on the ground
surface obtained with in-house code and ABAQUS\label{fig:Ground_history}}
\end{figure}

\begin{figure}
\hfill{}\subfloat[$t=0.15\,\unit{s}$]{\includegraphics[width=0.4\textwidth]{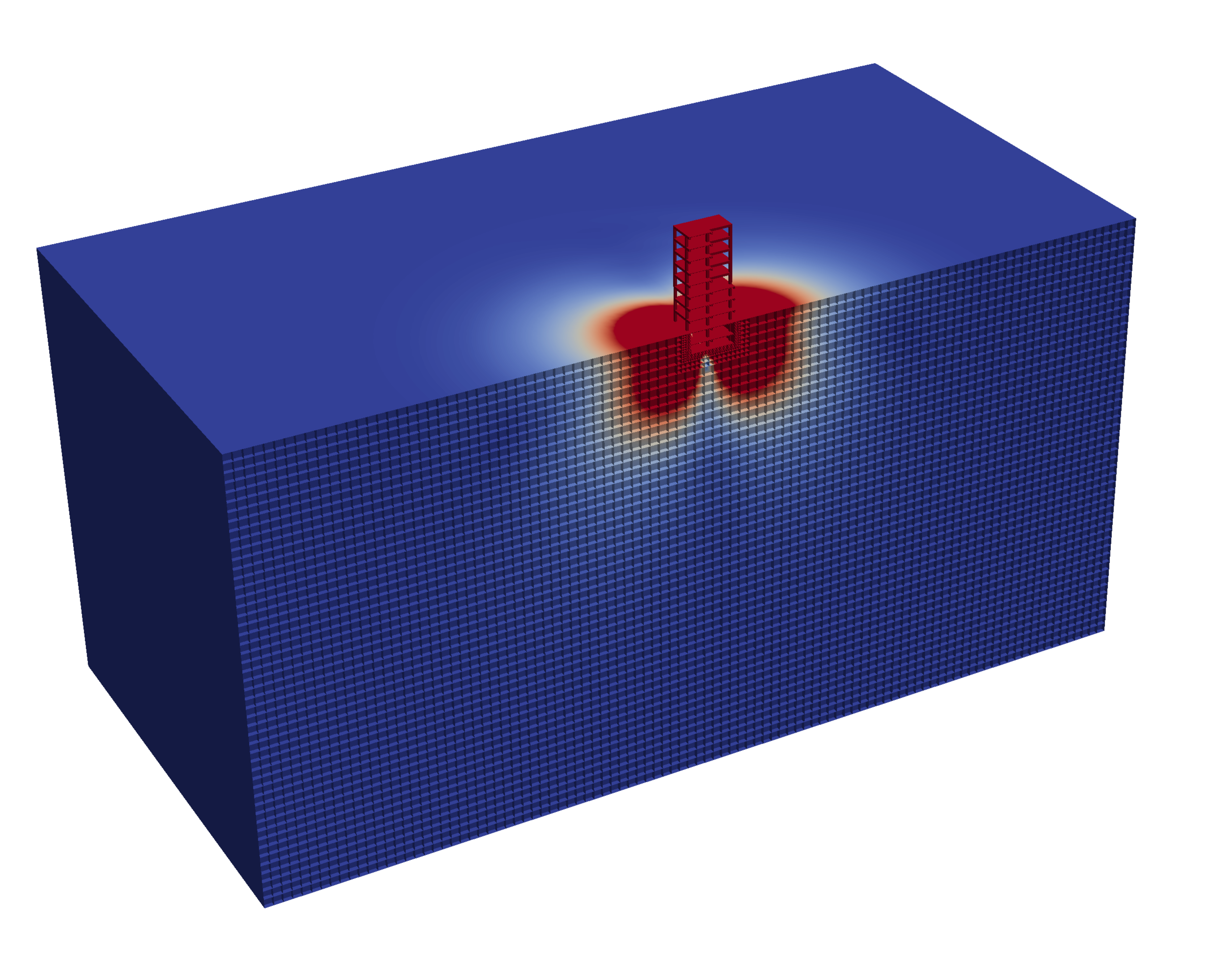}}\hfill{}\subfloat[$t=0.3\,\unit{s}$]{\includegraphics[width=0.4\textwidth]{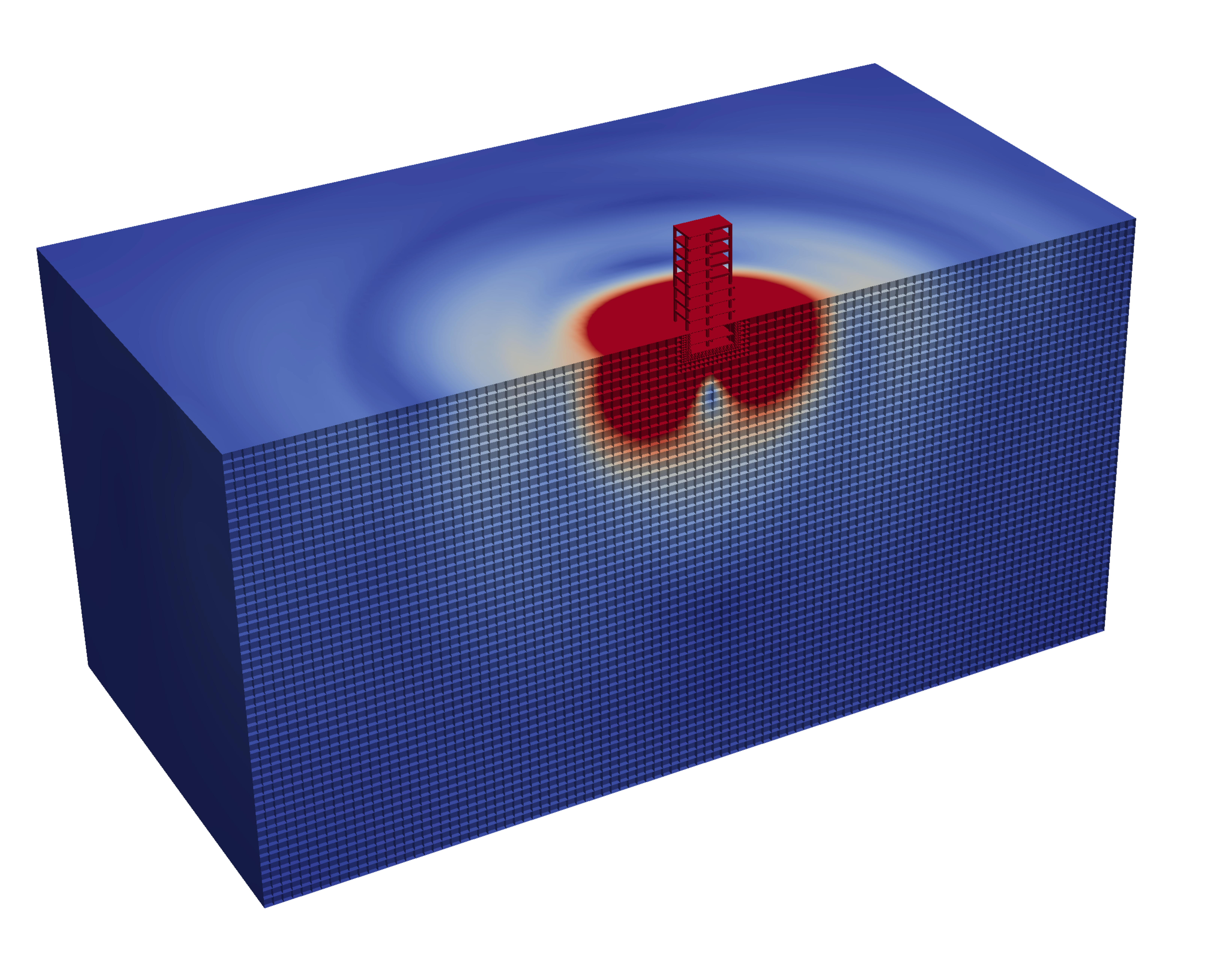}}\hspace{1cm}\hfill{}

\hfill{}\subfloat[$t=0.45\,\unit{s}$]{\includegraphics[width=0.4\textwidth]{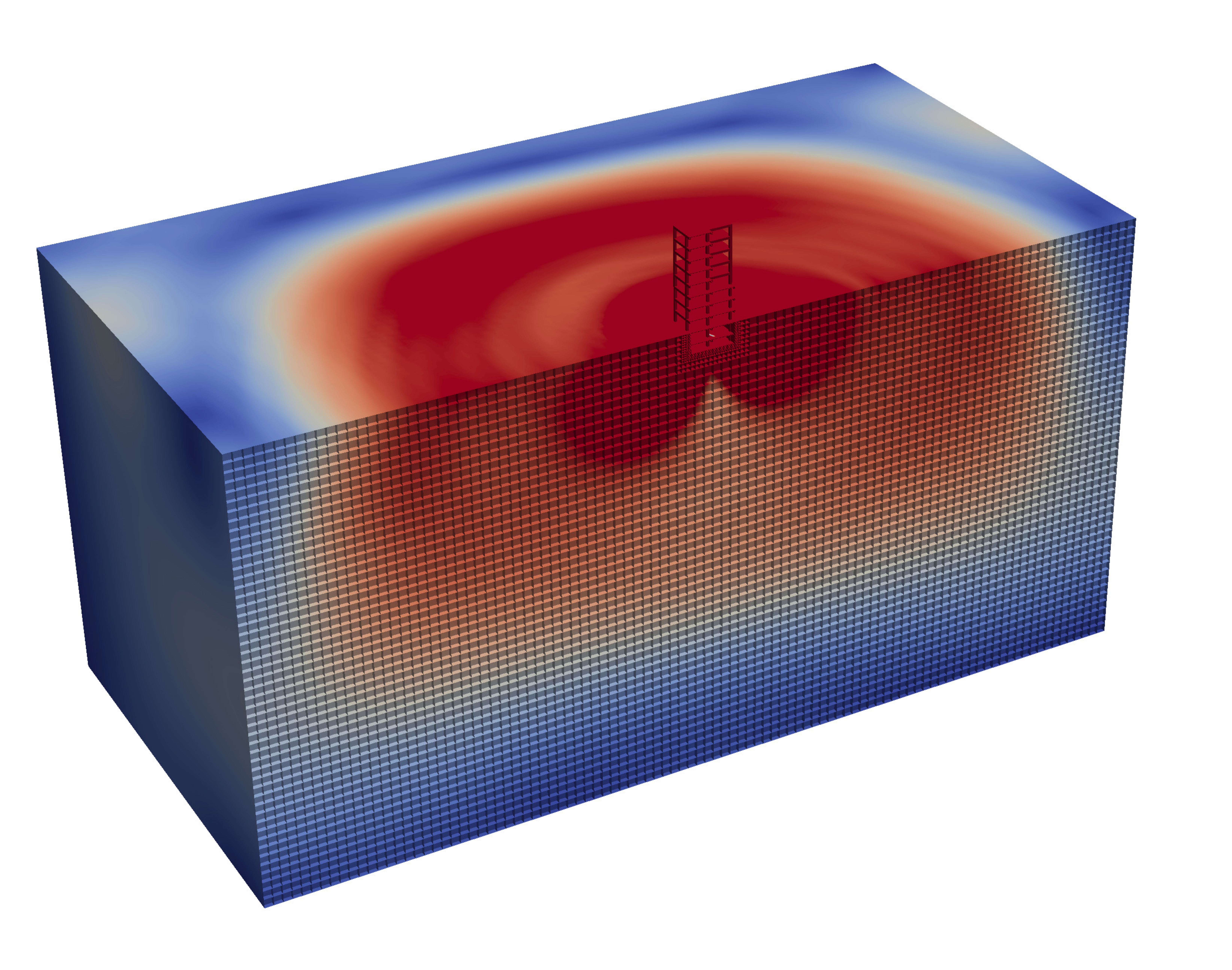}}\hfill{}\subfloat[$t=0.6\,\unit{s}$]{\includegraphics[width=0.4\textwidth]{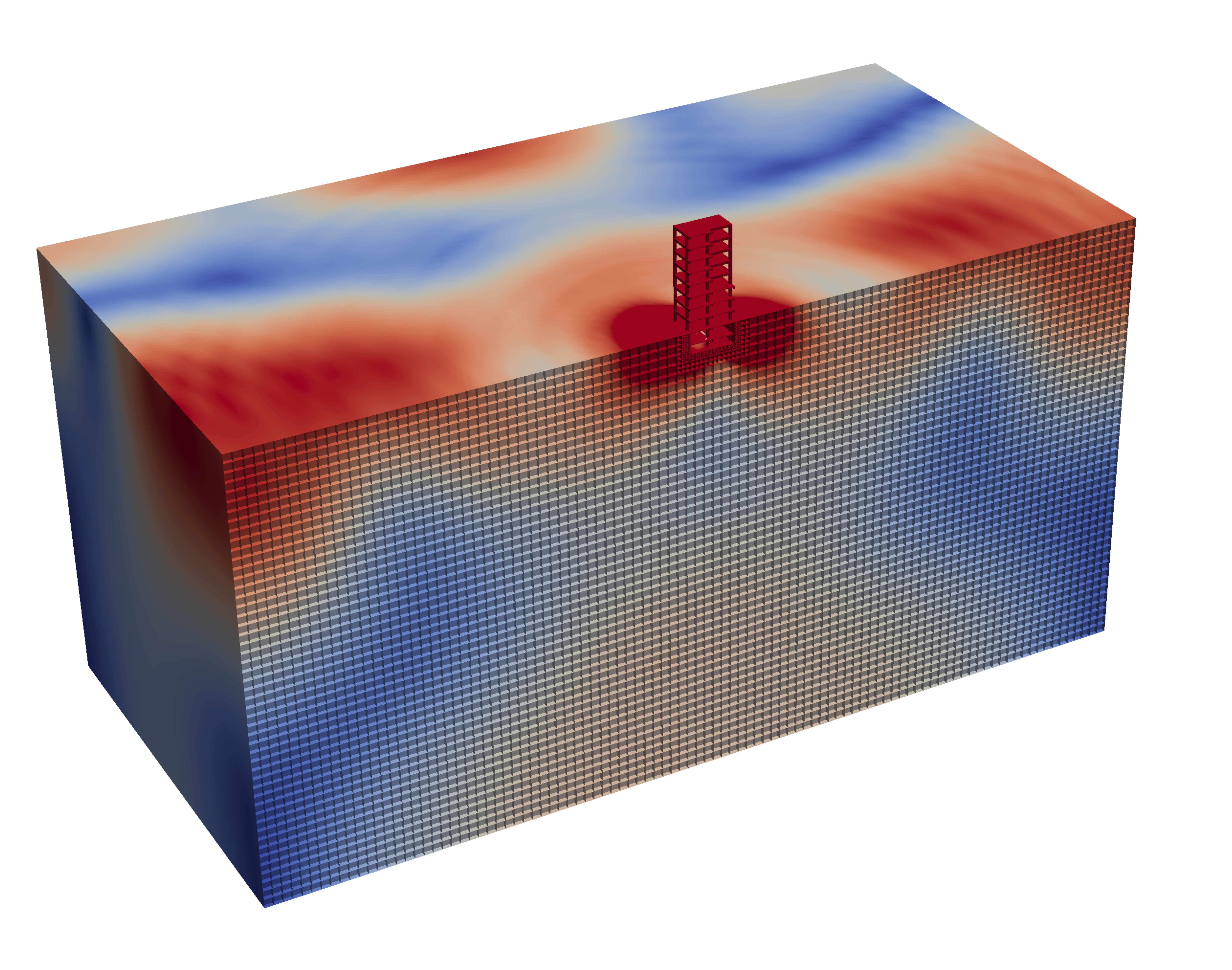}}\includegraphics{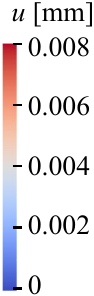}\hfill{}

\caption{Contour of displacement magnitude in the ground at various time instances
$t$\label{fig:Frame_disp}}
\end{figure}

Three points $E\!\left(0\,\unit{m},-30\,\unit{m},0\,\unit{m}\right)$,
$F\!\left(0\,\unit{m},-60\,\unit{m},0\,\unit{m}\right)$ and $G\!\left(0\,\unit{m},-90\,\unit{m},0\,\unit{m}\right)$
indicated in Fig.~\ref{fig:Frame_geo} and located at a certain distance
from the building are selected for comparison purposes. The corresponding
displacement and acceleration response histories are plotted in Fig.~\ref{fig:Ground_history}.
It is observed that similar results are obtained with ABAQUS using
the tetrahedral mesh and with our in-house code utilizing the octree
mesh. The process of wave propagation in the ground is illustrated
in Fig.~\ref{fig:Frame_disp} at four selected time instances.

\begin{figure}
\hfill{}\subfloat[Computational time taken by ABAQUS and the in-house code with different
number of cores\label{fig:Frame_time_1}]{\includegraphics{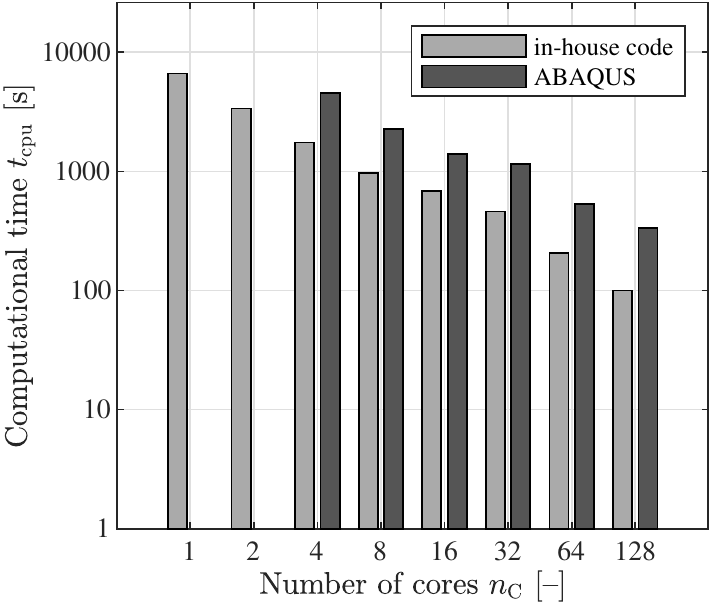}}\hfill{}\subfloat[Speedup and efficiency of in-house code at different number of cores\label{fig:Frame_time_2}]{\includegraphics{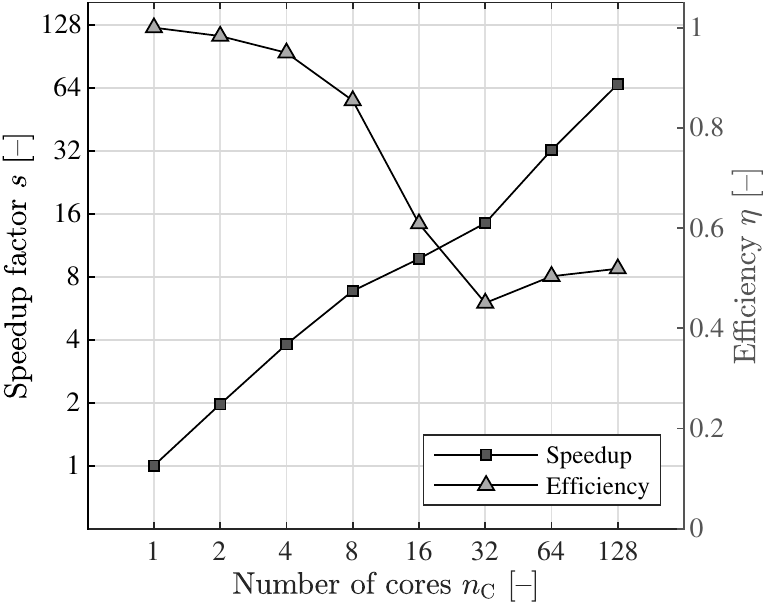}}\hfill{}

\caption{Comparison of computational time of the in-house code using octree
mesh and ABAQUS using tetrahedral mesh \label{fig:Frame_time}}
\end{figure}

Since the computational time increases linearly with the number of
steps of time integration, the CPU time required to complete 1000
steps is selected in measuring the computational performance. This
choice is made to facilitate the evaluation of the speedup of parallel
computing, starting from a small number of cores. The computational
times taken by the in-house code implementing the present technique
and by ABAQUS are listed in Table~\ref{tab:Frame_time} and are plotted
in a log-log scale in Fig.~\ref{fig:Frame_time_1}. To limit the
wall-clock time required by the ABAQUS analyses, we opt for using
four or more computer cores. It is observed that our in-house code
is significantly faster than the commercial software ABAQUS when using
a tetrahedral mesh to obtain an accuracy similar to our octree-based
solver. The speedup compared to a single core simulation and the efficiency
of the proposed method are shown in Fig.~\ref{fig:Frame_time_2}.

\begin{table}
\caption{Computational time, speedup and efficiency of in-house code and ABAQUS\label{tab:Frame_time}}

\hfill{}%
\begin{tabular}{|c|c|c|c|c|c|c|c|}
\hline 
\multirow{2}{*}{$N_{\mathrm{cores}}$} & ABAQUS & \multicolumn{5}{c|}{in-house code} & \multirow{2}{*}{Ratio $t_{\mathrm{T\textrm{-ABAQUS}}}/t_{\mathrm{T\textrm{-in-house}}}$}\tabularnewline
\cline{2-7} \cline{3-7} \cline{4-7} \cline{5-7} \cline{6-7} \cline{7-7} 
 & $t_{\mathrm{T}}$ {[}s{]} & $t_{\mathrm{C}}$ {[}s{]} & $t_{\mathrm{W}}$ {[}s{]} & $t_{\mathrm{T}}$ {[}s{]} & $s$ & $\eta$ & \tabularnewline
\hline 
\hline 
1 & - & 6565.47 & - & 6566.12 & 1.00 & 1.00 & -\tabularnewline
\hline 
2 & - & 3078.62 & 155.46 & 3234.37 & 2.03 & 1.02 & -\tabularnewline
\hline 
4 & 4550 & 1651.08 & 92.94 & 1744.20 & 3.76 & 0.94 & 2.61\tabularnewline
\hline 
8 & 2264 & 894.91 & 58.47 & 953.49 & 6.89 & 0.86 & 2.37\tabularnewline
\hline 
16 & 1407 & 634.49 & 45.90 & 680.48 & 9.65 & 0.60 & 2.07\tabularnewline
\hline 
32 & 1154 & 355.03 & 94.84 & 449.92 & 14.59 & 0.46 & 2.56\tabularnewline
\hline 
64 & 534 & 141.43 & 63.28 & 204.73 & 32.07 & 0.50 & 2.61\tabularnewline
\hline 
128 & 334 & 52.18 & 41.66 & 93.85 & 69.97 & 0.55 & 3.56\tabularnewline
\hline 
\multicolumn{8}{l}{$t_{\mathrm{C}}$: Calculation time}\tabularnewline
\multicolumn{8}{l}{$t_{\mathrm{W}}$: Waiting and communication time}\tabularnewline
\multicolumn{8}{l}{$t_{\mathrm{T}}$: Total time}\tabularnewline
\multicolumn{8}{l}{$s$: Speedup factor}\tabularnewline
\multicolumn{8}{l}{$\eta$: Efficiency}\tabularnewline
\end{tabular}\hfill{}
\end{table}

We observe that with an increasing number of cores the computational
time is significantly reduced in both our in-house research code and
in ABAQUS. Additionally, it is noted that the proposed explicit solver
outperforms ABAQUS by a factor of at least two, meaning that for this
example only half of the computational time or less is required (Table~\ref{tab:Frame_time}).
The parallel efficiency for the frame example stays in a reasonable
range around 50\% for 128 cores. For 128 cores, only around 230,000
DOFs are distributed to each core, which might be one source for the
decreasing efficiency with an increasing number of cores. The results
depicted in Fig.~\ref{fig:Frame_time_2} indicate that it is preferable
to partition the mesh such that more than one million DOFs are distributed
to each core. This is related to a more favorable ratio between time
spent on calculating nodal forces and displacements in contrast to
the communication and waiting time; therefore, the computational resources
can be used more efficiently.

\FloatBarrier

\subsection{Mountainous region\label{subsec:Everest}}

\begin{figure}[tb]
\hfill{}\subfloat[Satellite image \label{fig:Everest_geo_1}]{\includegraphics[width=0.3\textwidth]{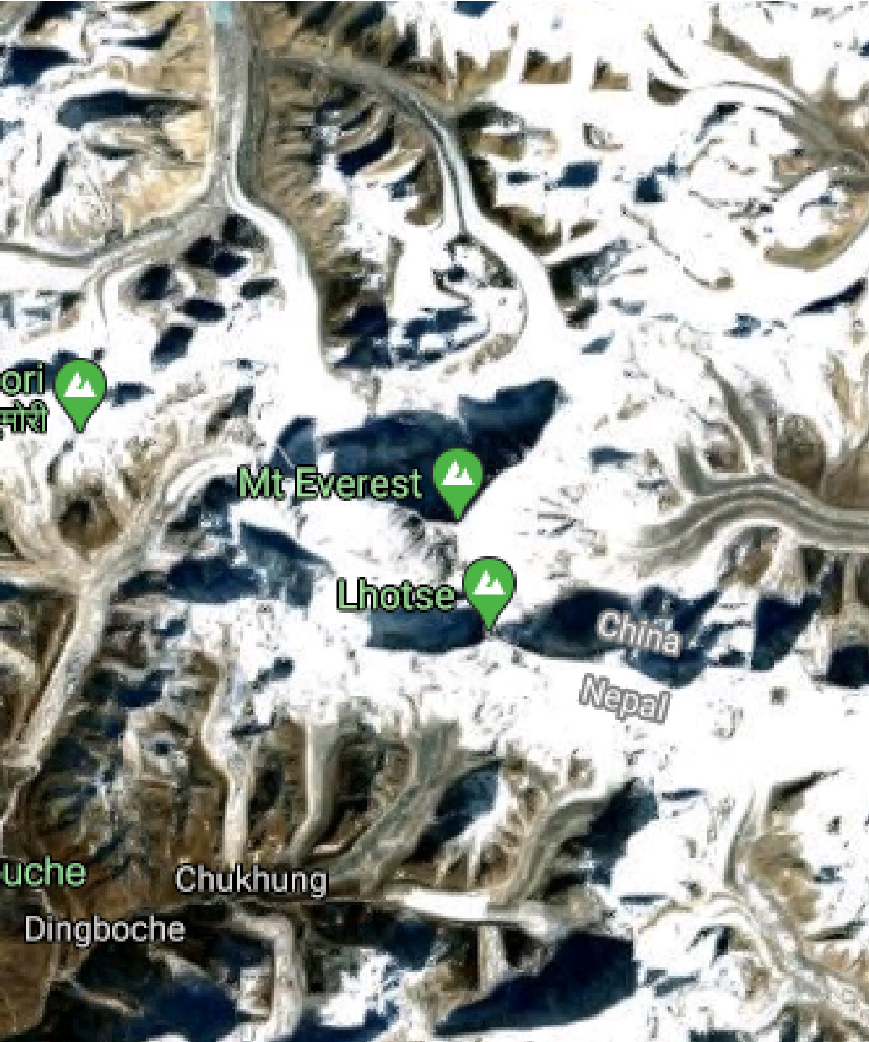}}\hfill{}\subfloat[Terrain \label{fig:Everest_geo_2}]{\includegraphics[width=0.3\textwidth]{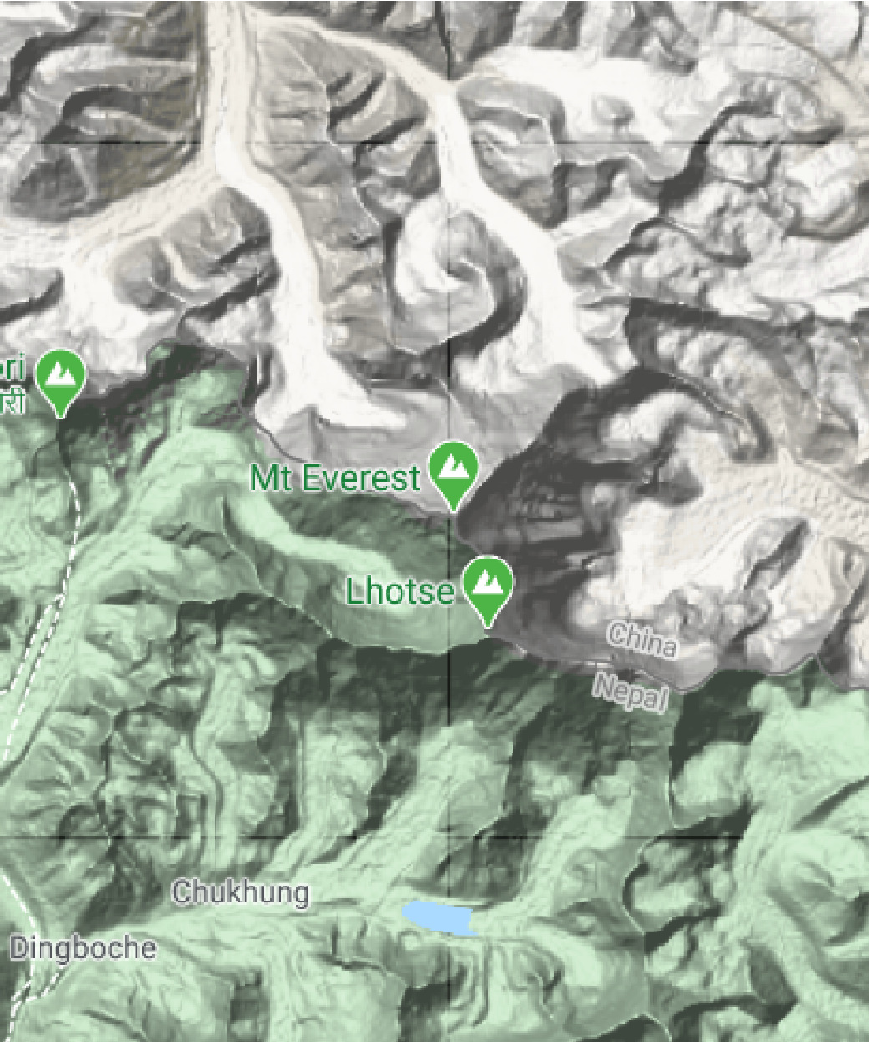}}\hfill{}\subfloat[STL model \label{fig:Everest_geo_3}]{\includegraphics[width=0.3\textwidth]{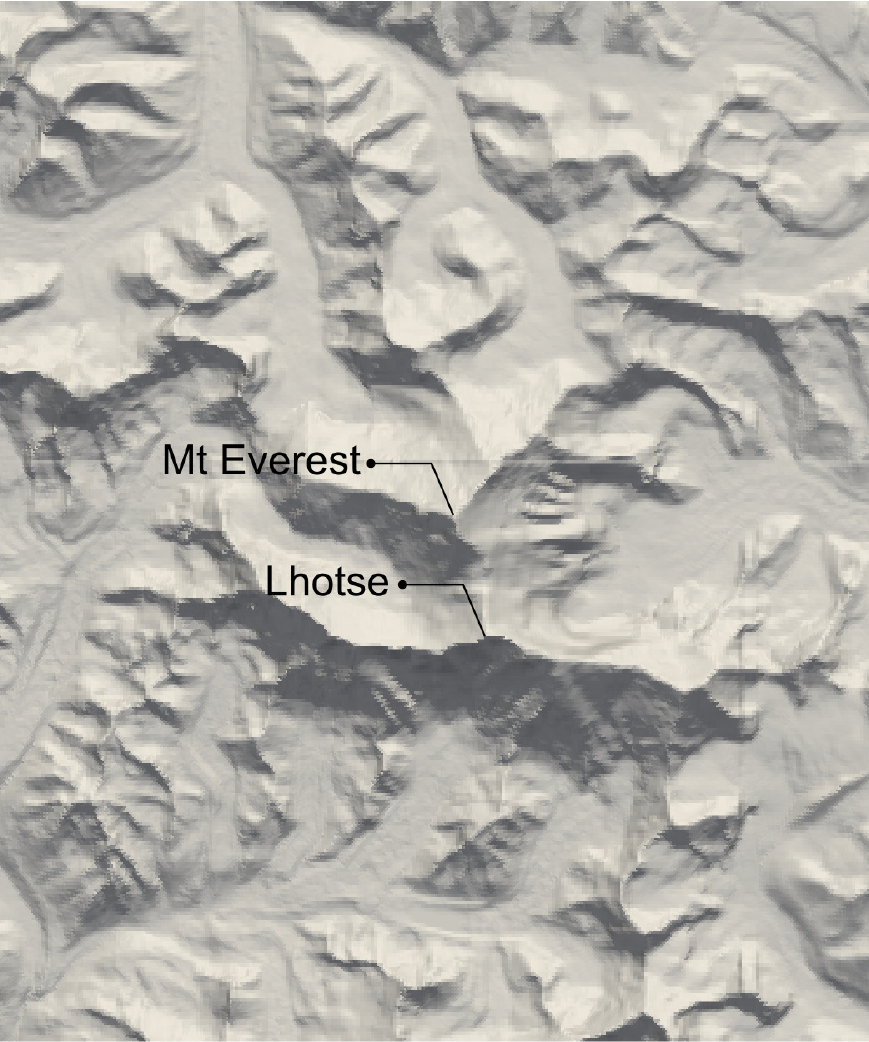}}\hfill{}\caption{Geometry of the mountainous region\label{fig:Everest_geo}}
\end{figure}

\begin{figure}[tb]
\hfill{}\includegraphics{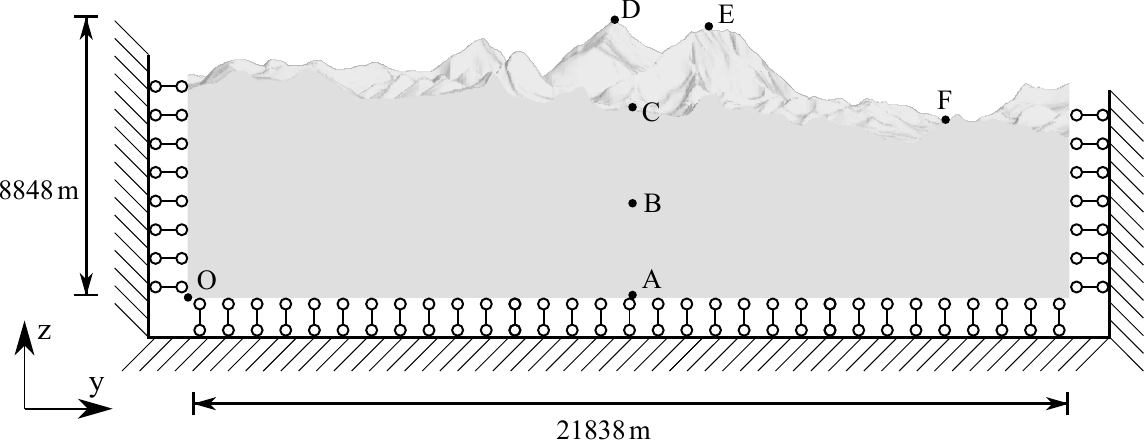}\hfill{}

\caption{Boundary condition and dimensions of the mountainous region\label{fig:Everest_bc}}
\end{figure}

In this example, wave propagation in a mountainous region of size
$21.8\,\unit{km}\times18.0\,\unit{km}$ is investigated as an indication
of the practical applicability of the proposed parallel explicit solver.
The location of the region is shown in a satellite image from Google
Maps in Fig.~\ref{fig:Everest_geo_1}, while the terrain information
(Fig.~\ref{fig:Everest_geo_2}) and the STL model (Fig.~\ref{fig:Everest_geo_3})
are generated using the online service Terrain2STL at \texttt{http://jthatch.com/Terrain2STL/}~\citep{JThatch}.
For illustration purposes, the model is assumed to be homogeneous
with a Young's modulus $E=55\,\unit{GPa}$, Poisson's ratio $\nu=0.2$
and mass density $\rho=2400\,\unitfrac{kg}{m^{3}}$. The dilatational
wave speed is $5046\,\unitfrac{m}{s}$, while the shear wave speed
is $3090\,\unitfrac{m}{s}$. The boundary conditions of the model
are shown in Fig.~\ref{fig:Everest_bc}, where the displacements
perpendicular to the side faces and the bottom are constrained. An
excitation in the form of a Ricker wavelet (Fig.~\ref{fig:Ricker})
with $t_{1}=0.5\,{\textstyle \mathrm{s}}$ and $P_{0}=1.157\times10^{15}\,\unit{N}$
is applied at the center of the bottom of the model (point $A$) in
the $y$ direction. The maximum frequency of interest is $5\,\unit{Hz}$,
therefore, the wavelength of the shear wave is $618\,\unit{m}$.

\begin{figure}
\hfill{}\subfloat[Displacements \label{fig:Everest_his_1}]{\includegraphics[scale=0.8]{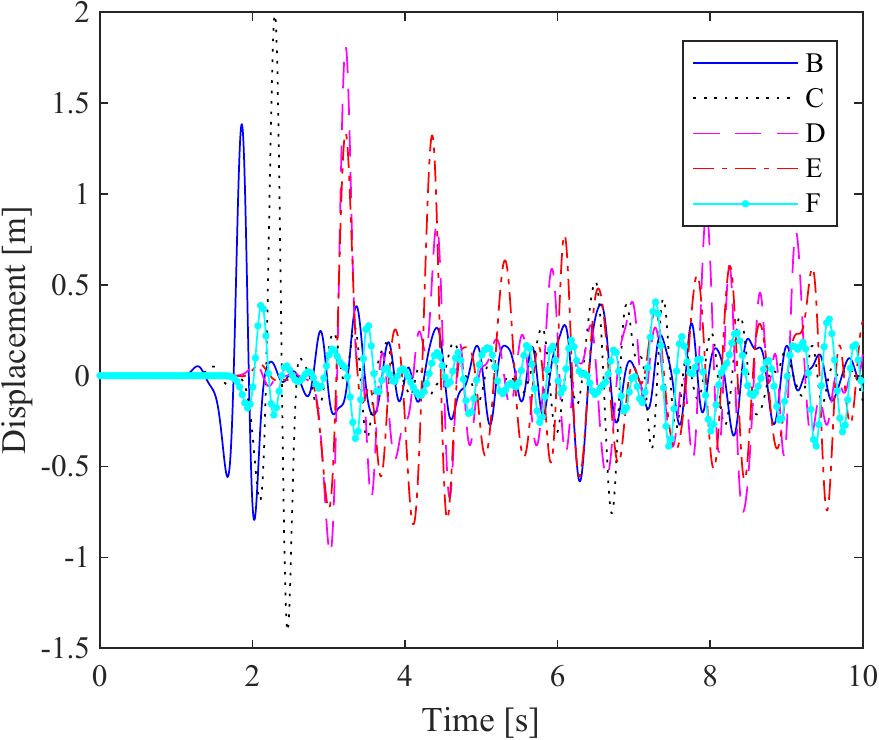}}\hfill{}\subfloat[Accelerations \label{fig:Everest_his_2}]{\includegraphics[scale=0.8]{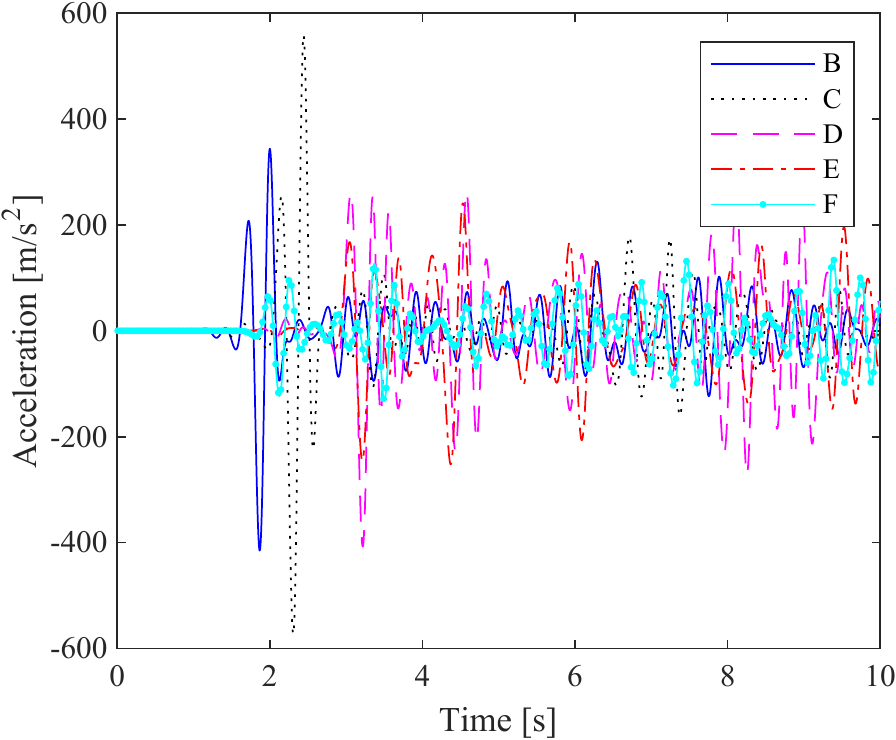}}\hfill{}

\caption{Displacements and accelerations in $y$ direction at five selected
points in the mountainous region model \label{fig:Everest_history}}
\end{figure}

\begin{figure}
\hfill{}\subfloat[$t=0.5\,\unit{s}$]{\includegraphics[width=0.5\textwidth]{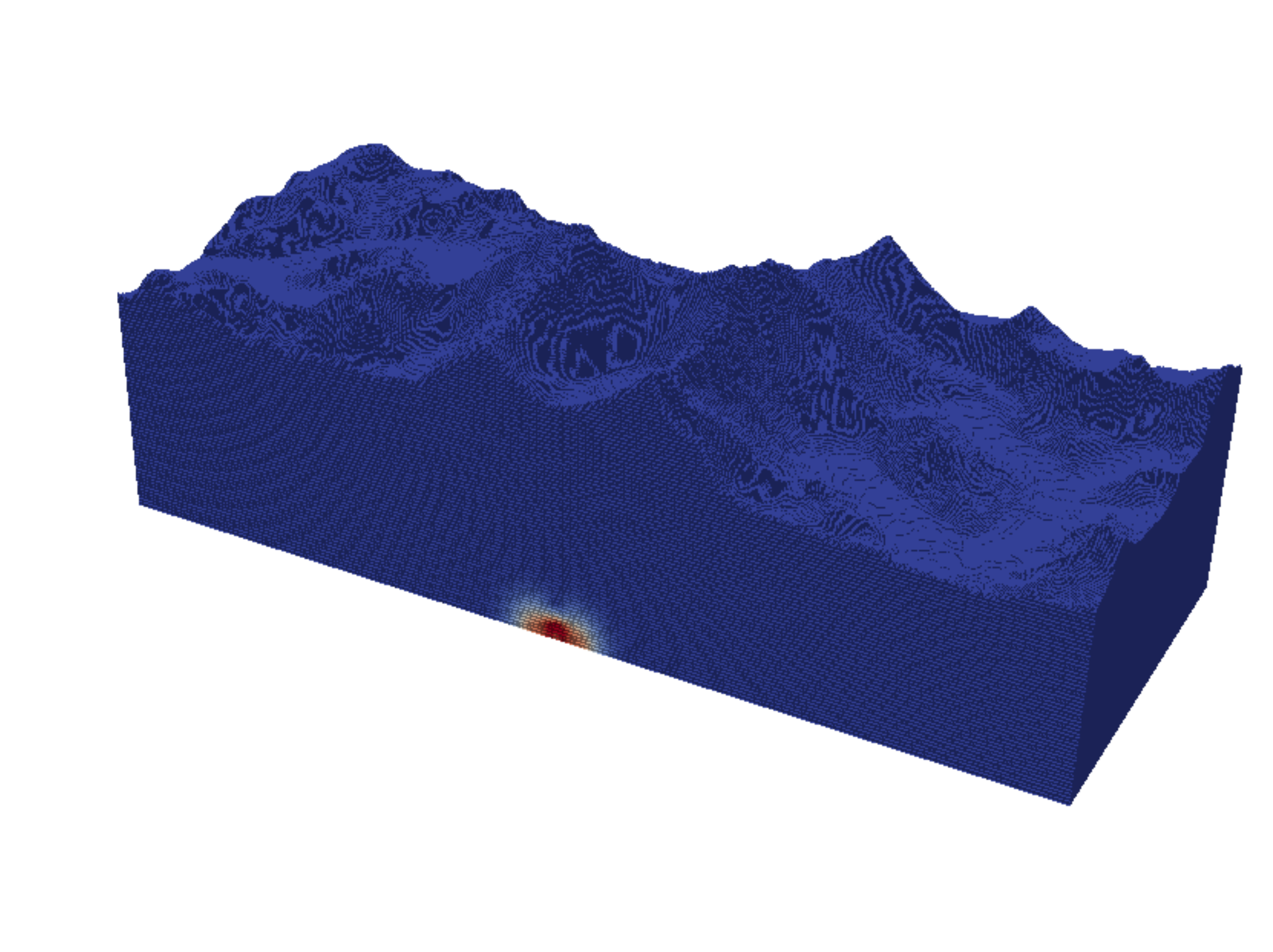}}\hfill{}\subfloat[$t=1\,\unit{s}$]{\includegraphics[width=0.5\textwidth]{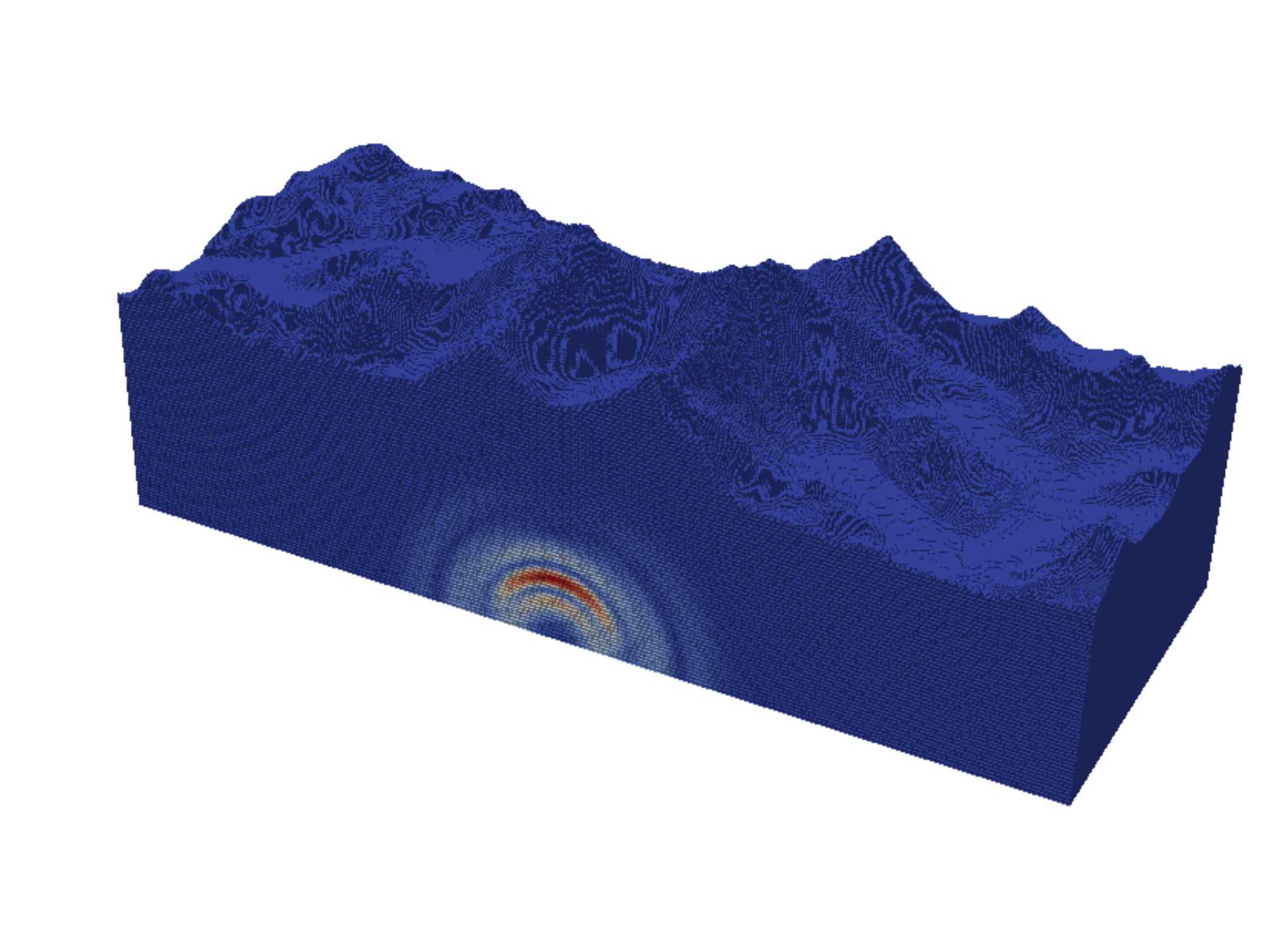}}\hfill{}

\hfill{}\subfloat[$t=1.5\,\unit{s}$]{\includegraphics[width=0.5\textwidth]{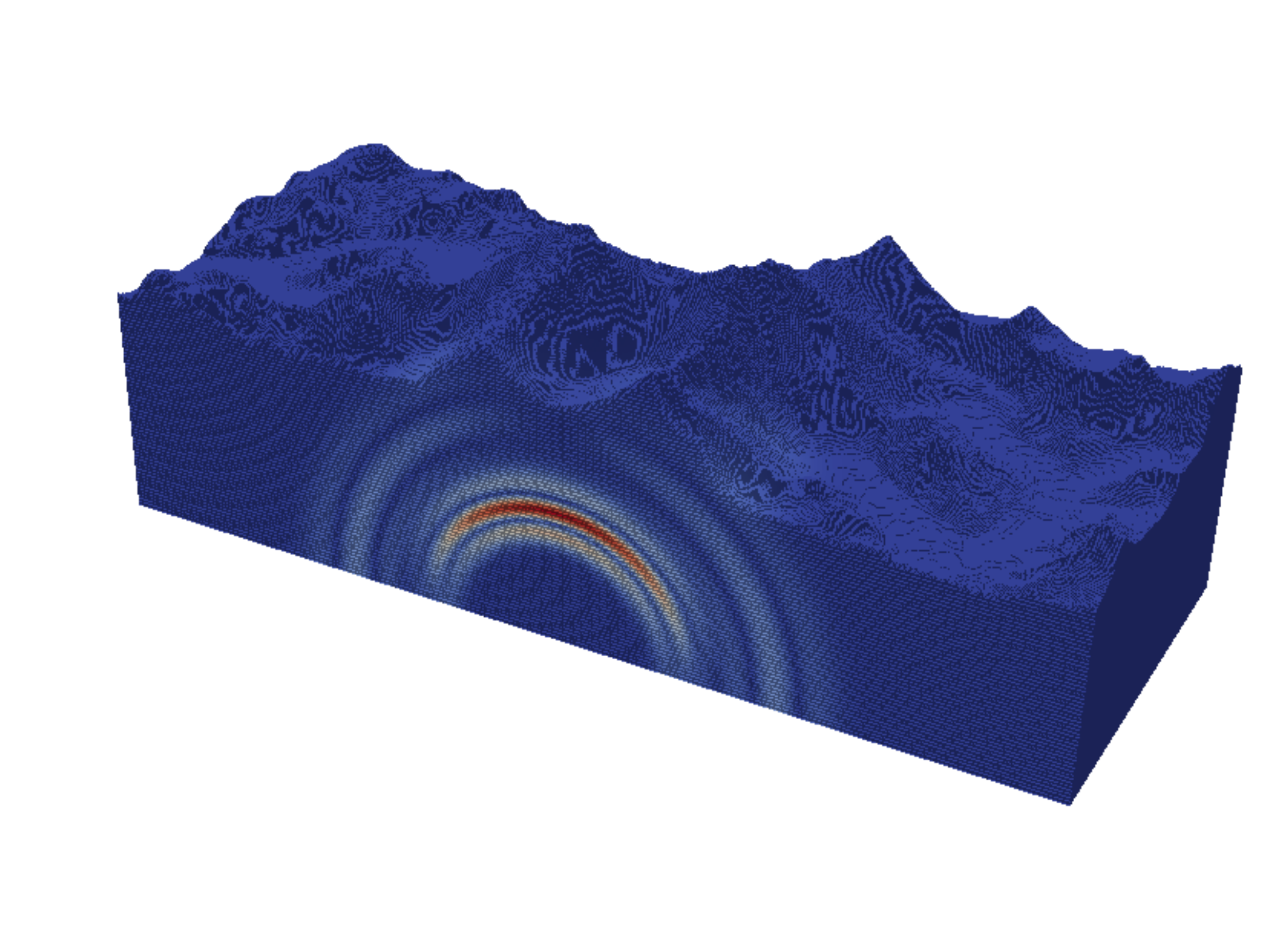}}\hfill{}\subfloat[$t=2\,\unit{s}$]{\includegraphics[width=0.5\textwidth]{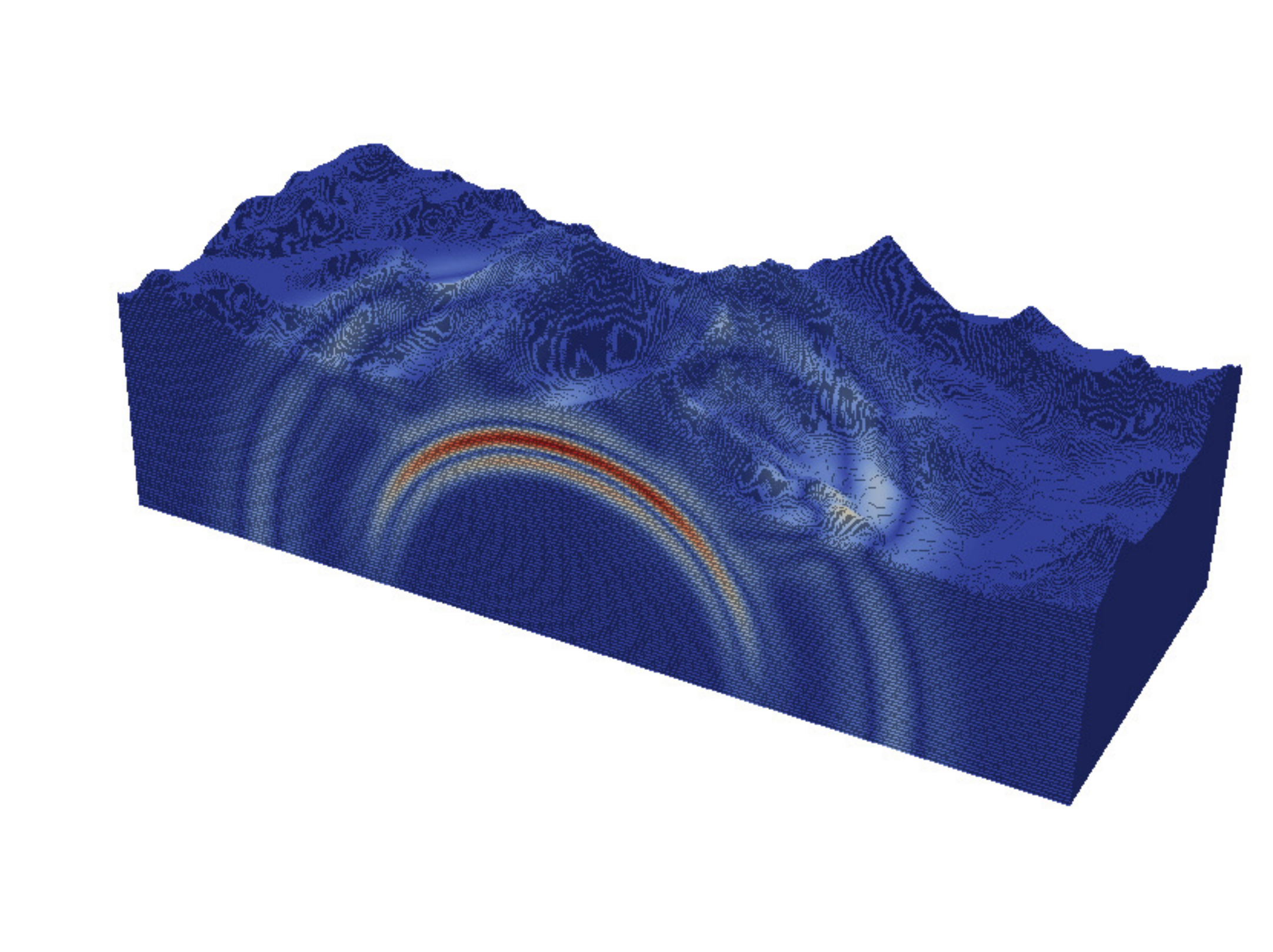}}\hfill{}

\caption{Contour of displacement magnitude of the mountain region model at
four selected time instances $t$\label{fig:Everest_disp}}
\end{figure}

An octree mesh is generated with a maximum element size of $68.125\,\unit{m}$
and a minimum element size of $8.516\,\unit{m}.$ Hence, the mesh
consists of 7,608,933 elements and 32,052,075 DOFs. The mesh is partitioned
using the algorithm proposed in Ref.~\citep{Gilbert1998}. The obtained
mesh partitioning is depicted in \ref{sec:Everest_part}. The time
step is chosen as $\Delta t=1.38\,\unit{ms}$. The response is analyzed
for a duration of $10\,\unit{s}$ using 7246 time steps. The results
at the points $B\!\left(9000\,\unit{m},10000\,\unit{m},4000\,\unit{m}\right)$,
$C\!\left(7848\,\unit{m},11090\,\unit{m},5460\,\unit{m}\right)$,
$D\!\left(11431\,\unit{m},11772\,\unit{m},7848\,\unit{m}\right)$,
$E\!\left(11772\,\unit{m},9042\,\unit{m},7336\,\unit{m}\right)$ and
$F\!\left(9895\,\unit{m},4095\,\unit{m},4265\,\unit{m}\right)$ are
selected to evaluate the results and to illustrate the time-dependent
displacements. The origin of the coordinate system is point $O$ in
Fig.~\ref{fig:Everest_bc}. The displacement and acceleration in
$y$ direction of these points are plotted against time in Fig.~\ref{fig:Everest_history}.
The contour of displacement magnitude at four time instances are presented
in Fig.~\ref{fig:Everest_disp}, illustrating the wave propagation
process.

\begin{figure}
\hfill{}\subfloat[Computational time \label{fig:Everest_time_1}]{\includegraphics{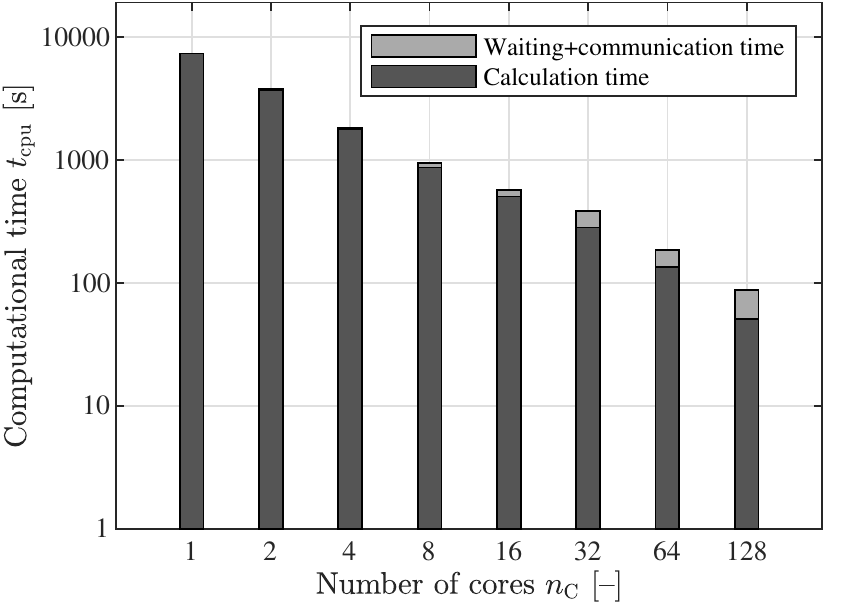}}\hfill{}\subfloat[Speedup and efficiency \label{fig:Everest_time_2}]{\includegraphics{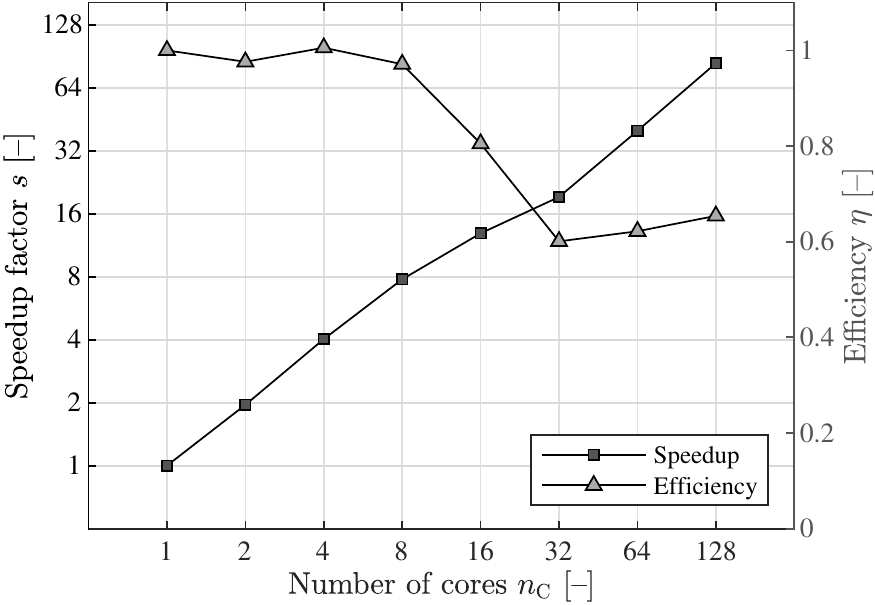}}\hfill{}

\caption{Computational time, speedup and efficiency of the mountainous region
simulation using different numbers of cores\label{fig:Everest_time}}
\end{figure}

The computational time and speedup for 1000 steps using different
numbers of cores are compared in Fig.~\ref{fig:Everest_time} and
Table~\ref{tab:Everest_time}. The presented results in terms of
the speedup factor and the parallel efficiency are in good agreement
with those reported for the frame example in Section~\ref{subsec:Benchmark}.
We again observe that the efficiency of the parallel explicit solver
decreases when fewer DOFs are distributed to each core. However, in
this example, a higher parallel efficiency of roughly 60\% is achieved
due to the shape of the model, which is more regular and allows for
a better partitioning.

\FloatBarrier

\begin{table}
\caption{Computational time, speedup and efficiency of the mountainous region\label{tab:Everest_time}}

\hfill{}%
\begin{tabular}{|c|c|c|c|c|c|}
\hline 
$N_{\mathrm{cores}}$ & $t_{\mathrm{C}}$ {[}s{]} & $t_{\mathrm{W}}$ {[}s{]} & $t_{\mathrm{T}}$ {[}s{]} & $s$ & $\eta$\tabularnewline
\hline 
\hline 
1 & 7273.36 & - & 7273.94 & 1.00 & 1.00\tabularnewline
\hline 
2 & 3650.19 & 46.03 & 3696.50 & 1.97 & 0.99\tabularnewline
\hline 
4 & 1788.73 & 36.26 & 1825.15 & 3.99 & 1.00\tabularnewline
\hline 
8 & 868.96 & 74.28 & 943.34 & 7.71 & 0.96\tabularnewline
\hline 
16 & 504.91 & 65.58 & 570.55 & 12.7 & 0.79\tabularnewline
\hline 
32 & 280.71 & 101.89 & 382.62 & 19.0 & 0.59\tabularnewline
\hline 
64 & 136.02 & 47.31 & 183.34 & 39.7 & 0.62\tabularnewline
\hline 
128 & 51.72 & 34.58 & 86.30 & 84.3 & 0.66\tabularnewline
\hline 
\multicolumn{6}{l}{$t_{\mathrm{C}}$: Calculation time}\tabularnewline
\multicolumn{6}{l}{$t_{\mathrm{W}}$: Waiting and communication time}\tabularnewline
\multicolumn{6}{l}{$t_{\mathrm{T}}$: Total time}\tabularnewline
\multicolumn{6}{l}{$s$: Speedup factor}\tabularnewline
\multicolumn{6}{l}{$\eta$: Efficiency}\tabularnewline
\end{tabular}\hfill{}
\end{table}

\FloatBarrier

\subsection{Sandwich panel\label{subsec:Panel}}

\begin{figure}[tb]
\hfill{}\subfloat[Geometry of the sandwich panel\label{fig:Panel_geo_1}]{\includegraphics[width=0.8\textwidth]{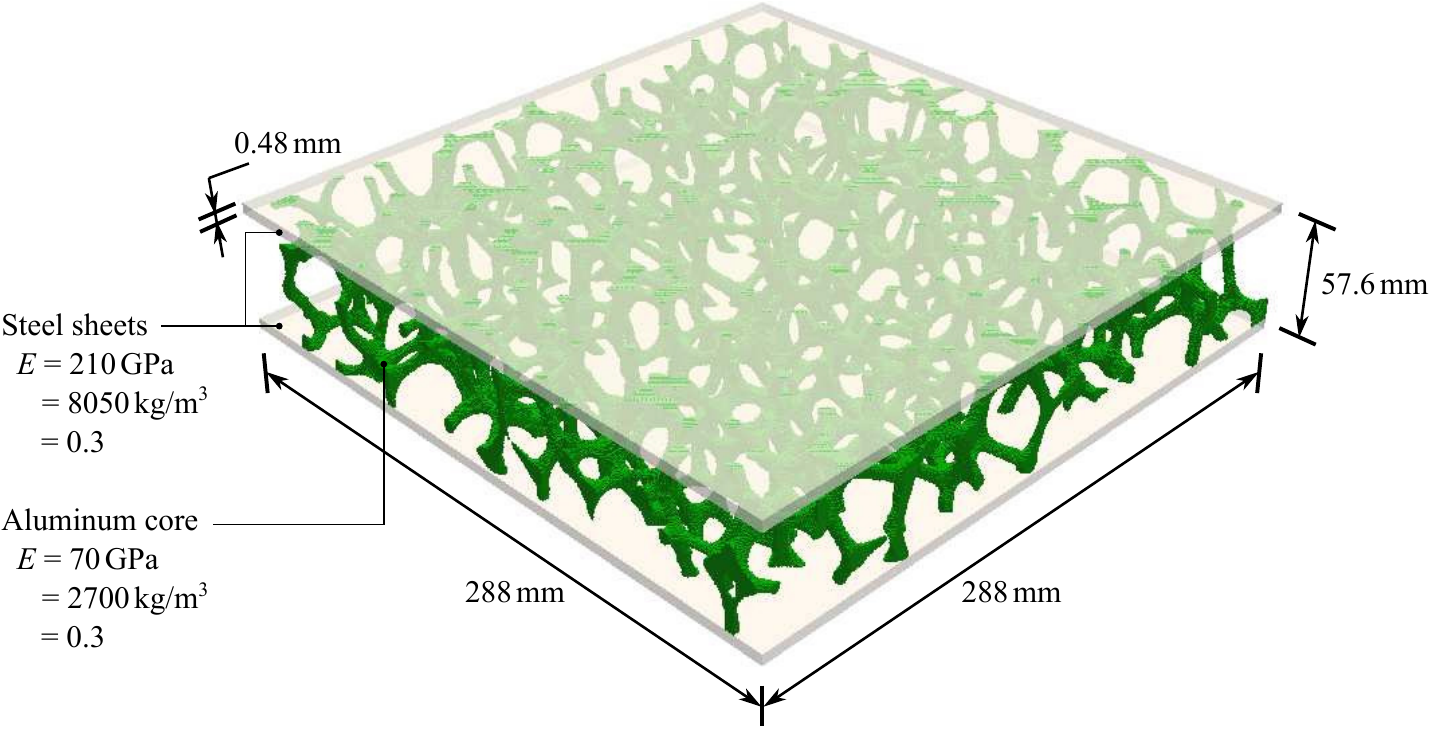}}\hfill{}

\hfill{}\subfloat[Mesh of the sandwich panel\label{fig:Panel_geo_2}]{\includegraphics[width=0.8\textwidth]{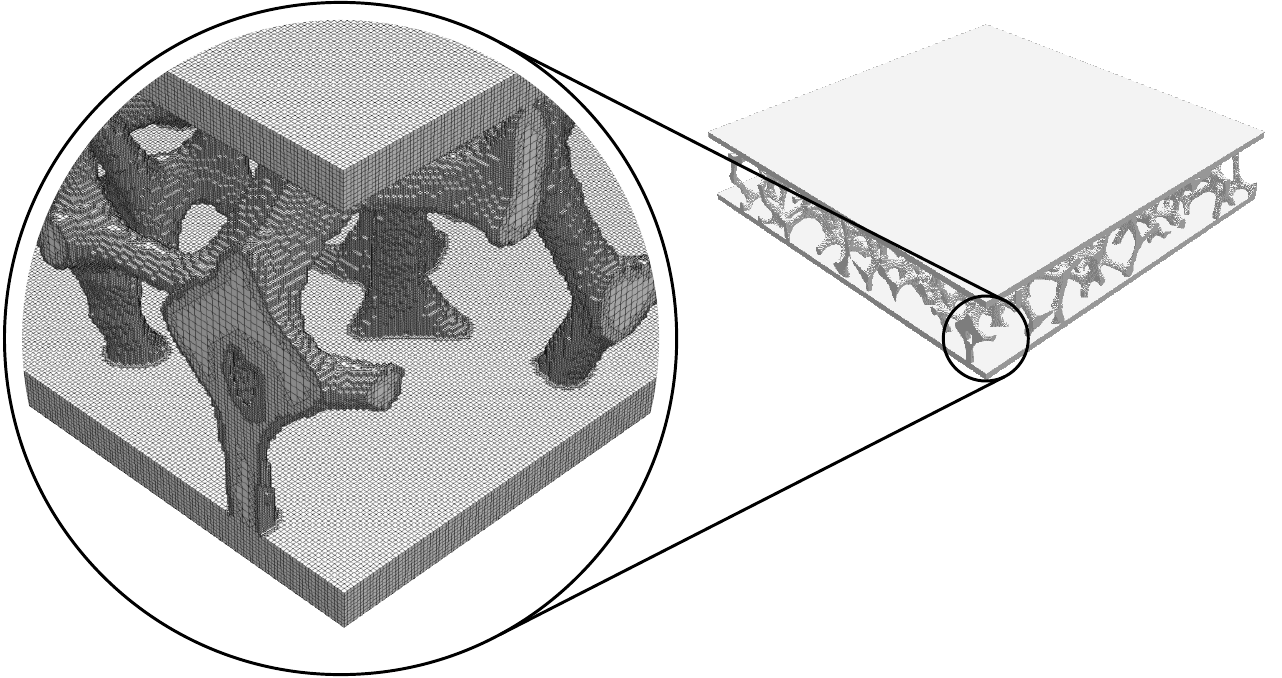}}\hfill{}

\caption{Geometry and mesh of the sandwich panel\label{fig:Panel_geo}}
\end{figure}

\begin{figure}[tb]
\hfill{}\subfloat[$2\times2$ panel\label{fig:Panel_mirror_1}]{\includegraphics[width=0.8\textwidth]{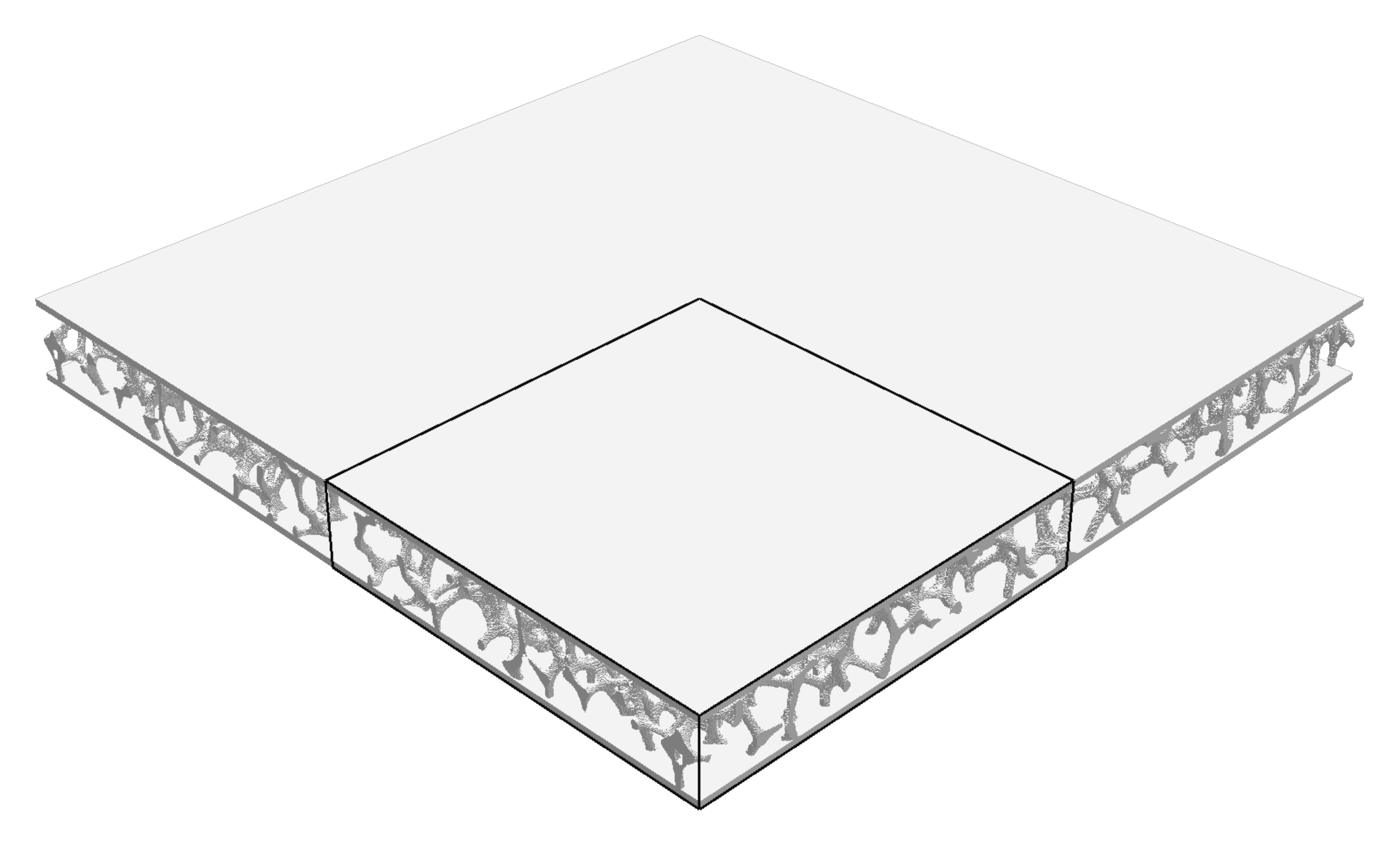}}\hfill{}

\hfill{}\subfloat[$4\times4$ panel\label{fig:Panel_mirror_2}]{\includegraphics[width=0.8\textwidth]{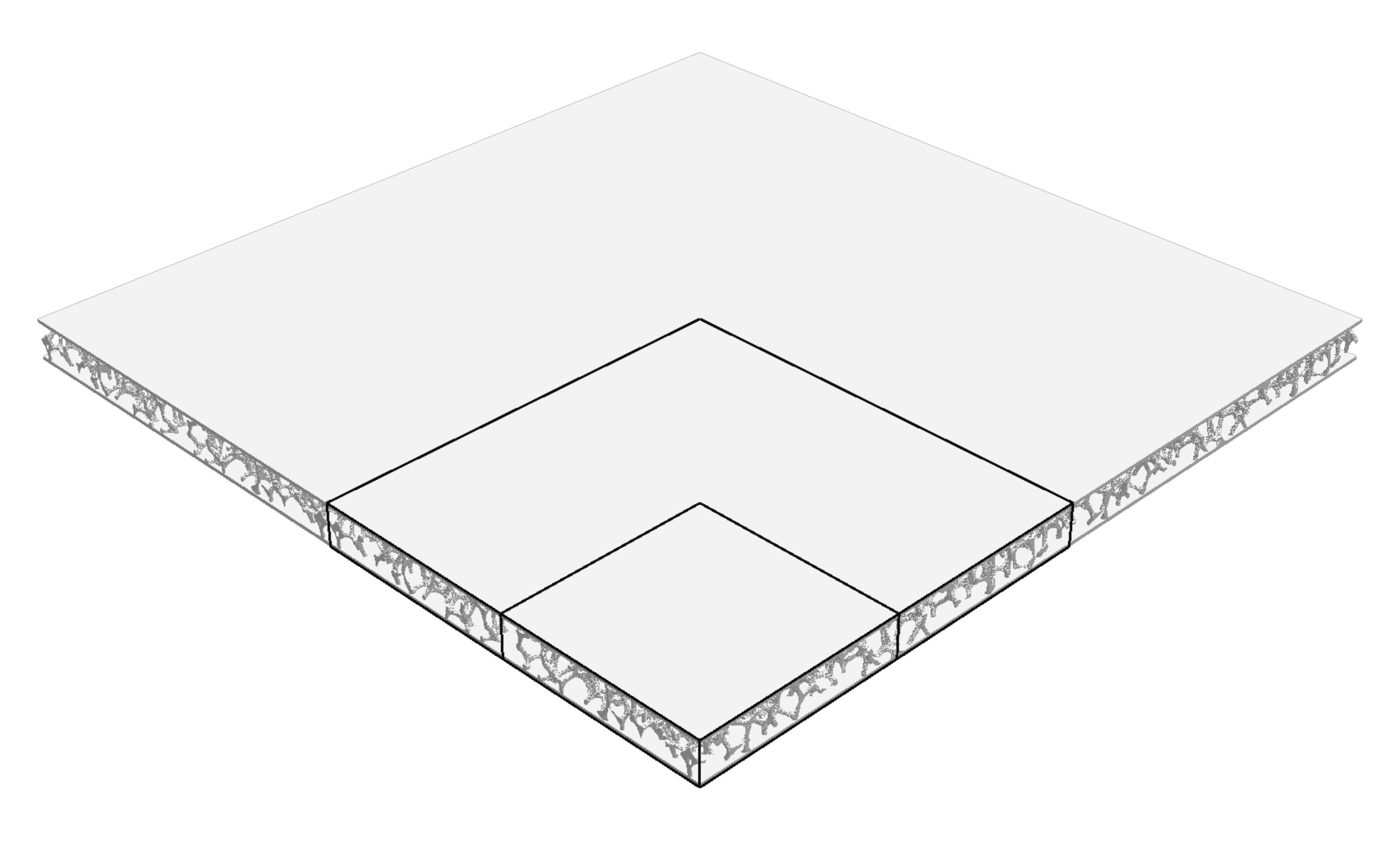}}\hfill{}

\caption{Mirroring operation of the sandwich panel\label{fig:Panel_mirror}}
\end{figure}

In the last example, a sandwich panel with two steel cover sheets
and a foam-like aluminum core is modeled. The digital image, shown
in Fig.~\ref{fig:Panel_geo_1}, is obtained by X-ray CT scans\footnote{The CT data of the aluminum foam core has been provided to the authors
by Prof.~A.~D\"uster and Dr.~Meysam Joulaian~\citep{PhDJoulaian2017,Mossaiby2019}
from Hamburg University of Technology.}. The size of the panel is $288\,\unit{mm}\times288\,\unit{mm}\times57.6\,\unit{mm}$,
while the thickness of the cover sheet is $4.8\,\unit{mm}$. Young's
modulus, Poisson's ratio and the mass density of aluminum are $E_{\mathrm{a}}=70\,\unit{GPa}$,
$\nu_{\mathrm{a}}=0.3$ and $\rho_{\mathrm{a}}=2700\,\unitfrac{kg}{m^{3}}$,
while the corresponding material parameters of steel are taken as
$E_{\mathrm{s}}=210\,\unit{GPa}$, $\nu_{\mathrm{s}}=0.3$ and $\rho_{\mathrm{s}}=8050\,\unitfrac{kg}{m^{3}}$.
An octree mesh is generated with a maximum element size of $0.96\,\unit{mm}$
and a minimum element size of $0.24\,\unit{mm}$, as shown in Fig.~\ref{fig:Panel_geo_2}.
The mesh consists of 13,520,194 elements and 66,645,588 DOFs (for
comparison, a uniform voxel based mesh would contain 89,862,518 elements
and 288,049,833 DOFs). In order to test our approach for massively
parallel computing of large-scale wave propagation problems and examine
the increase of CPU time with the problem-size, we mirror the panel
in the $x$ and $y$ directions twice. Hence, the edge length of the
panel becomes $576\,\unit{mm}$ ($2\times2$ panel, Fig.~\ref{fig:Panel_mirror_1})
and $1152\,\unit{mm}$ ($4\times4$ panel, Fig.~\ref{fig:Panel_mirror_2}),
respectively. In the $2\times2$ panel, there are 54,080,776 elements
and 266,295,885 DOFs, and in the $4\times4$ panel the number of elements
is 216,323,104, and the number of DOFs is 1,064,602,902.

\begin{figure}[tb]
\hfill{}\includegraphics{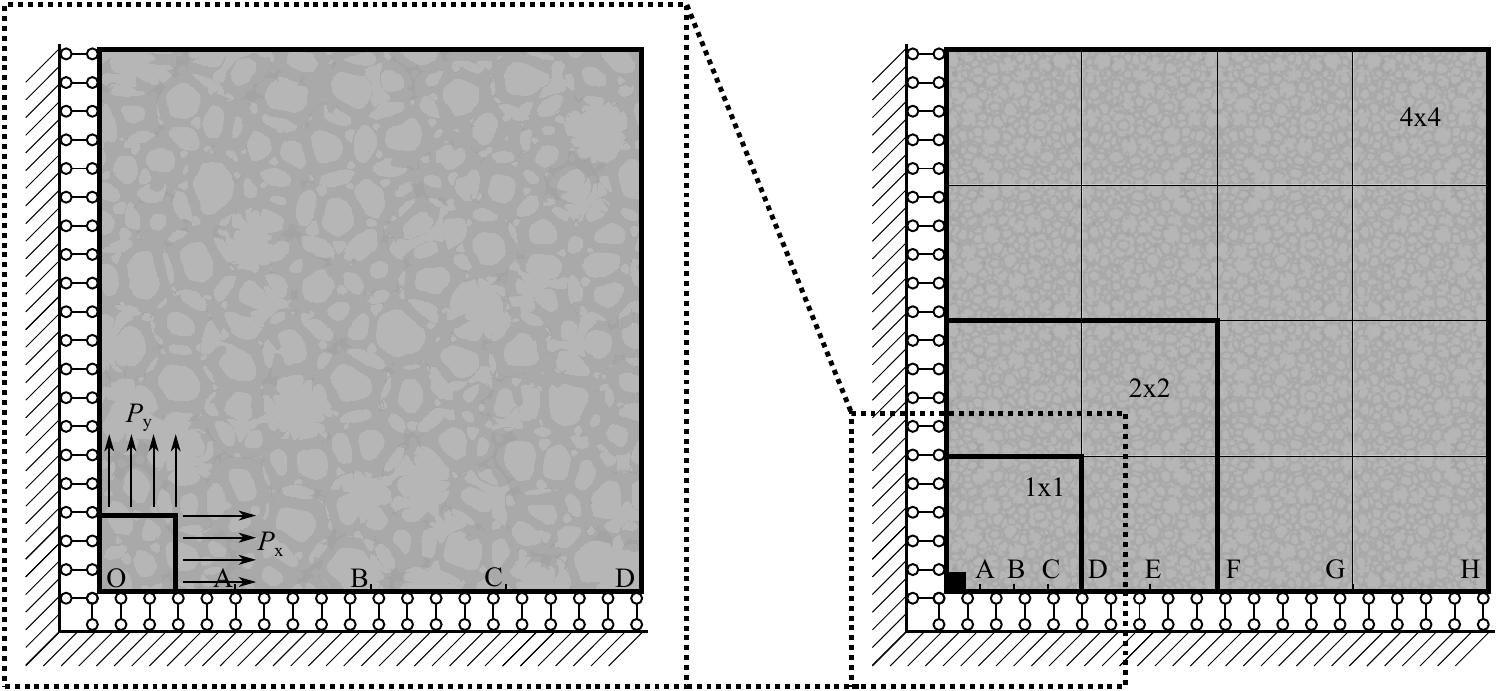}\hfill{}

\caption{Boundary condition of the sandwich panel with an enlarged view of
the loading area (left)\label{fig:Panel_bc}}
\end{figure}

In the simulation, we fix the displacement DOFs of the side faces
at $x=0$ and $y=0$ in orthogonal direction, as indicated in Fig.~\ref{fig:Panel_bc}.
Additionally, Neumann boundary conditions are applied to approximate
the wave excitation that is usually achieved by means of piezoelectric
transducers being attached to the plate. To this end, two distributed
line loads, $P_{\mathrm{x}}$ and $P_{\mathrm{y}}$, are applied along
two line segments to simulate the excitation of a shear transducer.
This excitation is used to avoid the adverse effects of a concentrated
force and to achieve an excitation that is more realistic without
the need of including multi-physics elements with piezoelectric capabilities.
The fundamental idea for this excitation is based on the pin-force
model which is often adopted in the simulation of guided waves in
the context of structural health monitoring~\citep{ArticleNiewenhius2005,InproceedingsGreve2005,BookGiurgiutiu2008}.
$P_{\mathrm{x}}$ is applied in $x$ direction on the line segment
($x=19.2\,\unit{mm}$, $0\,\unit{mm}\leq y\leq19.2\,\unit{mm}$ and
$z=57.6\,\unit{mm}$), while $P_{\mathrm{y}}$ is applied in $y$
direction, and the coordinate range is ($0\,\unit{mm}\leq x\leq19.2\,\unit{mm}$,
$y=19.2\,\unit{mm}$ and $z=57.6\,\unit{mm}$). These lines describe
the outer contour of a piezoelectric shear transducer which is often
applied in wave propagation analyses. Both of the distributed loads
vary over time following the sine-burst signal depicted in Fig.~\ref{fig:Lamb}
with $t_{1}=0.050\,{\textstyle \mathrm{ms}}$, $n=5$ and $P_{0}=2.1\times10^{4}\,\unit{\unitfrac{N}{mm}}$.
In aluminum, the dilatational wave speed is $6198\,\unitfrac{m}{s}$,
and the shear wave speed is $3122\,\unitfrac{m}{s}$. In steel, the
dilatational wave speed is $5926\,\unitfrac{m}{s}$, and the shear
wave speed is $3168\,\unitfrac{m}{s}$. The center frequency is $100\,\unit{kHz}$
while the maximum frequency of interest is $116\,\unit{kHz}$, and
the shear wavelengths at the maximum frequency in the aluminum and
steel parts are $26.9\,\unit{mm}$ and $27.3\,\unit{mm}$, respectively.
The time step used in this example is $\Delta t=2.98\times10^{-5}\,{\textstyle \mathrm{ms}}$
according to Eq.~(\ref{eq:tcr}), and a total duration of $0.4\,{\textstyle \mathrm{ms}}$
is analyzed. Note that in the cover plate Lamb waves are excited which
will interact with the core and hence, energy will leak gradually
from the top plate into the bottom plate. For more detailed discussions
on guided waves and Lamb waves in particular the interested reader
is referred to Refs.~\citep{BookVictorov1967,BookRose1999,BookGiurgiutiu2008}.

\begin{figure}[tb]
\hfill{}\subfloat[Displacement of four points in the sandwich panel\label{fig:Panel_his_1}]{\includegraphics{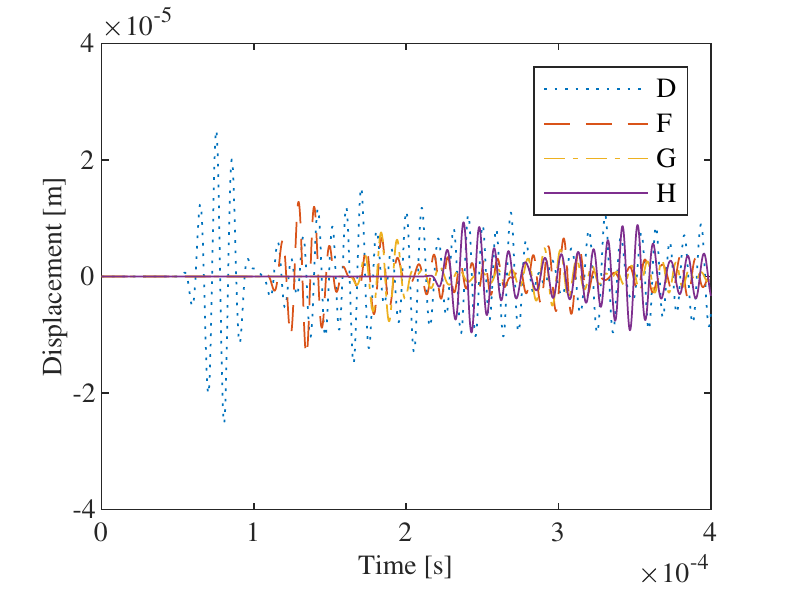}}\hfill{}\subfloat[Acceleration of four points in the sandwich panel\label{fig:Panel_his_2}]{\includegraphics{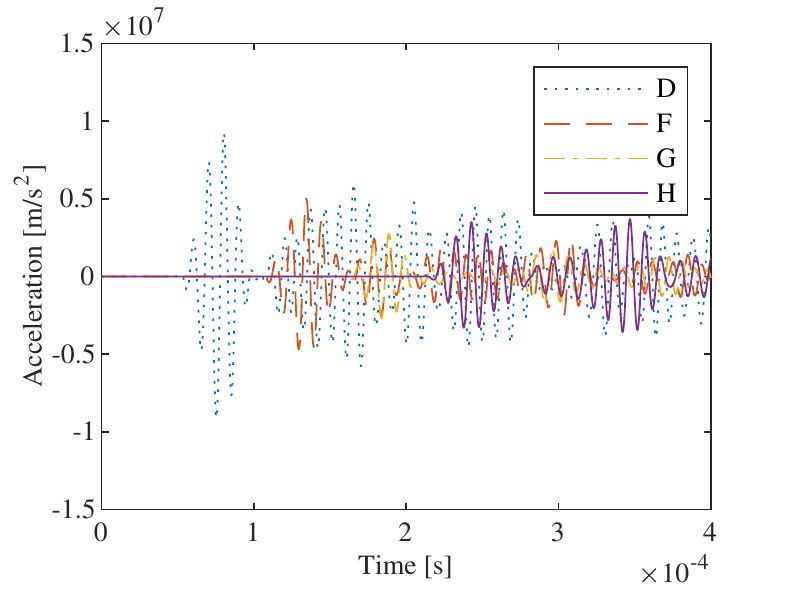}}\hfill{}

\caption{Response history of selected points of the sandwich panel\label{fig:Panel_history}}
\end{figure}

\begin{figure}[tb]
\hfill{}\subfloat[Computational time of $1\times1$ panel\label{fig:Panel_time_1}]{\includegraphics{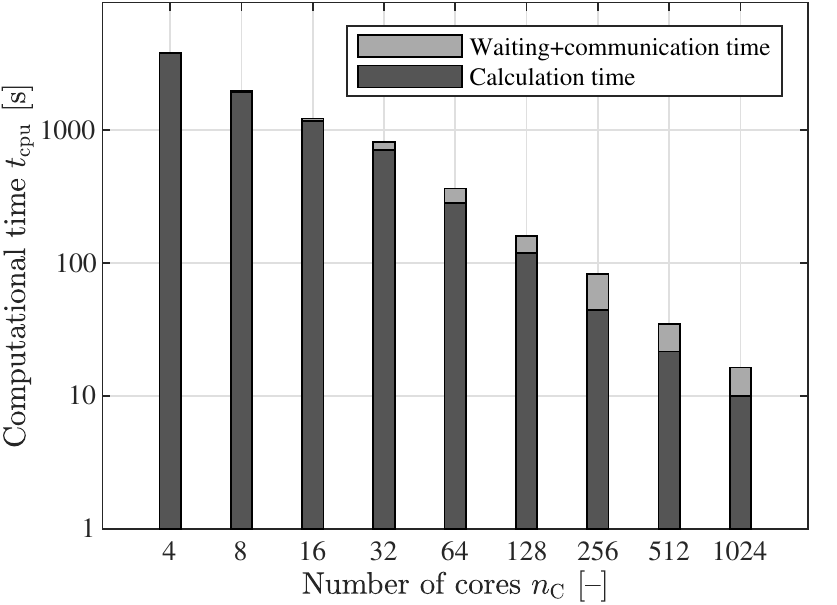}}\hfill{}\subfloat[Speedup and efficiency of $1\times1$ panel\label{fig:Panel_time_2}]{\includegraphics{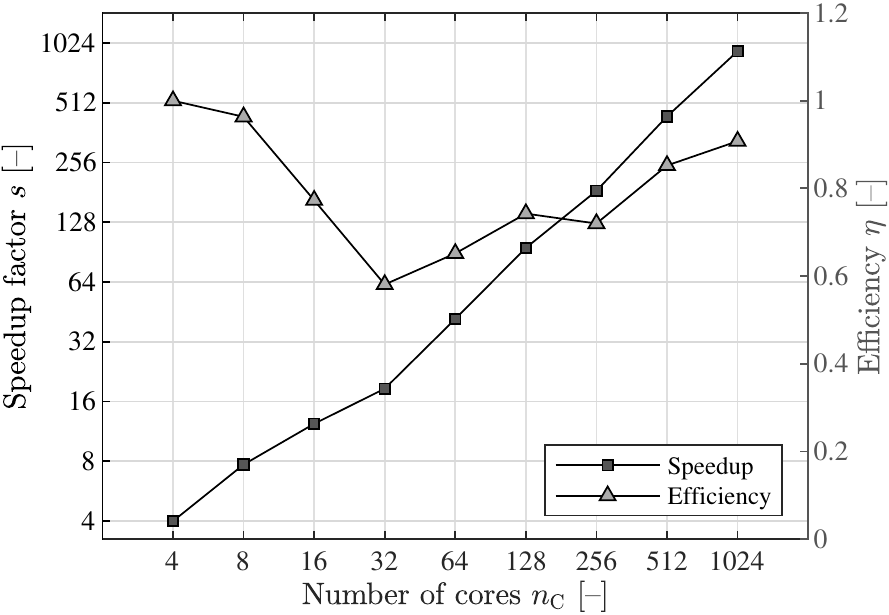}}\hfill{}

\hfill{}\subfloat[Computational time of $2\times2$ panel\label{fig:Panel_time_1-1}]{\includegraphics{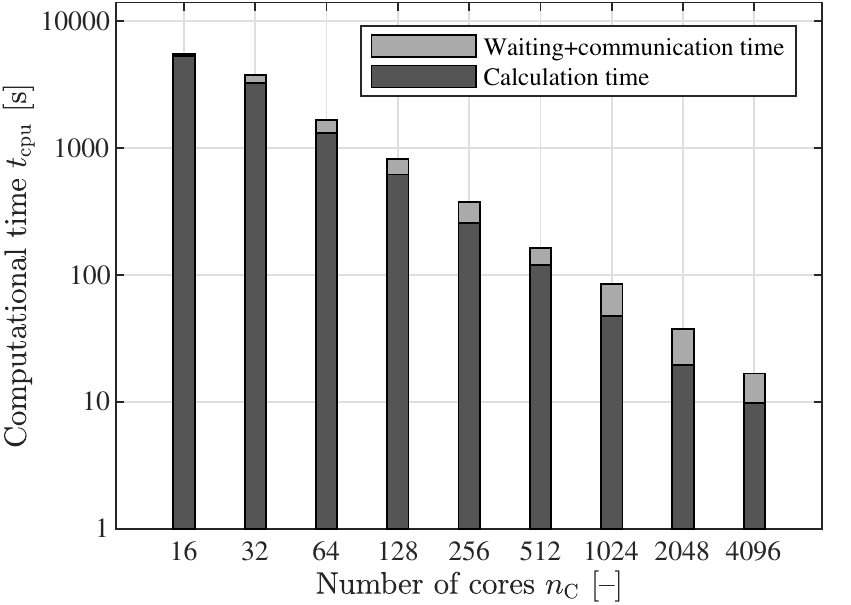}}\hfill{}\subfloat[Speedup and efficiency of $2\times2$ panel\label{fig:Panel_time_2-1}]{\includegraphics{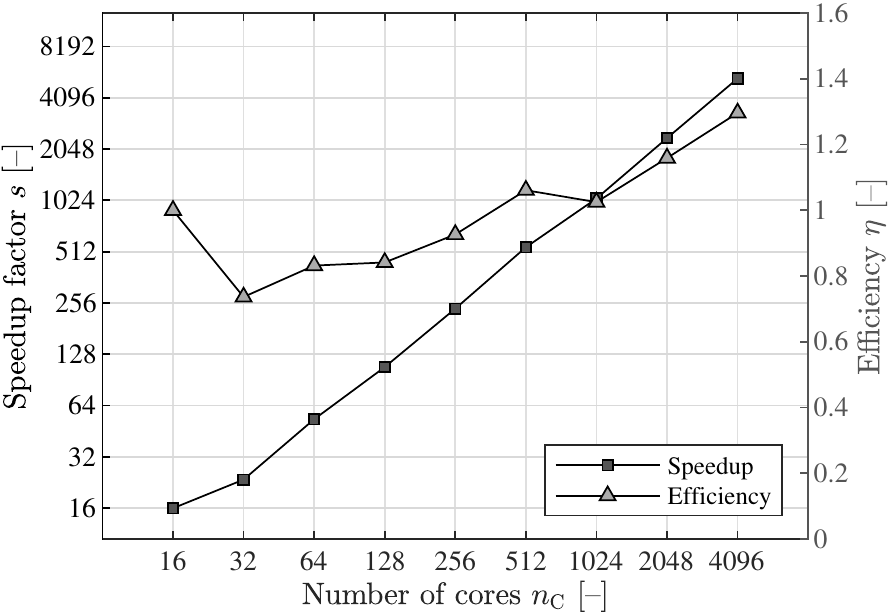}}\hfill{}

\hfill{}\subfloat[Computational time of $4\times4$ panel\label{fig:Panel_time_1-2}]{\includegraphics{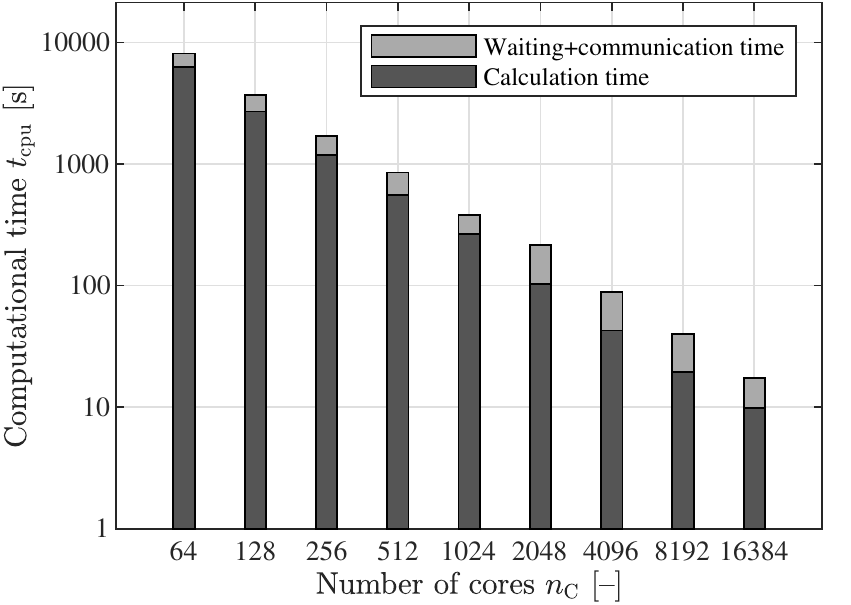}}\hfill{}\subfloat[Speedup and efficiency of $4\times4$ panel\label{fig:Panel_time_2-2}]{\includegraphics{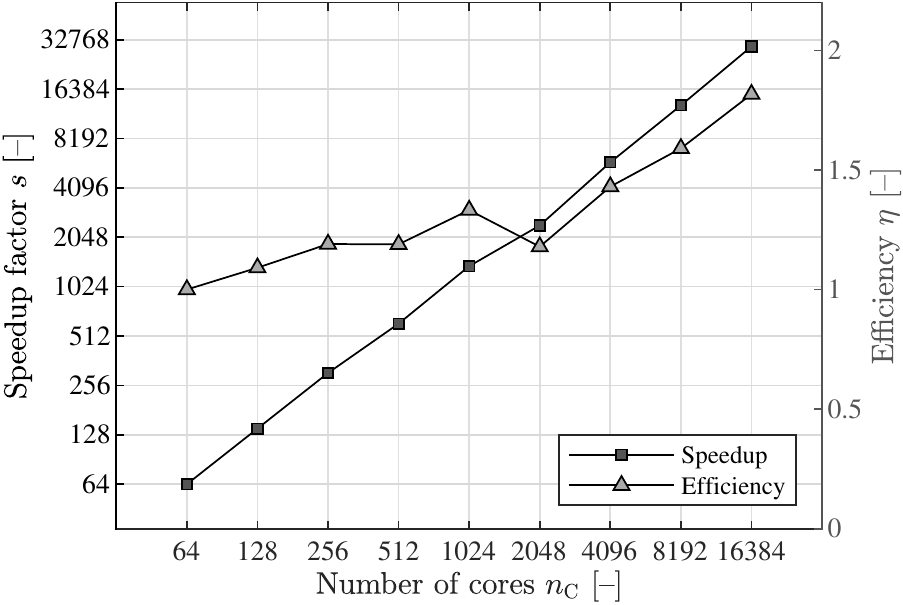}}\hfill{}

\caption{Time and speedup using multiple cores for the sandwich panel\label{fig:Panel_time}}
\end{figure}

The mesh corresponding to the $4\times4$ panel can be naturally divided
into 16 parts by the mirror planes, and each part is further divided
into 1024 parts, as shown in \ref{sec:Panel_part}. The top and bottom
views of the magnitude of the displacement response at five time instances
are presented in Fig.~\ref{fig:Panel_disp}, illustrating the wave
propagation process. It can be observed that the wave in the top panel
disperses into the core and eventually to the bottom panel. The displacement
and acceleration of four selected points shown in Fig.~\ref{fig:Panel_bc}
(coordinates are $D\!\left(288,0,57.6\right)$, $F\!\left(576,0,57.6\right)$,
$G\!\left(864,0,57.6\right)$ and $H\!\left(1152,0,57.6\right)$,
unit: mm) are plotted in Fig.~\ref{fig:Panel_history}. The speedup
of the multiple cores is illustrated in Fig.~\ref{fig:Panel_time}
and Table~\ref{tab:Panel_time}.

\begin{figure}[tb]
\hfill{}\subfloat[Top view, $t=0.04\,\unit{ms}$]{\includegraphics[width=0.4\textwidth]{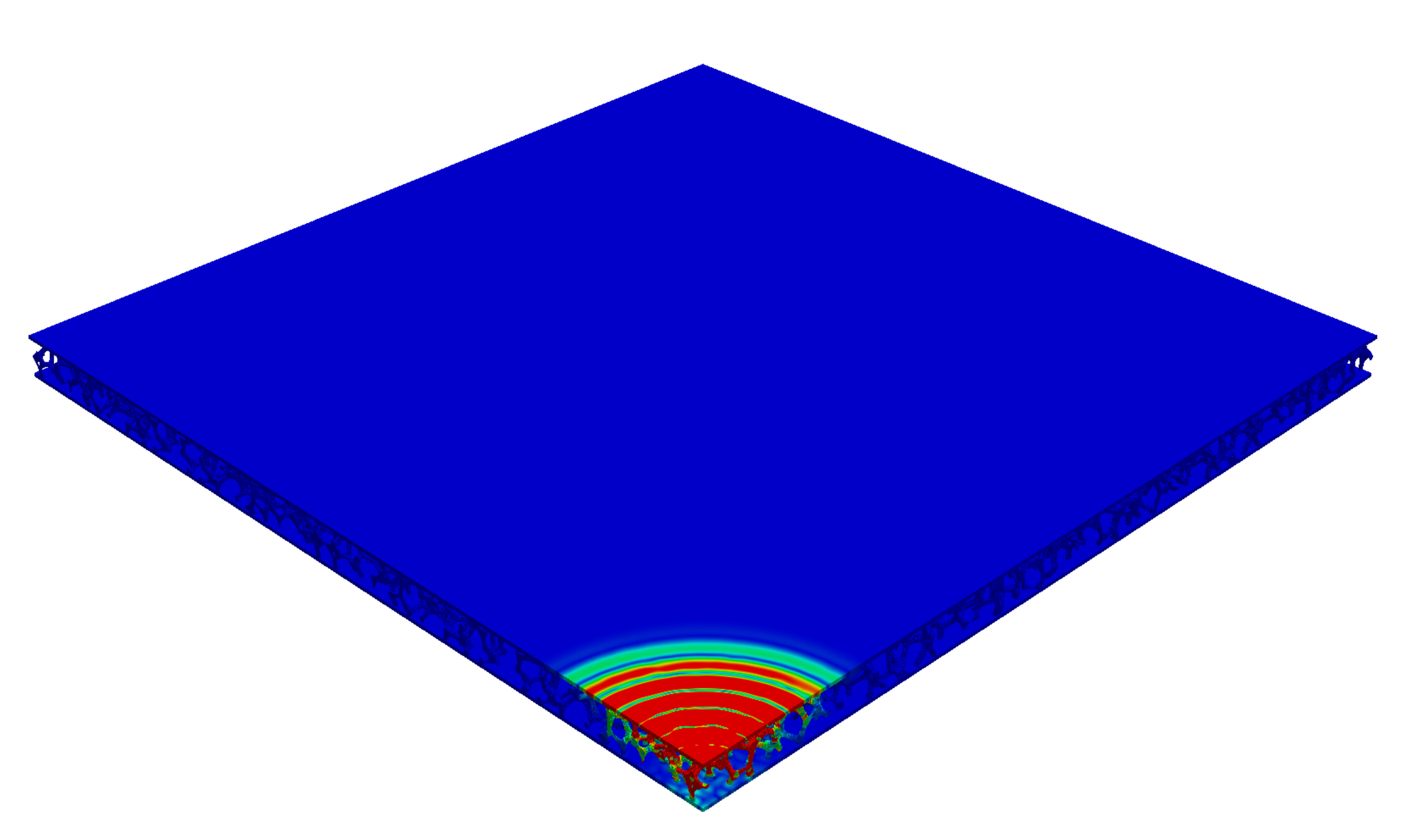}}\hfill{}\subfloat[Bottom view, $t=0.04\,\unit{ms}$]{\includegraphics[width=0.4\textwidth]{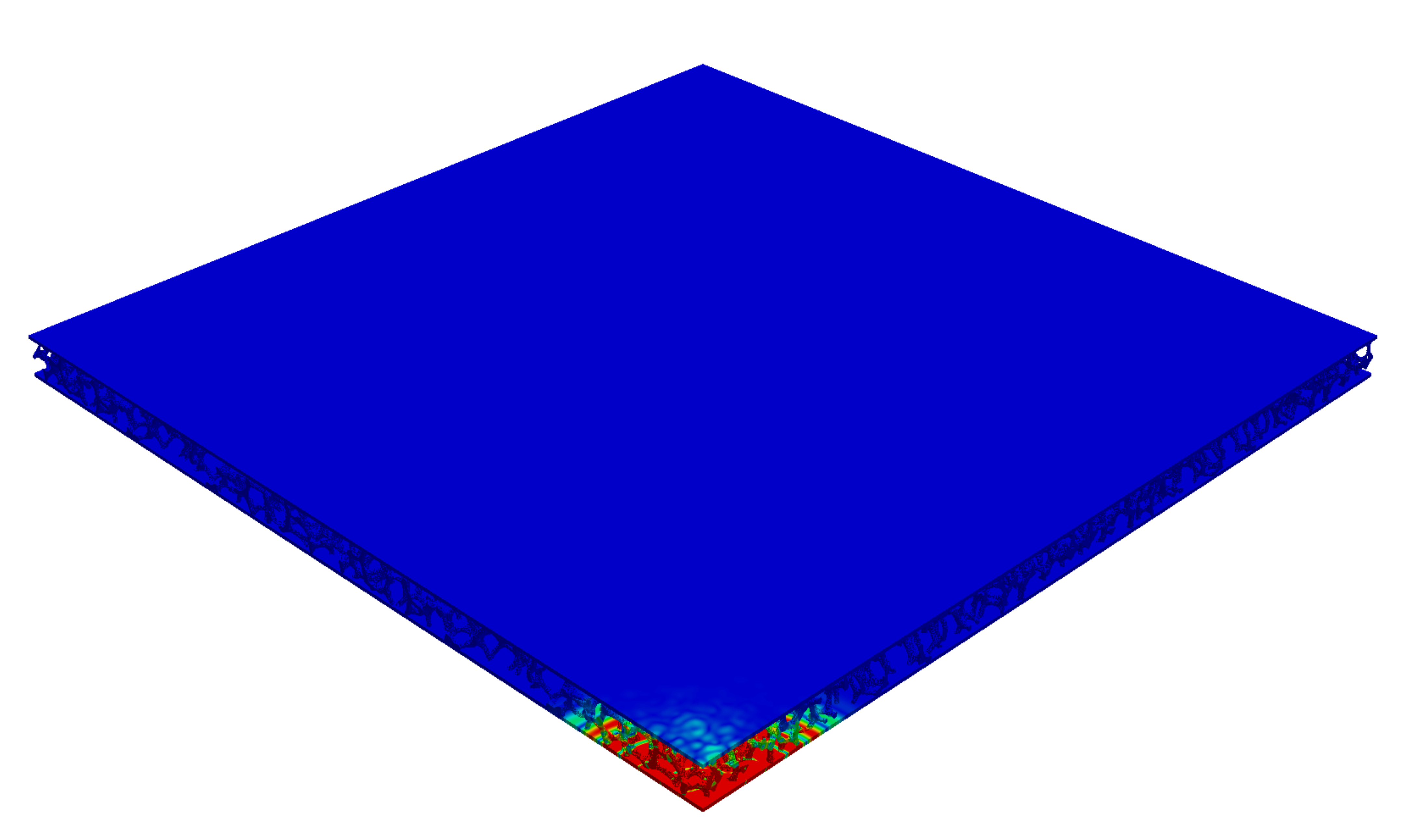}}\hspace{1cm}\hfill{}

\hfill{}\subfloat[Top view, $t=0.12\,\unit{ms}$]{\includegraphics[width=0.4\textwidth]{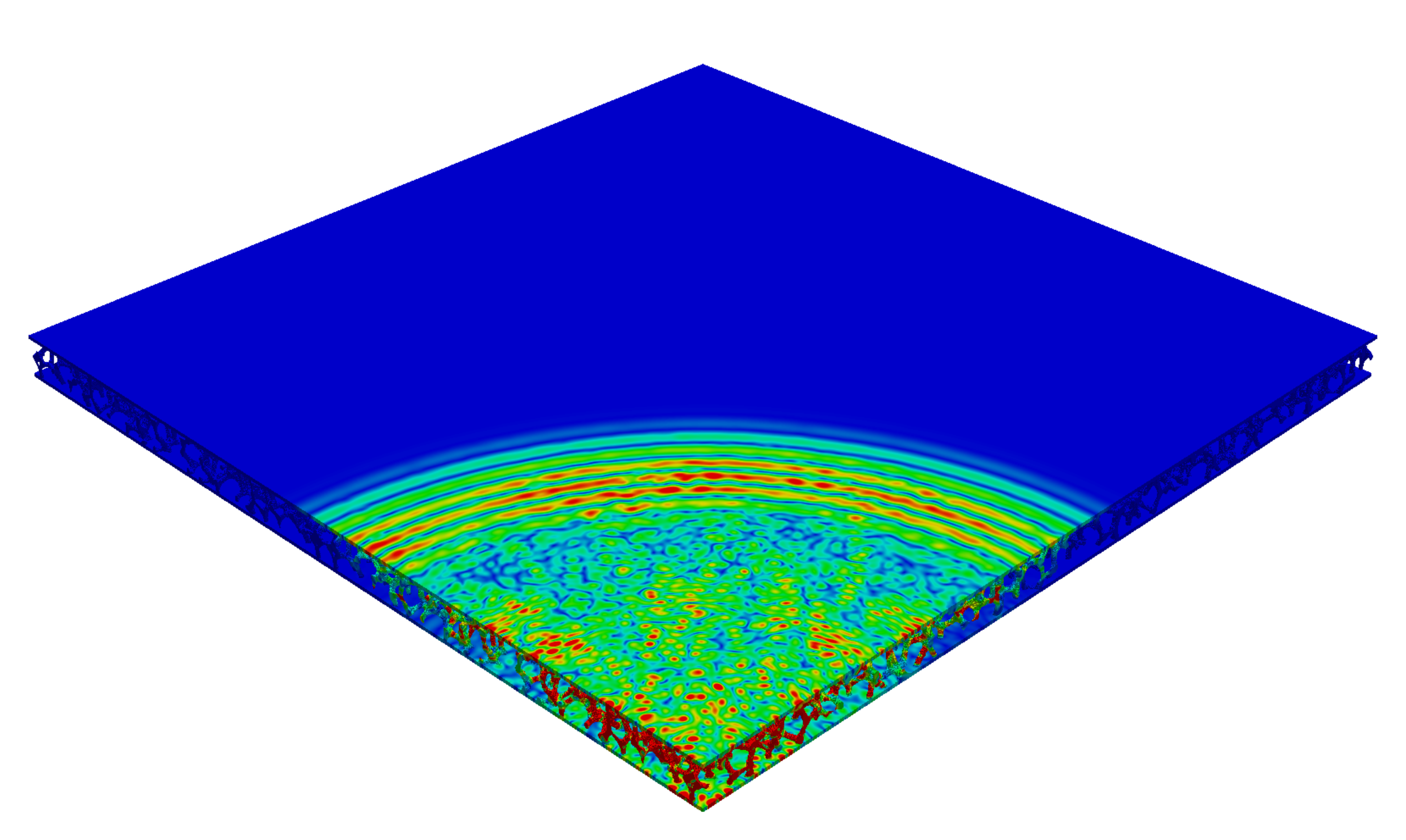}}\hfill{}\subfloat[Bottom view, $t=0.12\,\unit{ms}$]{\includegraphics[width=0.4\textwidth]{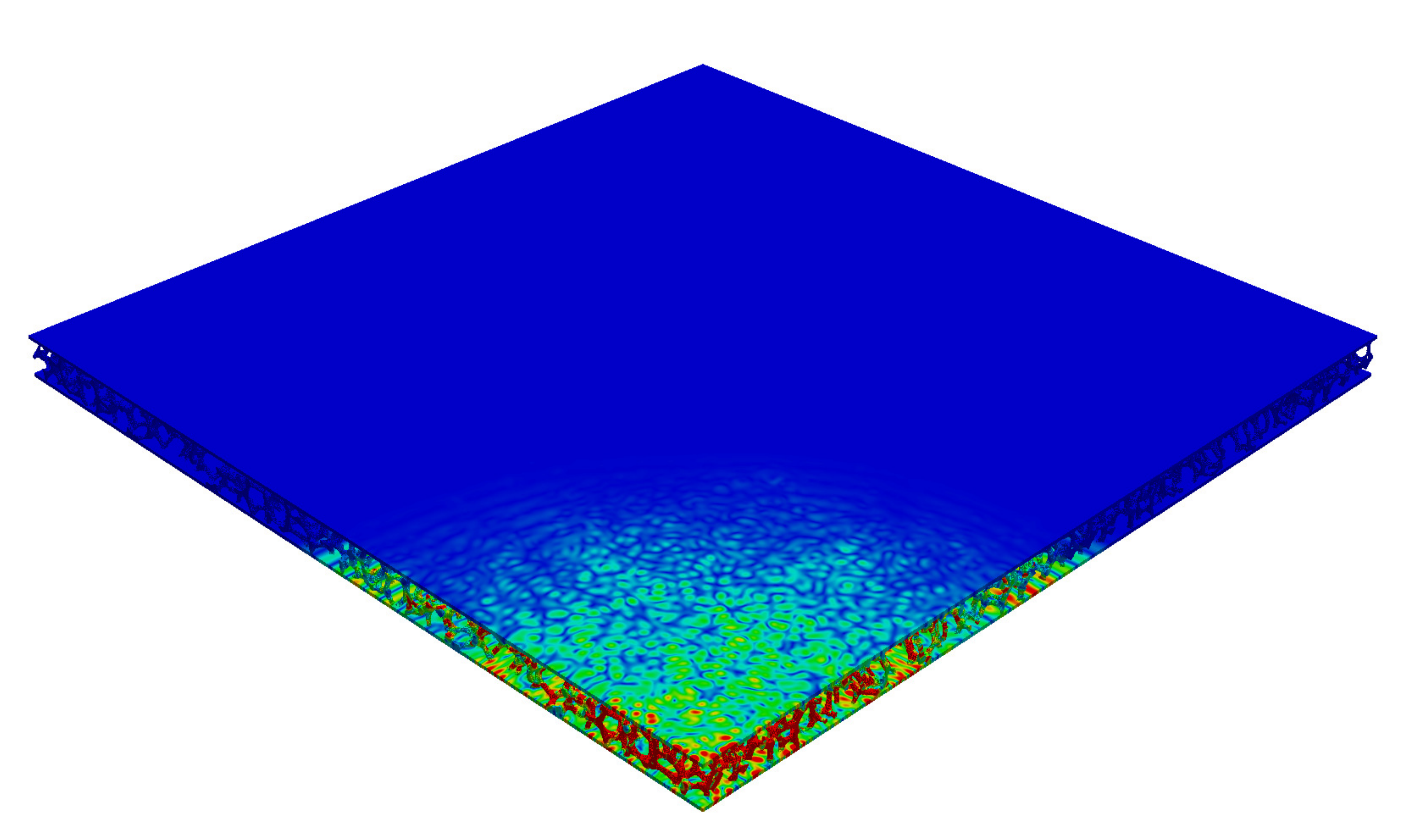}}\hspace{1cm}\hfill{}

\hfill{}\subfloat[Top view, $t=0.2\,\unit{ms}$]{\includegraphics[width=0.4\textwidth]{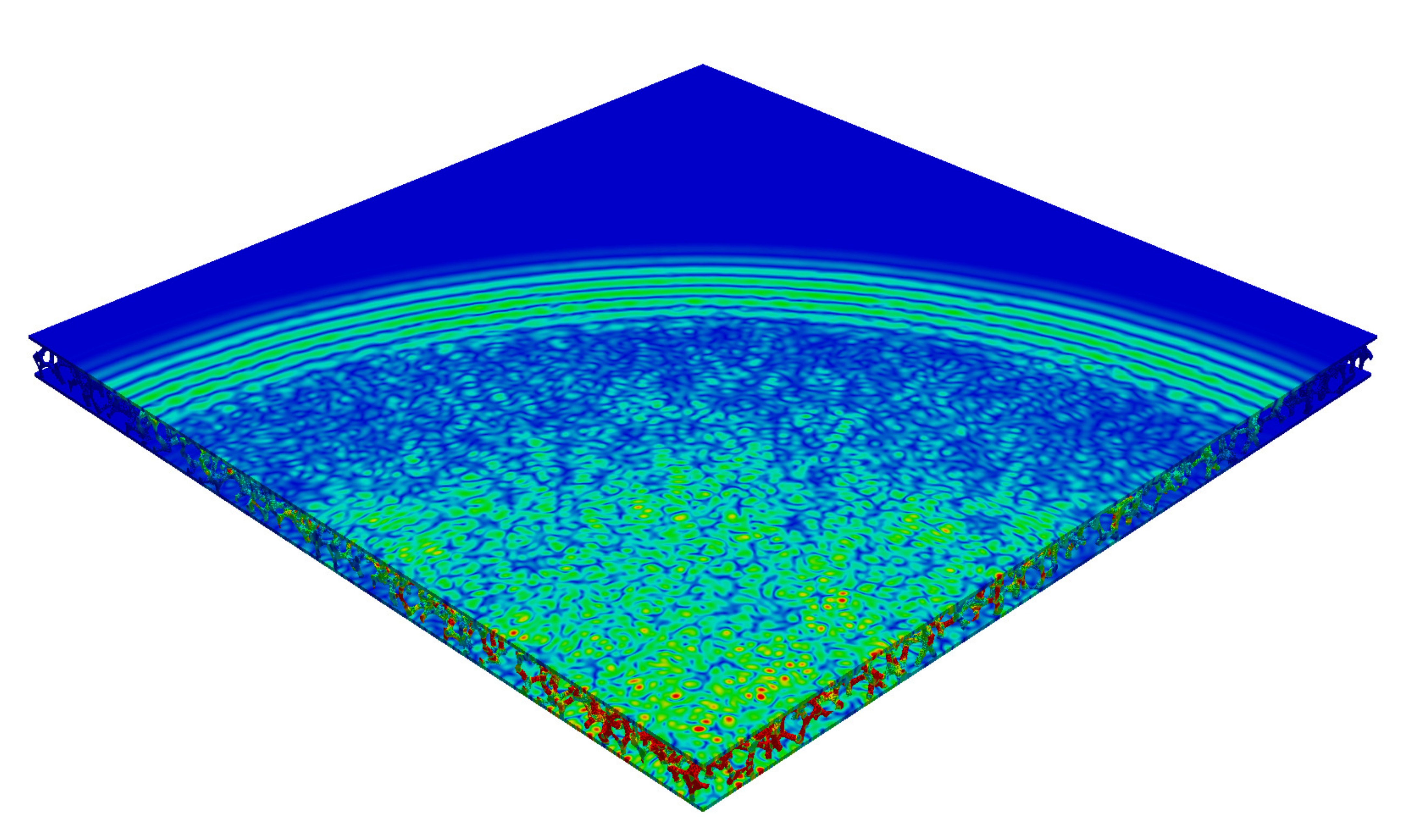}}\hfill{}\subfloat[Bottom view, $t=0.2\,\unit{ms}$]{\includegraphics[width=0.4\textwidth]{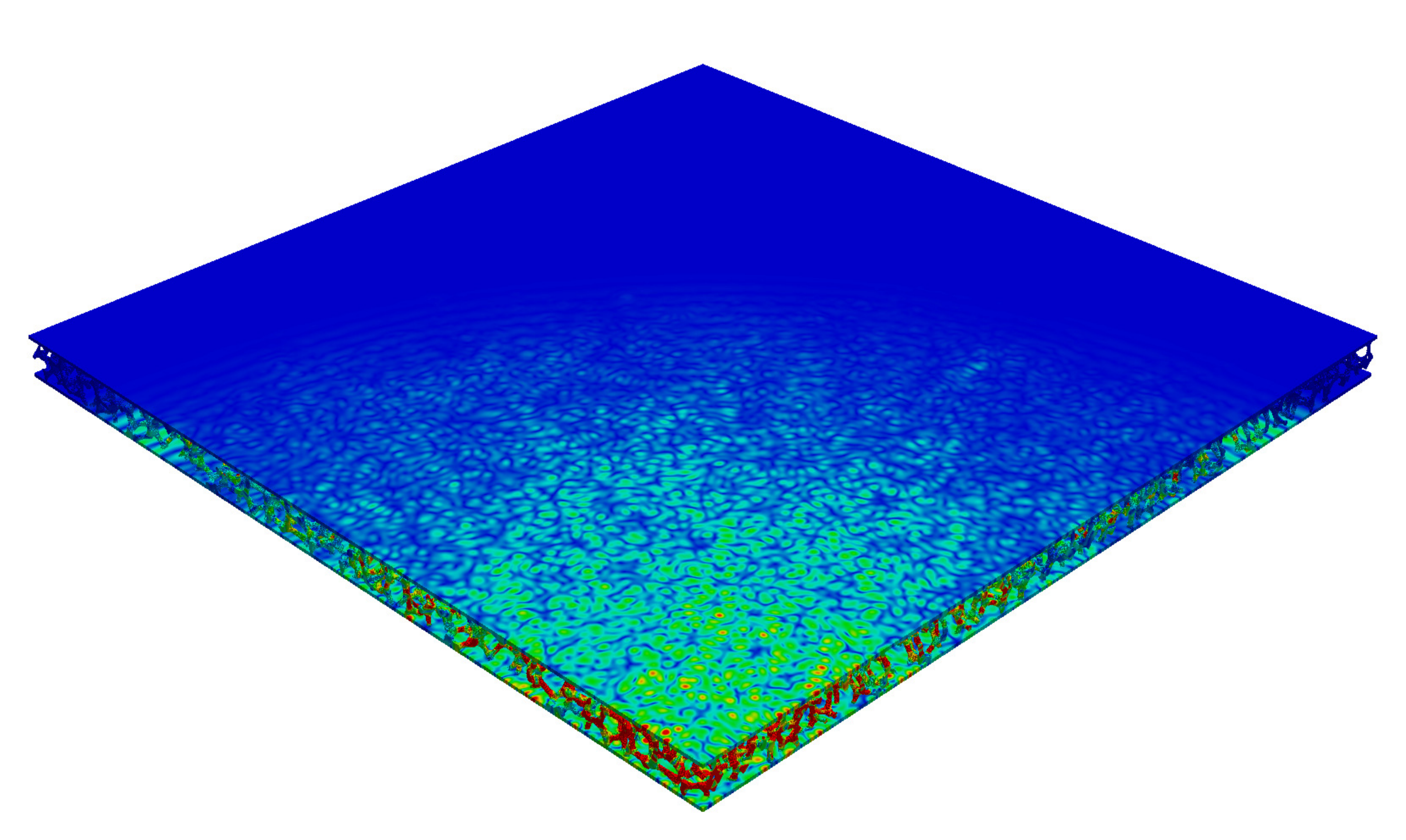}}\hspace{1cm}\hfill{}

\hfill{}\subfloat[Top view, $t=0.28\,\unit{ms}$]{\includegraphics[width=0.4\textwidth]{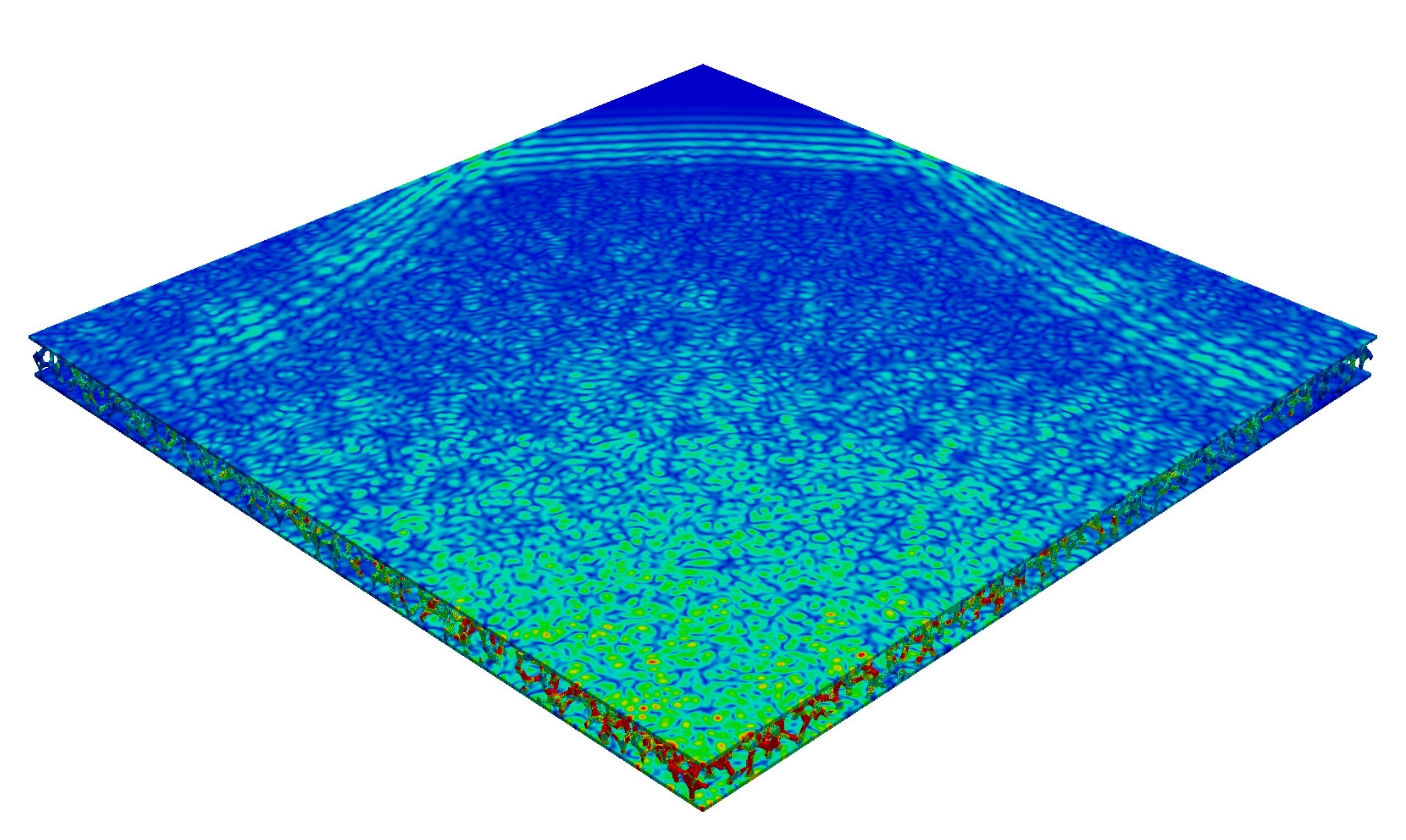}}\hfill{}\subfloat[Bottom view, $t=0.28\,\unit{ms}$]{\includegraphics[width=0.4\textwidth]{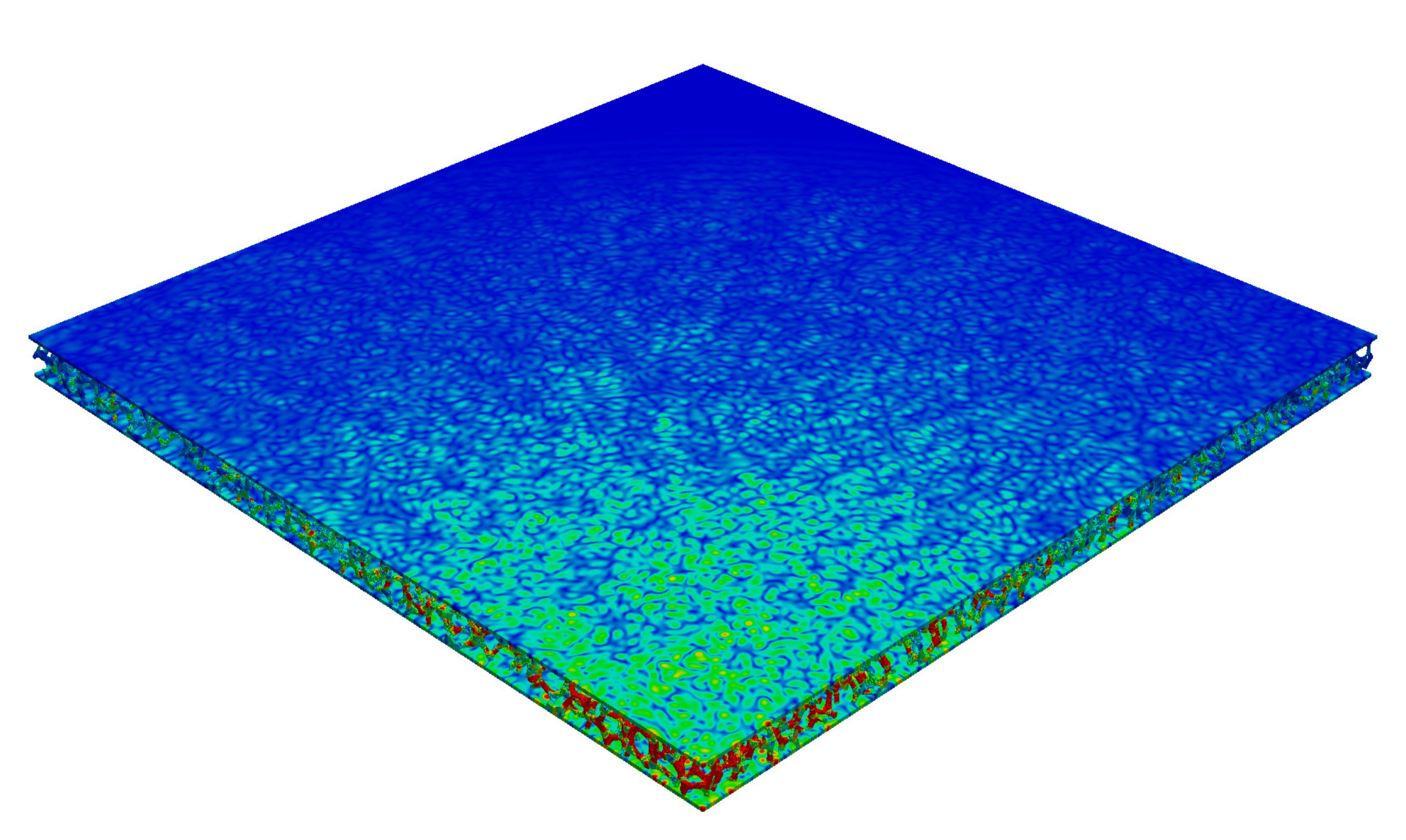}}\hspace{1cm}\hfill{}

\caption{Contour of displacement magnitude at various time instances $t$\label{fig:Panel_disp}}
\end{figure}

\begin{figure}[tb]
\ContinuedFloat

\hfill{}\subfloat[Top view, $t=0.36\,\unit{ms}$]{\includegraphics[width=0.4\textwidth]{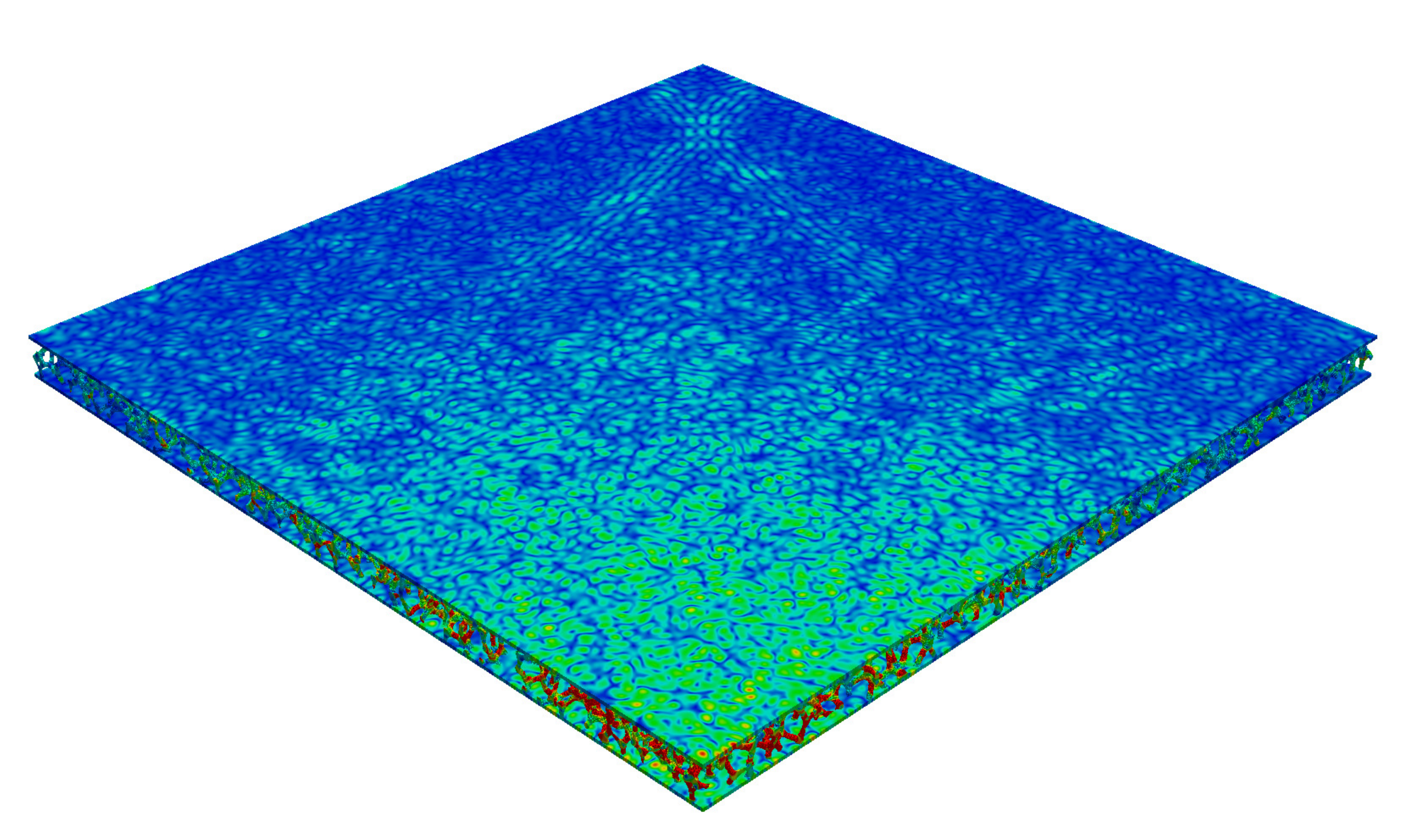}}\hfill{}\subfloat[Bottom view, $t=0.36\,\unit{ms}$]{\includegraphics[width=0.4\textwidth]{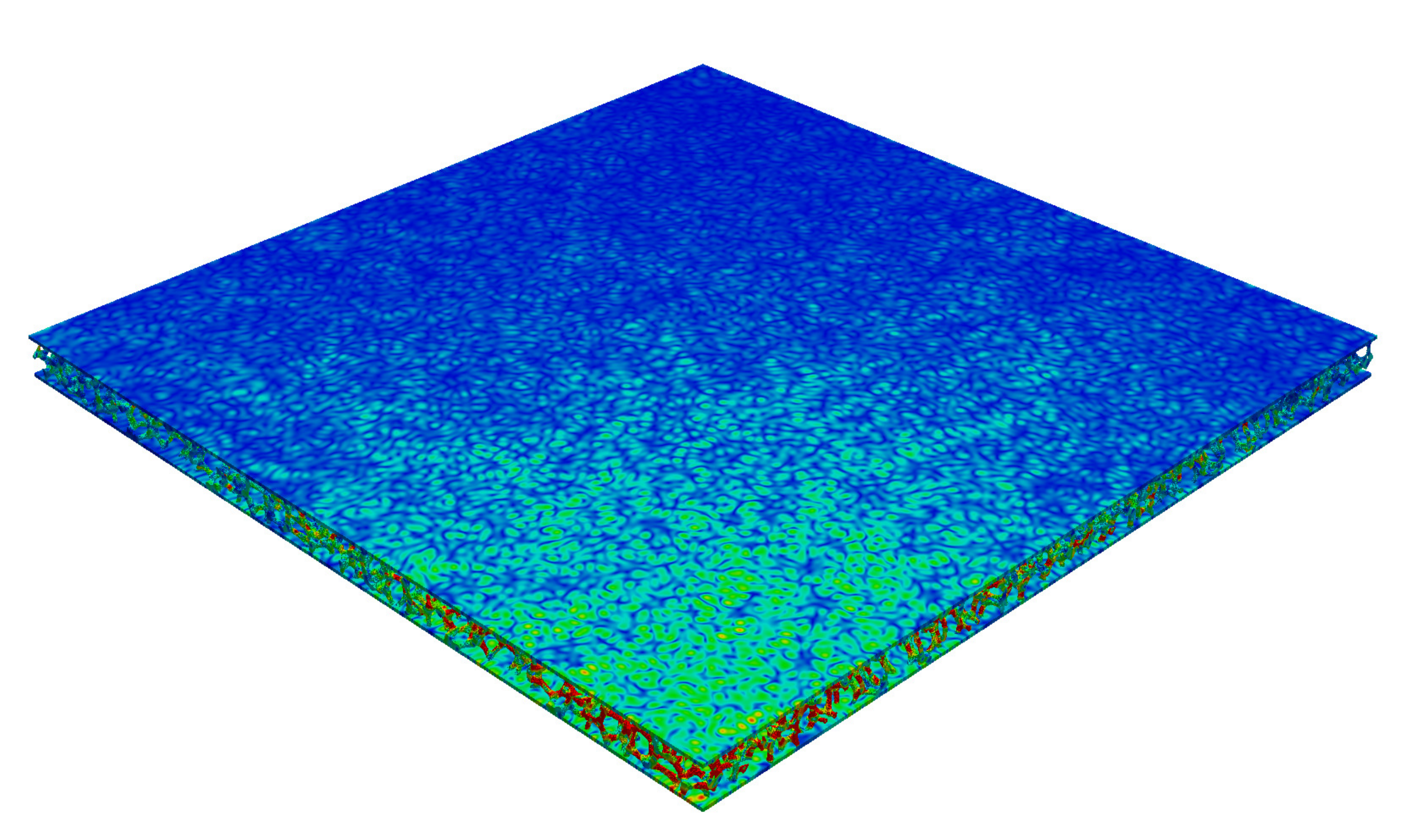}}\includegraphics{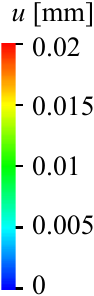}\hfill{}

\caption{Contour of displacement magnitude at various time instances $t$\label{fig:Panel_disp-1}}
\end{figure}

\FloatBarrier

\begin{table}
\caption{Computational time of the sandwich panel\label{tab:Panel_time}}

\hfill{}%
\begin{tabular}{|>{\centering}p{0.07\textwidth}|>{\centering}p{0.07\textwidth}|>{\centering}p{0.07\textwidth}|>{\centering}p{0.07\textwidth}|>{\centering}p{0.07\textwidth}|>{\centering}p{0.07\textwidth}|>{\centering}p{0.07\textwidth}|>{\centering}p{0.07\textwidth}|>{\centering}p{0.07\textwidth}|>{\centering}p{0.07\textwidth}|}
\hline 
\multirow{2}{0.07\textwidth}{$N_{\mathrm{cores}}$} & \multicolumn{3}{c|}{$1\times1$ panel} & \multicolumn{3}{c|}{$2\times2$ panel} & \multicolumn{3}{c|}{$4\times4$ panel}\tabularnewline
\cline{2-10} \cline{3-10} \cline{4-10} \cline{5-10} \cline{6-10} \cline{7-10} \cline{8-10} \cline{9-10} \cline{10-10} 
 & $t_{\mathrm{C}}$ {[}s{]} & $t_{\mathrm{W}}$ {[}s{]} & $t_{\mathrm{T}}$ {[}s{]} & $t_{\mathrm{C}}$ {[}s{]} & $t_{\mathrm{W}}$ {[}s{]} & $t_{\mathrm{T}}$ {[}s{]} & $t_{\mathrm{C}}$ {[}s{]} & $t_{\mathrm{W}}$ {[}s{]} & $t_{\mathrm{T}}$ {[}s{]}\tabularnewline
\hline 
\hline 
4 & 3790.78 & 12.54 & 3803.67 & - & - & - & - & - & -\tabularnewline
\hline 
8 & 1937.86 & 36.19 & 1974.24 & - & - & - & - & - & -\tabularnewline
\hline 
16 & 1175.07 & 55.04 & 1230.24 & 5317.17 & 232.99 & 5550.67 & - & - & -\tabularnewline
\hline 
32 & 709.82 & 108.81 & 818.73 & 3253.78 & 514.92 & 3769.13 & - & - & -\tabularnewline
\hline 
64 & 283.48 & 81.35 & 364.87 & 1320.66 & 347.40 & 1668.21 & 6316.70 & 1780.27 & 8097.55\tabularnewline
\hline 
128 & 119.05 & 41.04 & 160.11 & 620.23 & 204.36 & 824.66 & 2699.81 & 1010.44 & 3710.49\tabularnewline
\hline 
256 & 44.49 & 38.07 & 82.57 & 258.58 & 115.88 & 374.49 & 1184.53 & 516.15 & 1700.79\tabularnewline
\hline 
512 & 21.61 & 13.25 & 34.87 & 120.45 & 43.02 & 163.48 & 553.94 & 297.26 & 851.26\tabularnewline
\hline 
1024 & 9.92 & 6.46 & 16.37 & 47.45 & 37.05 & 84.63 & 264.50 & 115.35 & 379.87\tabularnewline
\hline 
2048 & - & - & - & 19.60 & 17.82 & 37.42 & 103.23 & 111.36 & 214.60\tabularnewline
\hline 
4096 & - & - & - & 9.81 & 6.92 & 16.73 & 42.76 & 45.70 & 88.46\tabularnewline
\hline 
8192 & - & - & - & - & - & - & 18.30 & 28.78 & 47.08\tabularnewline
\hline 
16384 & - & - & - & - & - & - & 9.81 & 7.60 & 17.41\tabularnewline
\hline 
\multicolumn{10}{l}{$t_{\mathrm{C}}$: Calculation time}\tabularnewline
\multicolumn{10}{l}{$t_{\mathrm{W}}$: Waiting and communication time}\tabularnewline
\multicolumn{10}{l}{$t_{\mathrm{T}}$: Total time}\tabularnewline
\end{tabular}\hfill{}

\caption{Speedup and efficiency of the sandwich panel\label{tab:Panel_speedup}}

\hfill{}%
\begin{tabular}{|>{\centering}p{0.07\textwidth}|>{\centering}p{0.07\textwidth}|>{\centering}p{0.07\textwidth}|>{\centering}p{0.07\textwidth}|>{\centering}p{0.07\textwidth}|>{\centering}p{0.07\textwidth}|>{\centering}p{0.07\textwidth}|}
\hline 
\multirow{2}{0.07\textwidth}{$N_{\mathrm{cores}}$} & \multicolumn{2}{c|}{$1\times1$ panel} & \multicolumn{2}{c|}{$2\times2$ panel} & \multicolumn{2}{c|}{$4\times4$ panel}\tabularnewline
\cline{2-7} \cline{3-7} \cline{4-7} \cline{5-7} \cline{6-7} \cline{7-7} 
 & $s$ & $\eta$ & $s$ & $\eta$ & $s$ & $\eta$\tabularnewline
\hline 
\hline 
4 & 4.00 & 1.00 & - & - & - & -\tabularnewline
\hline 
8 & 7.71 & 0.96 & - & - & - & -\tabularnewline
\hline 
16 & 12.37 & 0.77 & 16.00 & 1.00 & - & -\tabularnewline
\hline 
32 & 18.58 & 0.58 & 23.56 & 0.74 & - & -\tabularnewline
\hline 
64 & 41.70 & 0.65 & 53.24 & 0.83 & 64.00 & 1.00\tabularnewline
\hline 
128 & 95.03 & 0.74 & 107.69 & 0.84 & 139.67 & 1.09\tabularnewline
\hline 
256 & 184.27 & 0.72 & 237.15 & 0.93 & 304.71 & 1.19\tabularnewline
\hline 
512 & 436.36 & 0.85 & 543.25 & 1.06 & 608.80 & 1.19\tabularnewline
\hline 
1024 & 929.21 & 0.91 & 1049.42 & 1.02 & 1364.27 & 1.33\tabularnewline
\hline 
2048 & - & - & 2373.29 & 1.16 & 2414.93 & 1.18\tabularnewline
\hline 
4096 & - & - & 5309.59 & 1.30 & 5858.25 & 1.43\tabularnewline
\hline 
8192 & - & - & - & - & 13028.57 & 1.59\tabularnewline
\hline 
16384 & - & - & - & - & 29775.04 & 1.82\tabularnewline
\hline 
\multicolumn{7}{l}{$s$: Speedup factor}\tabularnewline
\multicolumn{7}{l}{$\eta$: Efficiency}\tabularnewline
\end{tabular}\hfill{}
\end{table}

It can be observed that the CPU time increases with the number of
DOFs slightly more than linearly. For a fixed number of cores, better
efficiency can be achieved when the scale of the problem increases.
In the analysis of the $1\times1$ panel, a speedup factor of 929
is obtained when 1024 cores are used. In the $2\times2$ panel, the
maximum speedup is 5308 when using 4096 cores. In the $4\times4$
panel, a speedup factor of 29760 is obtained with 16384 cores. It
is noted that these values are calculated based on the assumption
that $100\%$ efficiency is achieved when using the lowest number
of cores, e.g., 4, 16 and 64, respectively. It is observed that for
a fixed problem scale, the efficiency can increase with the number
of cores in some cases, and some of the values are even higher than
1. This super-linear speedup is caused by the cache effect, i.e.,
as the mesh is partitioned and distributed to more cores, more data
can be fit into the cache memory of the compute nodes, which has significantly
higher data processing speed than RAM.

\clearpage{}

\section{Summary\label{sec:Summary}}

In this article, a parallel explicit solver for structural dynamics
based on the CDM is developed, exploiting the inherent advantages
of octree meshes. In our implementation, the octree meshes are efficiently
handled by the SBFEM, which is our method of choice. However, it is
important to note that exactly the same approach can also be applied
to FEA when special transition elements with well-conditioned lumped
mass matrices are deployed. These could be based on transfinite mappings~\citep{BookProvatidis2019,Duczek2020},
virtual elements~\citep{Park2019}, etc. A special mass lumping technique
is deployed by extending the work reported in Ref.~\citep{Gravenkamp2020a}
to 3D, yielding a well-conditioned diagonal mass matrix. The advantages
of an explicit time integrator are leveraged, which enables the nodal
displacements to be computed without the need for solving a system
of linear equations. The pre-computation approach for the stiffness
and mass matrices, relying on a limited number of unique octree cells
in balanced meshes, further improves the efficiency of the algorithm.

The performance of the massively parallel explicit solver using the
octree mesh generation and the SBFEM is demonstrated by means of five
numerical examples, including academic benchmark problems with simple
geometry as well as image and STL based models. Large-scale problems
with up to one billion DOFs are simulated while up to 16,384 computing
cores are utilized. The octree algorithm provides a fully automatic
approach for mesh generation of complex geometric model. A Ricker
wavelet, a triangular pulse load and a sine-burst signal are considered
in these examples to excite the structures and to illustrate realistic
and high-resolution engineering applications. Compared to a conventional
(serial) implementation, a significant speedup is observed using the
parallelization approach on multi-processor machines, and the in-house
code can achieve an improved efficiency compared to the commercial
software ABAQUS. In the current approach, the imbalance of workload
of individual processors hinders a better scalability, as some processors
have to wait for the slower processors to finish the job so that the
results can be synchronized. In the future applications, a dynamic
partitioning approach and a load distribution of the job based on
the availability of individual processors may further improve the
performance of the proposed algorithm. In addition, non-linear analyses,
such as damage and contact problems can be studied making use of the
novel methodology.

\section*{Acknowledgments}

The work presented in this paper is partially supported by the Australian
Research Council through Grant Number DP180101538 and DP200103577.
This research includes results that have been obtained using the computational
cluster Katana supported by Research Technology Services at UNSW Sydney.
Furthermore, the provided resources from the National Computational
Infrastructure (NCI Australia), an NCRIS enabled capability supported
by the Australian Government, are gratefully acknowledged. The authors
would like to thank the NCI user support team for their technical
support regarding the use of NCI's supercomputing facility. The authors
would also like to thank Dr.~Meysam Joulaian and Prof.~Alexander
D�ster from Hamburg University of Technology for providing the X-ray
CT scan data, which was used in Section~\ref{subsec:Panel}.

\clearpage{}

\bibliographystyle{elsarticle-num-names}

\clearpage{}

\appendix

\section{Unique octree patterns in 3D\label{sec:Octree_pattern}}

\begin{figure}[H]
\hfill{}\includegraphics[scale=0.8]{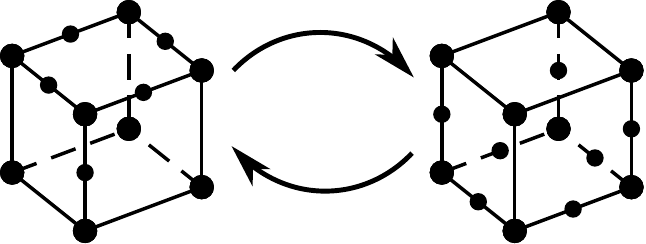}\hfill{}

\caption{Octree cells with hanging nodes on seven edges (right) can be obtained
from those with five edges (left)\label{fig:Oct144_7_5}}
\end{figure}

The unique octree patterns with hanging nodes on zero to six edges
are listed. The remaining cases (seven to twelve hanging nodes) can
be obtained from these cases, e.g., the first case of an octree cell
with seven hanging nodes can be obtained from the first case with
five hanging nodes by swapping the edges with and without hanging
nodes, as shown in Fig.~\ref{fig:Oct144_7_5}. The hanging nodes
at the center of faces are not shown, which will be added following
the discretization patterns.

\begin{minipage}[t]{0.45\textwidth}%
\begin{figure}[H]
\hfill{}\includegraphics[scale=0.8]{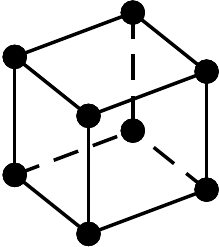}\hfill{}

\caption{Octree cell without hanging node}
\end{figure}
\end{minipage}%
\begin{minipage}[t]{0.45\textwidth}%
\begin{figure}[H]
\hfill{}\includegraphics[scale=0.8]{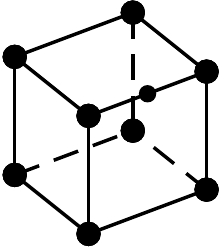}\hfill{}

\caption{Octree cell with hanging node on one edge}
\end{figure}
\end{minipage}

\begin{figure}[H]
\hfill{}\includegraphics[scale=0.8]{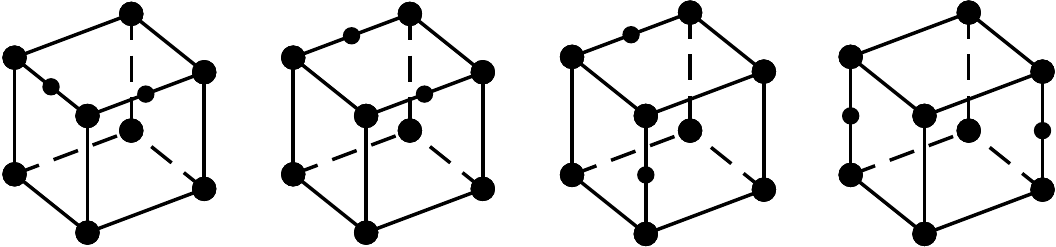}\hfill{}

\caption{Octree cells with hanging nodes on two edges}
\end{figure}

\begin{figure}[H]
\hfill{}\includegraphics[scale=0.8]{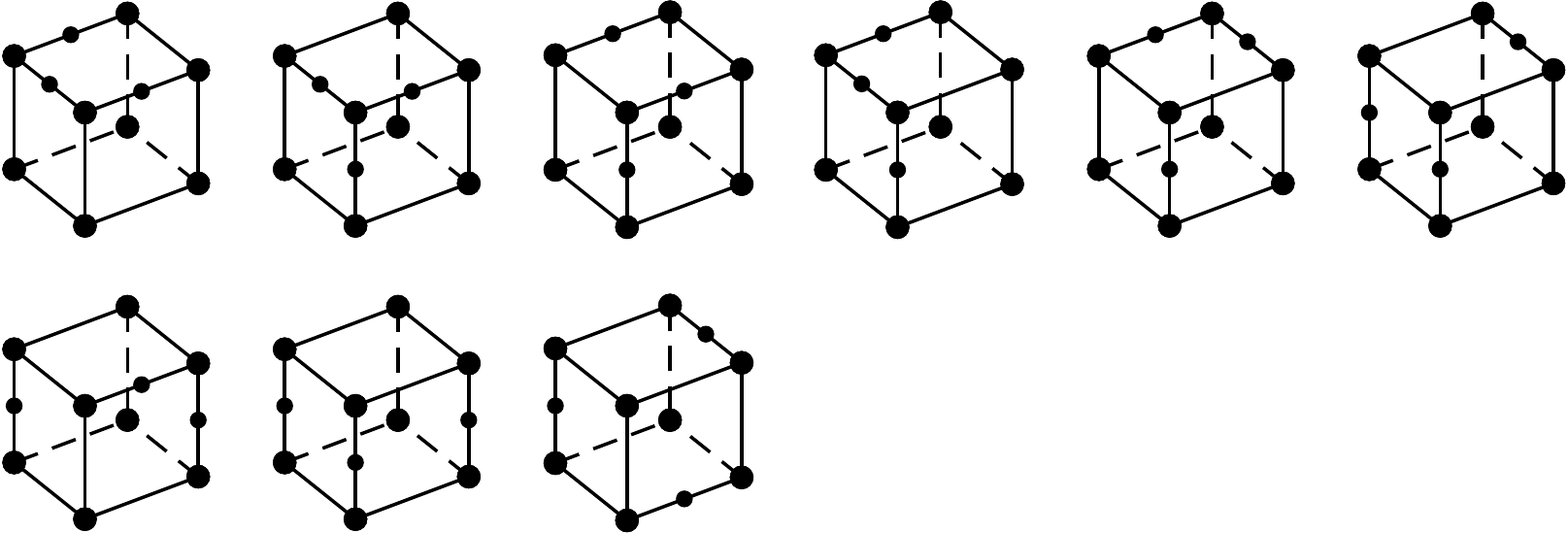}\hfill{}

\caption{Octree cells with hanging nodes on three edges}
\end{figure}

\begin{figure}[H]
\hfill{}\includegraphics[scale=0.8]{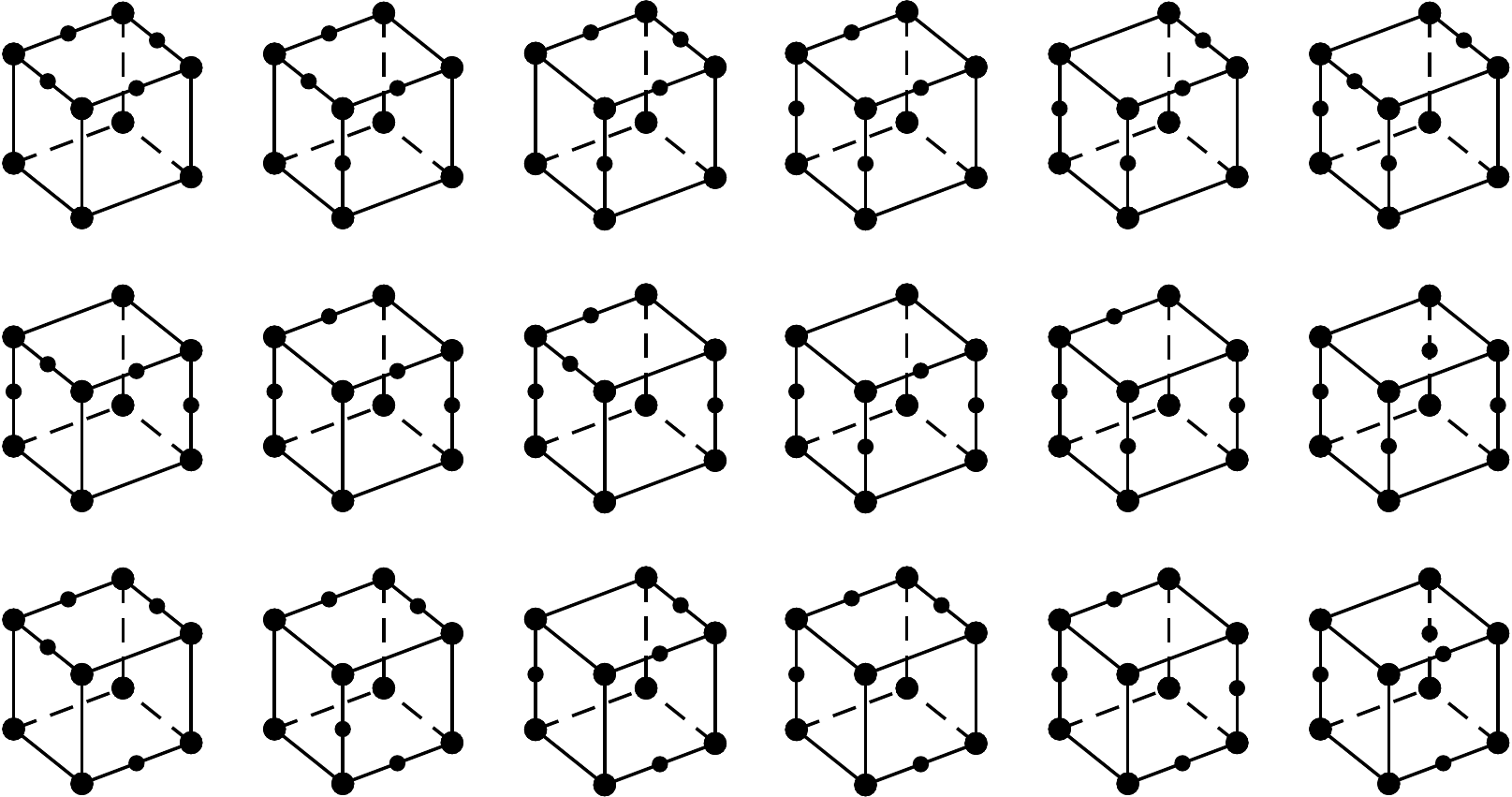}\hfill{}

\caption{Octree cells with hanging nodes on four edges}
\end{figure}

\begin{figure}[H]
\hfill{}\includegraphics[scale=0.8]{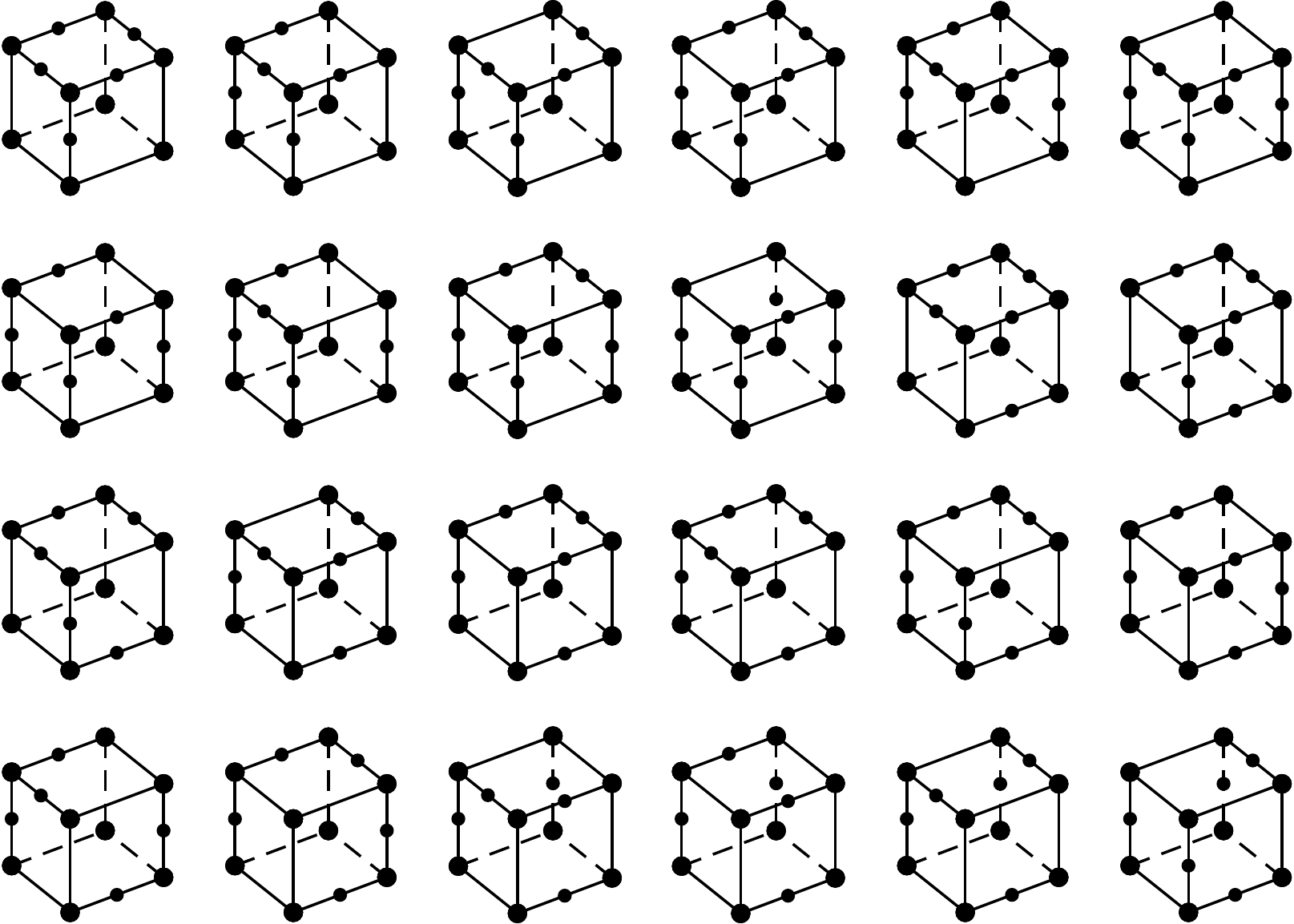}\hfill{}

\caption{Octree cells with hanging nodes on five edges}
\end{figure}

\begin{figure}[H]
\hfill{}\includegraphics[scale=0.8]{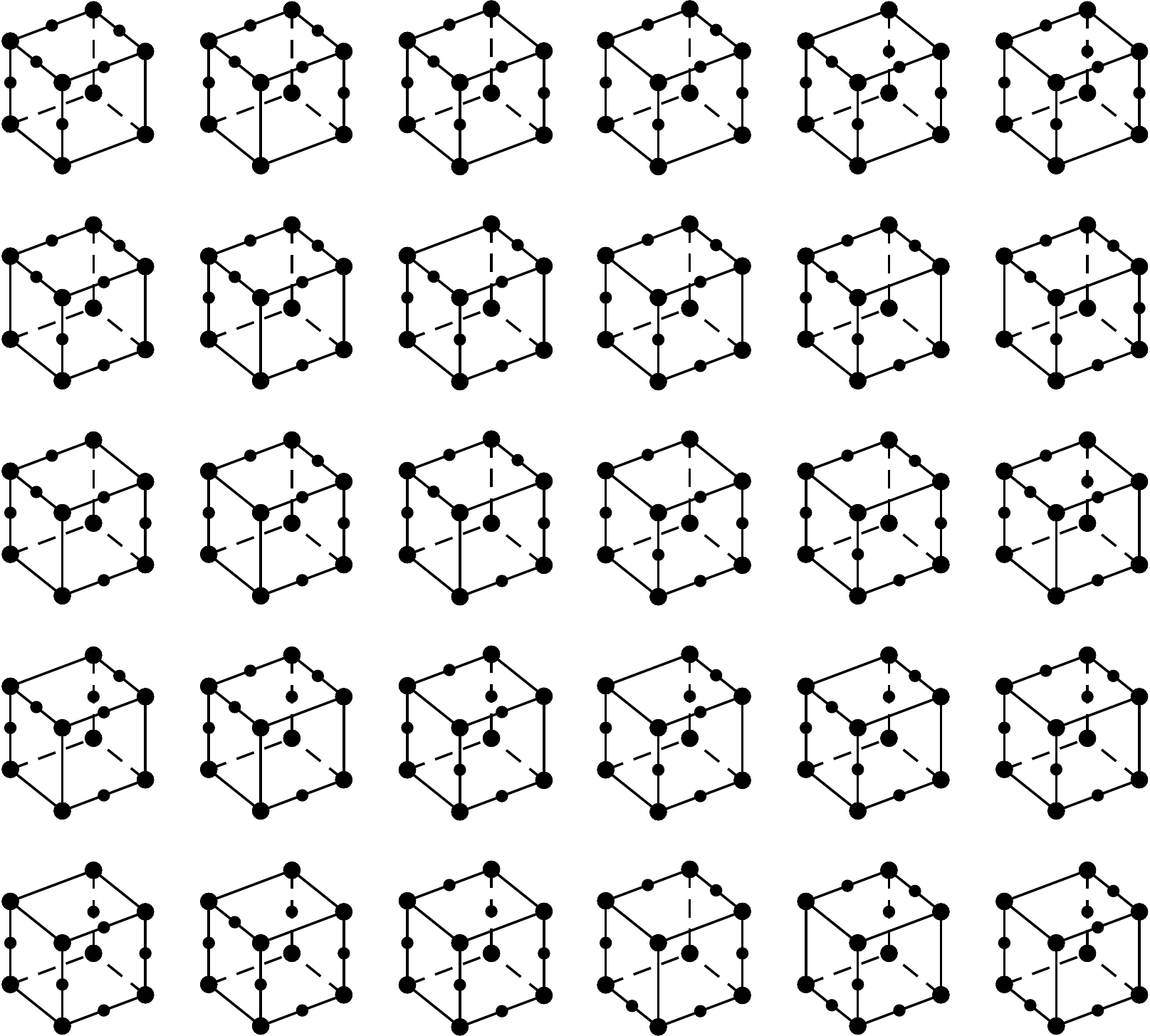}\hfill{}

\caption{Octree cells with hanging nodes on six edges}
\end{figure}

\clearpage{}

\section{Mesh partition of the mountainous region\label{sec:Everest_part}}

\begin{figure}[tb]
\hfill{}\subfloat[two parts]{\includegraphics[width=0.4\textwidth]{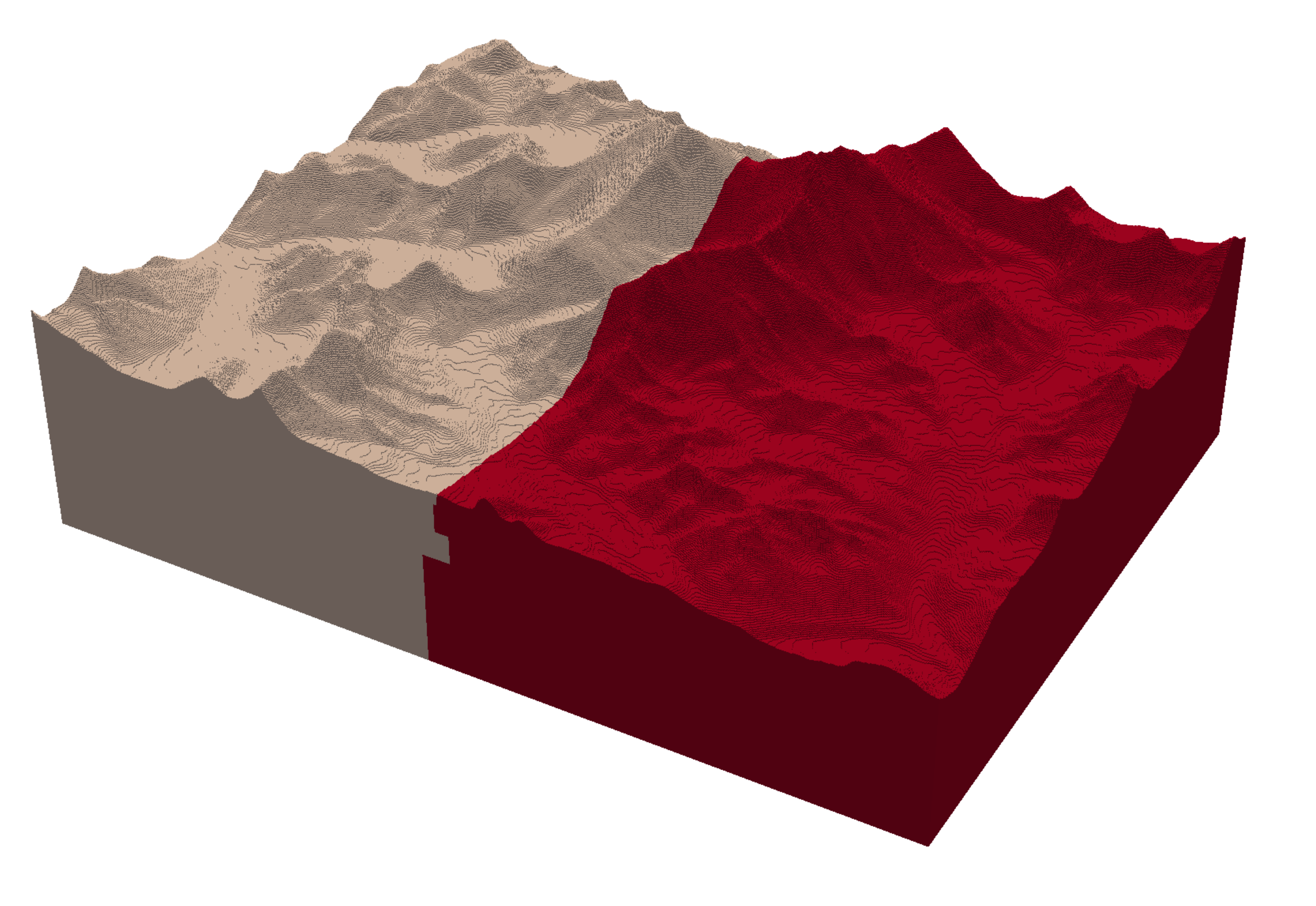}\hspace{-0.05\textwidth}\includegraphics[width=0.05\textwidth]{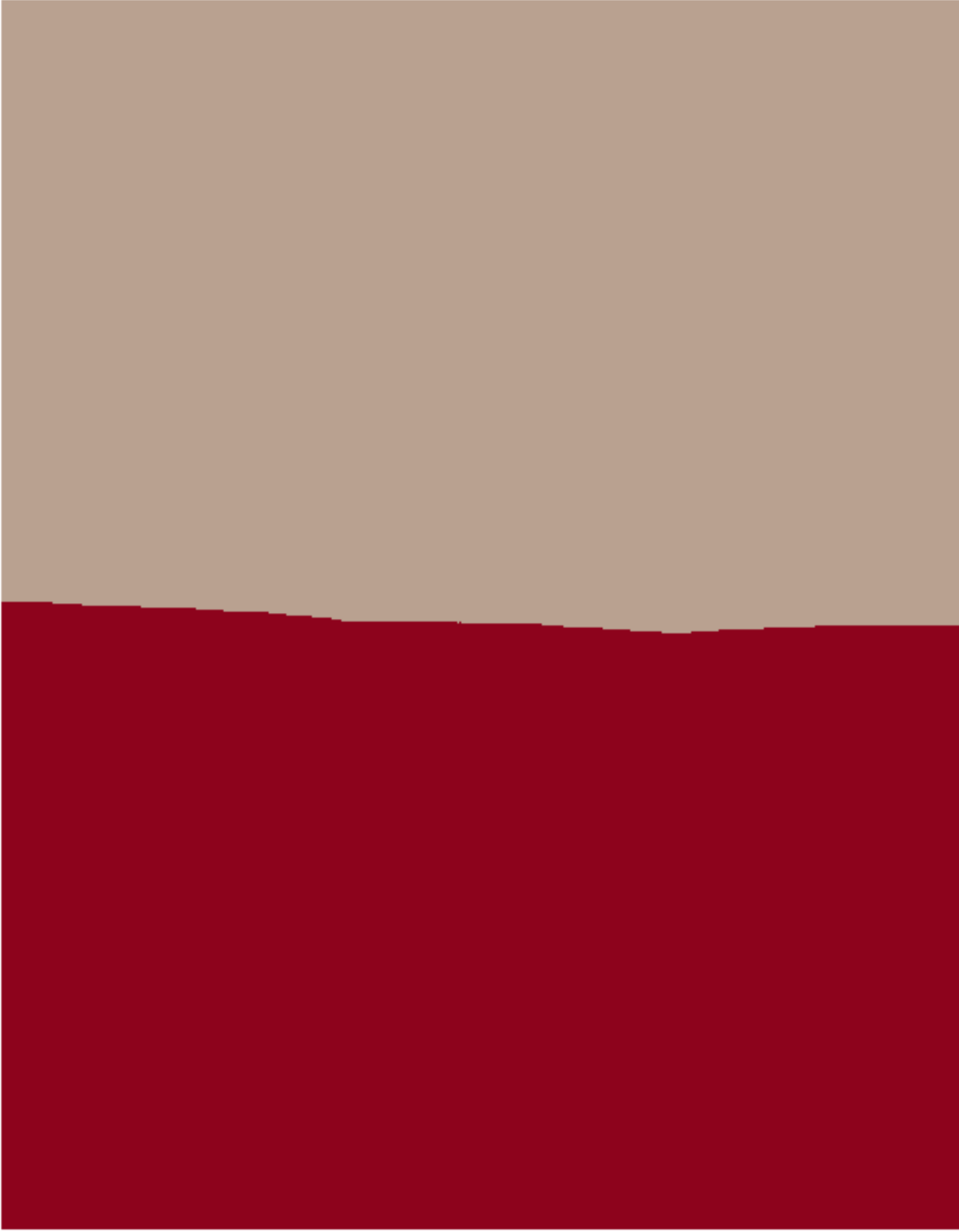}}\hfill{}\subfloat[four parts]{\includegraphics[width=0.4\textwidth]{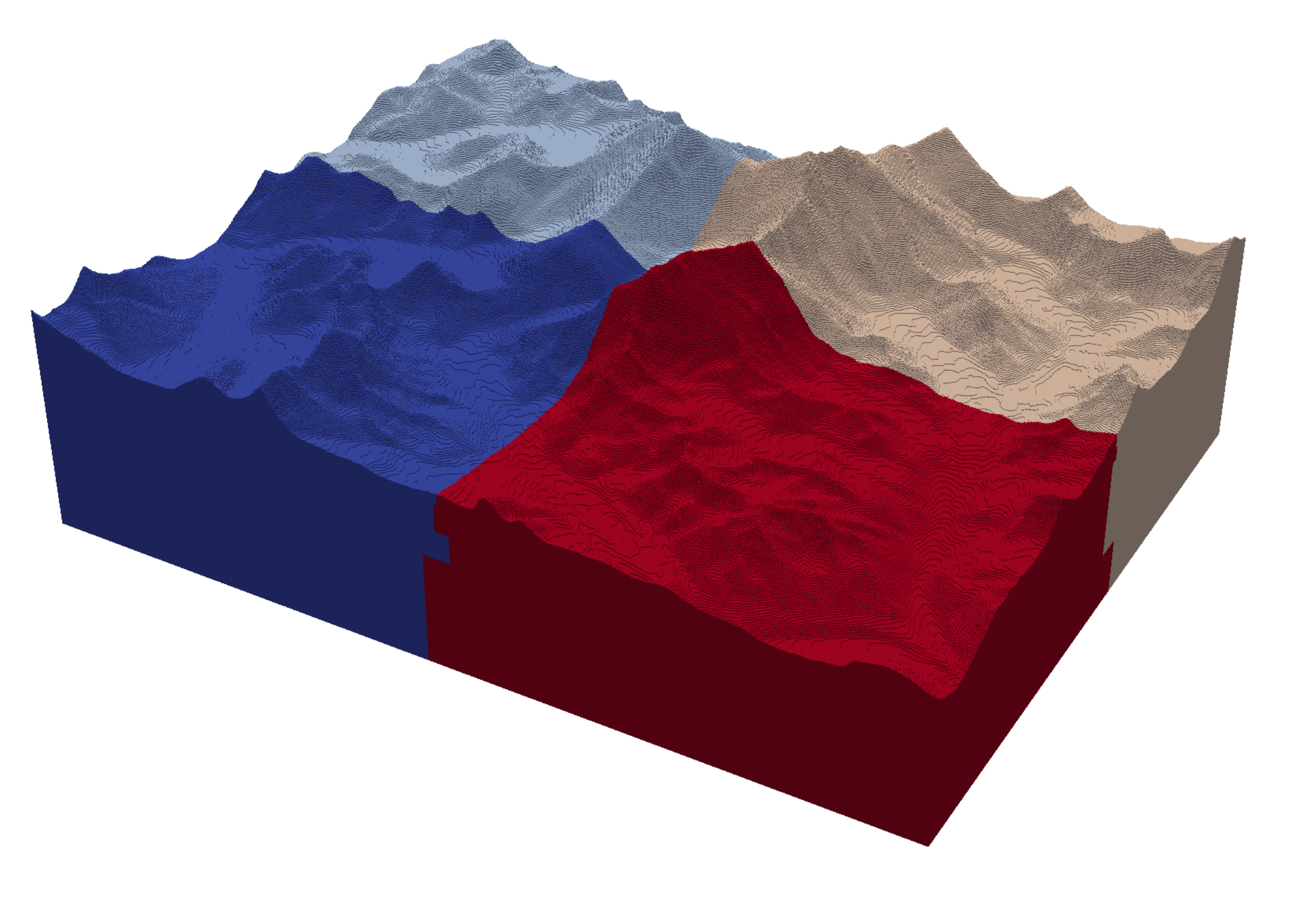}\hspace{-0.05\textwidth}\includegraphics[width=0.05\textwidth]{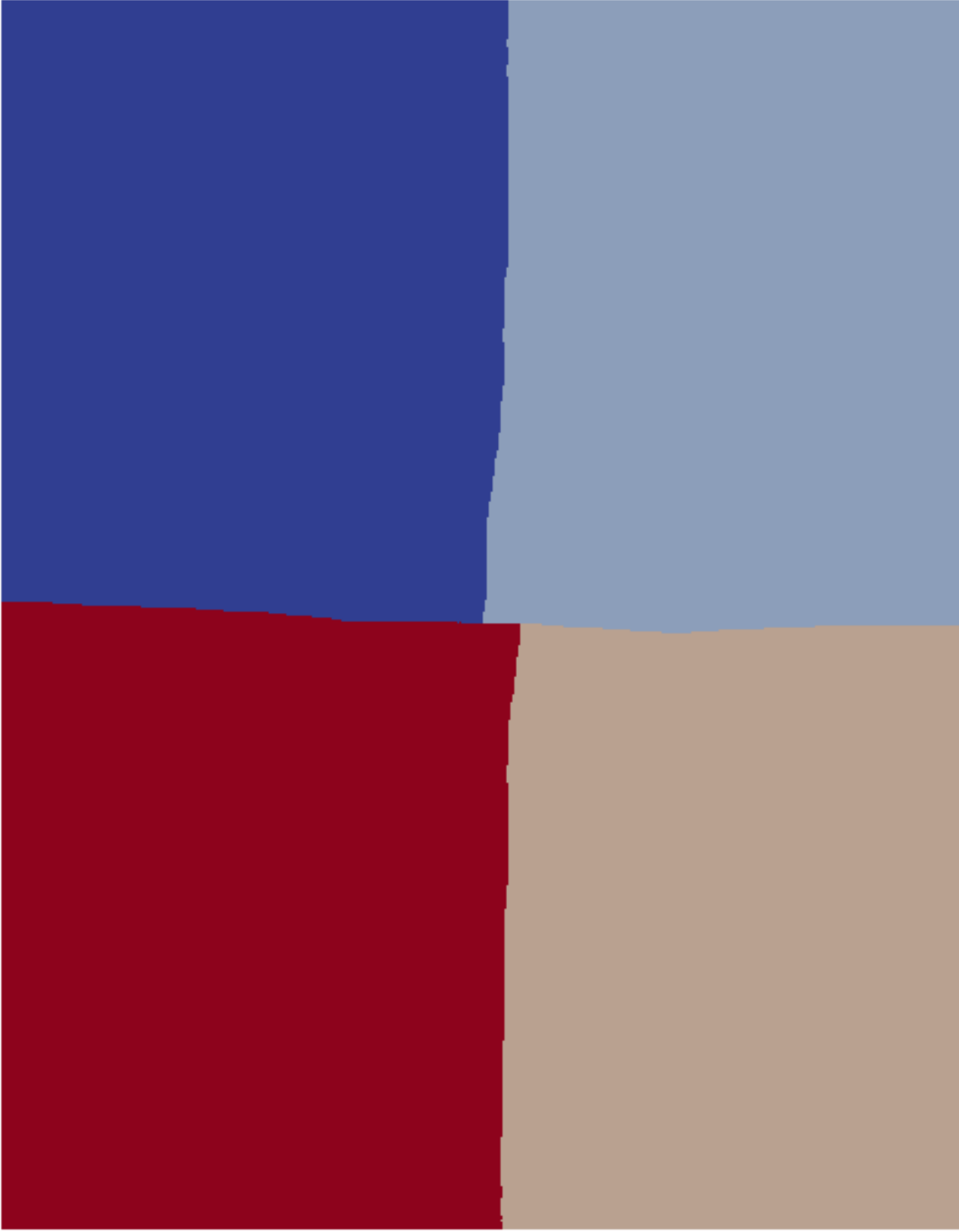}}\hfill{}

\hfill{}\subfloat[eight parts]{\includegraphics[width=0.4\textwidth]{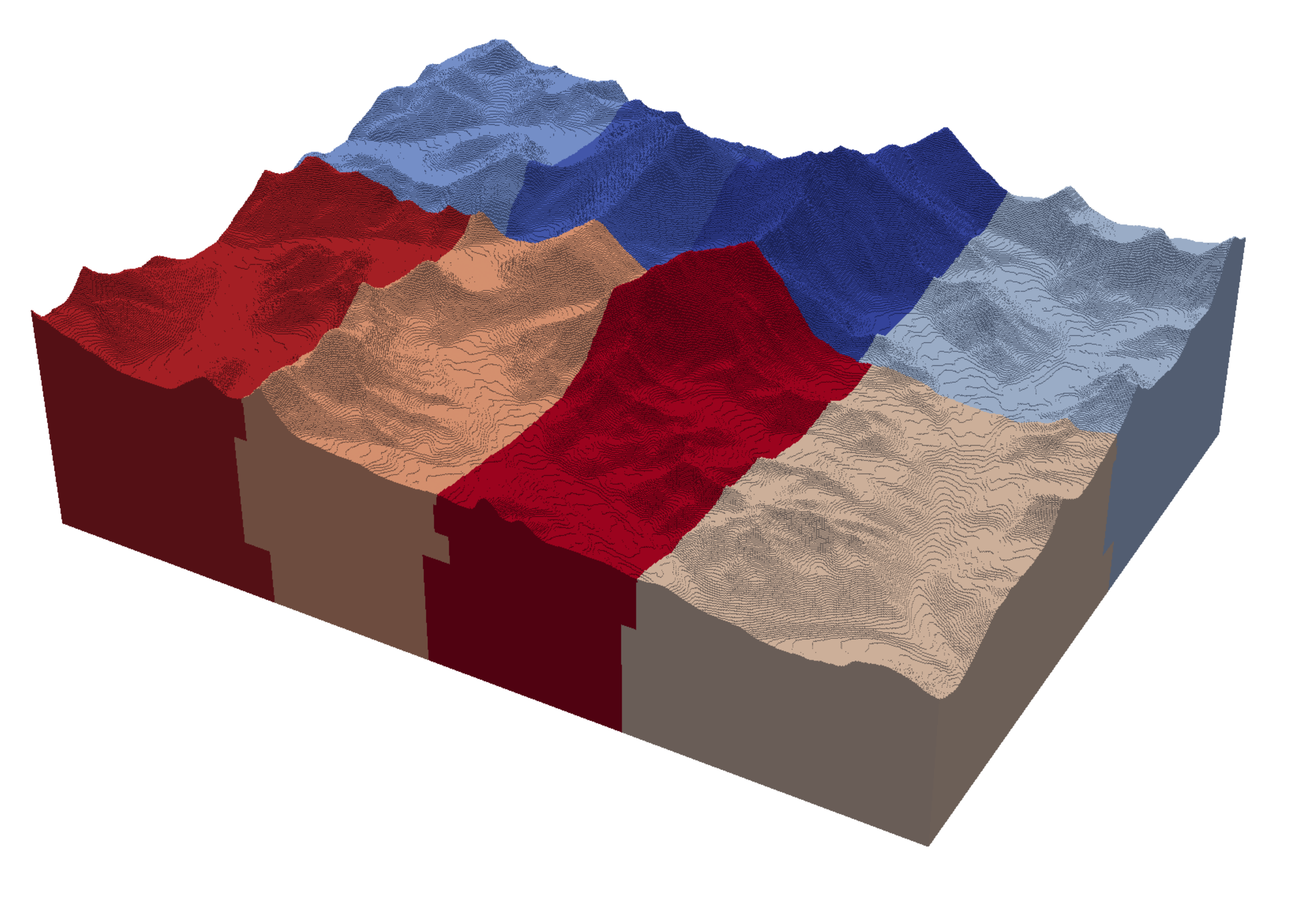}\hspace{-0.05\textwidth}\includegraphics[width=0.05\textwidth]{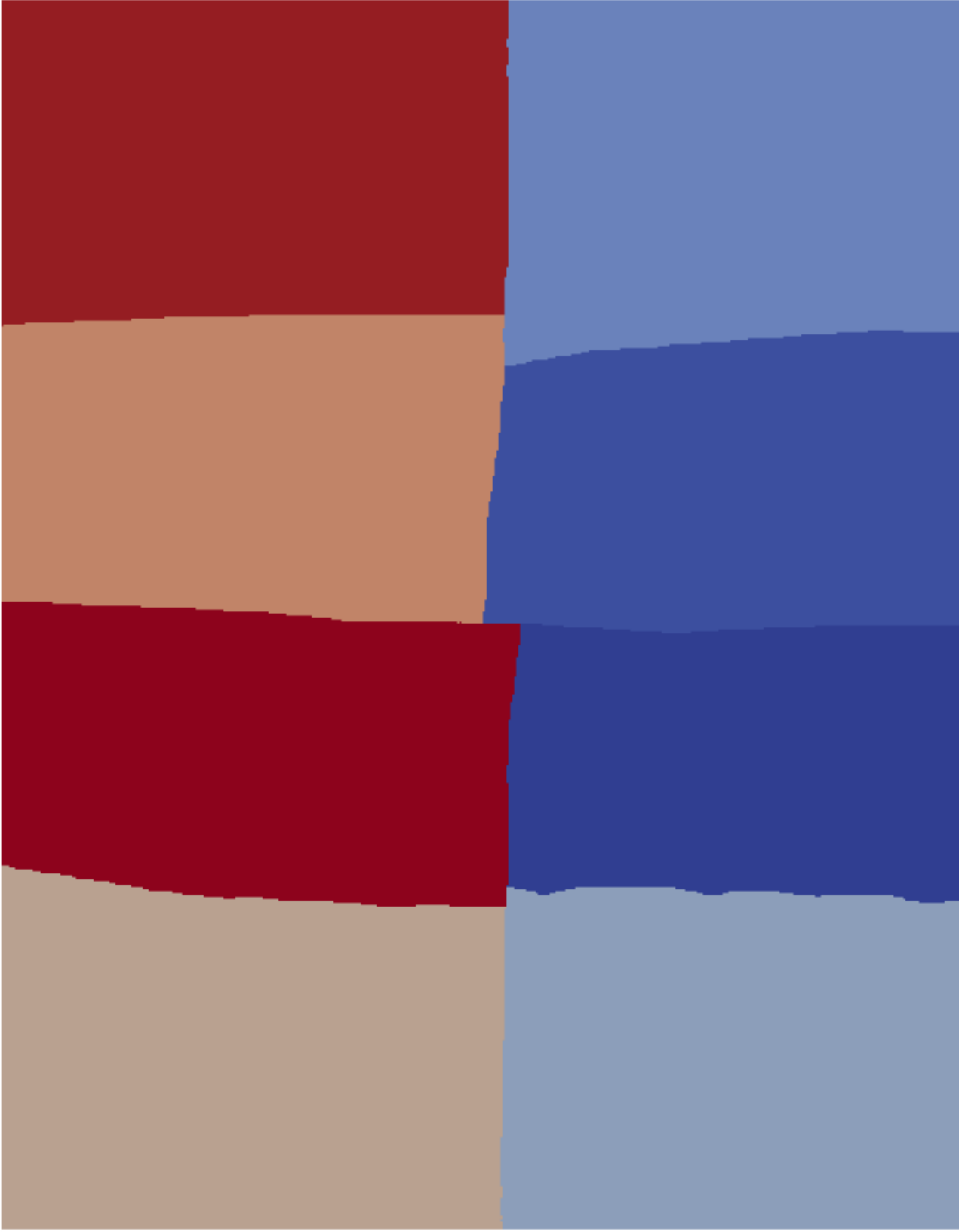}}\hfill{}\subfloat[16 parts]{\includegraphics[width=0.4\textwidth]{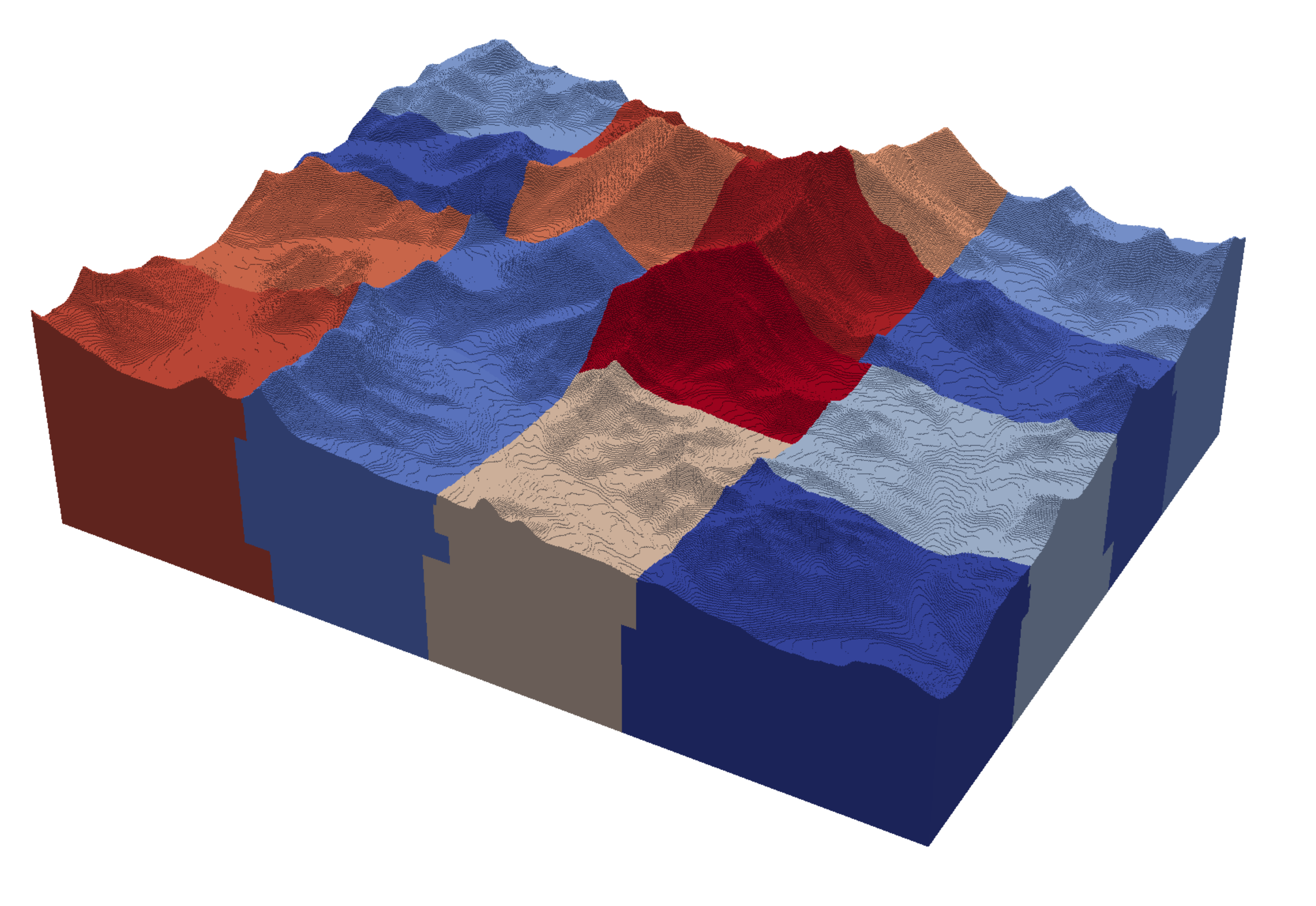}\hspace{-0.05\textwidth}\includegraphics[width=0.05\textwidth]{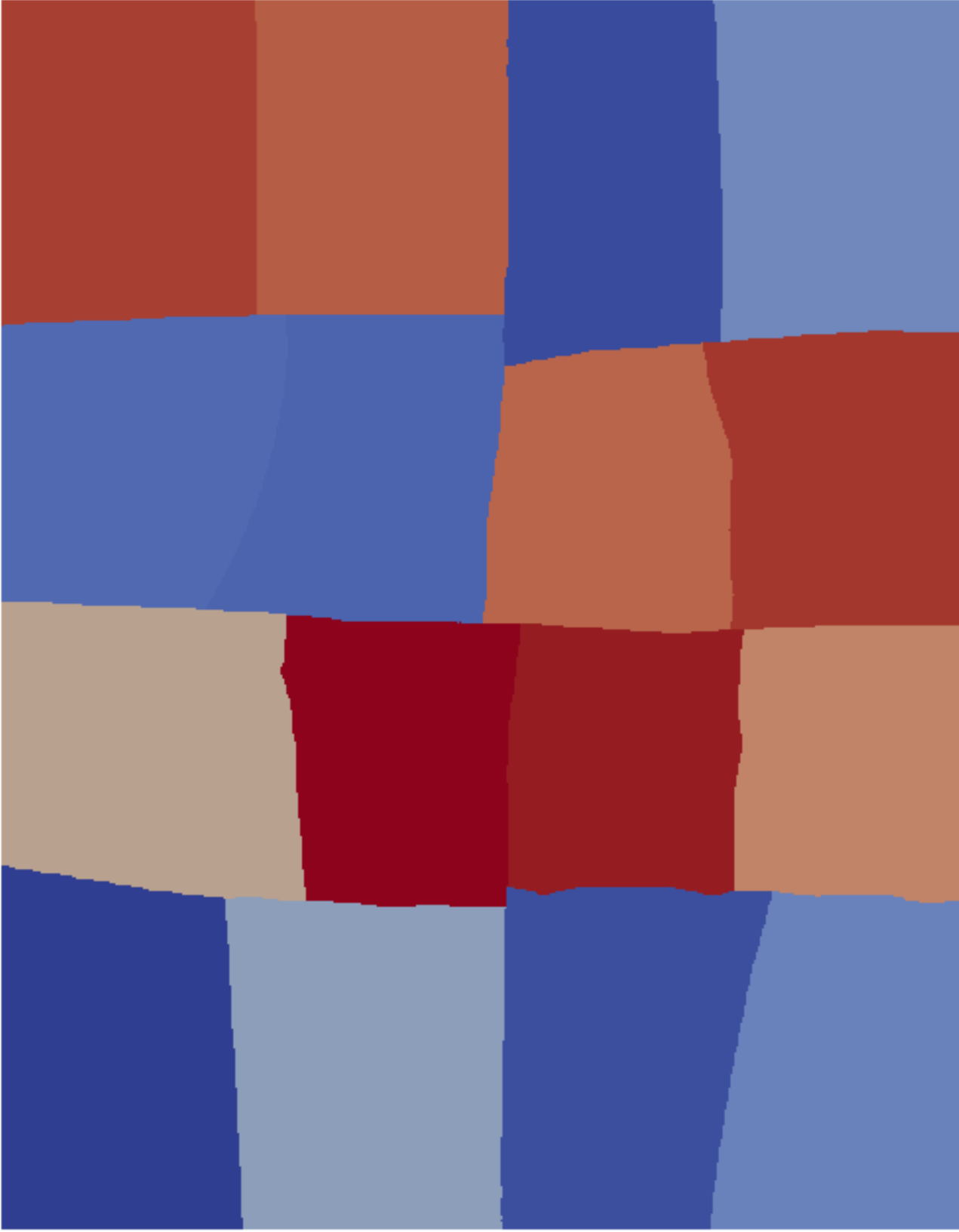}}\hfill{}

\hfill{}\subfloat[32 parts]{\includegraphics[width=0.4\textwidth]{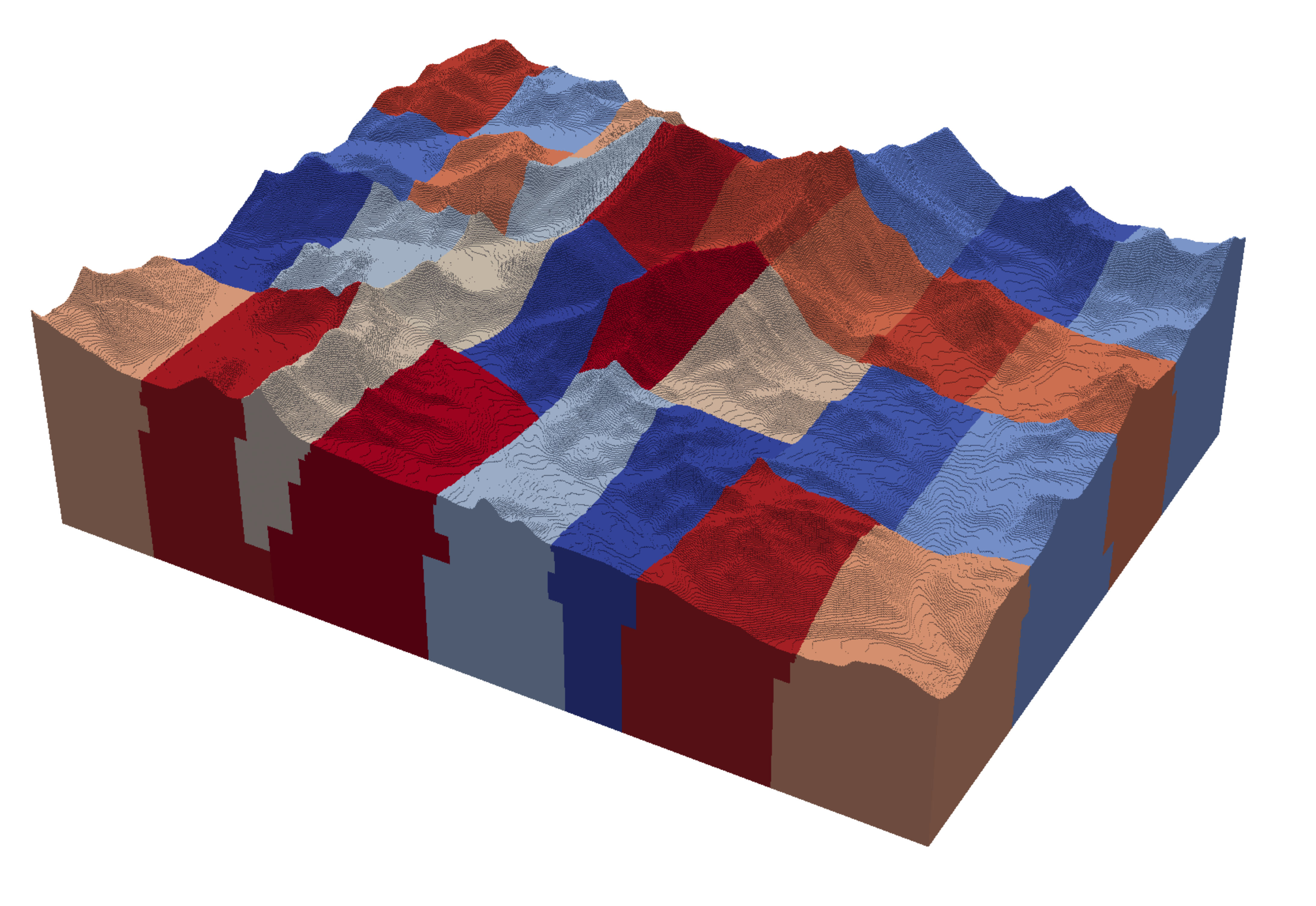}\hspace{-0.05\textwidth}\includegraphics[width=0.05\textwidth]{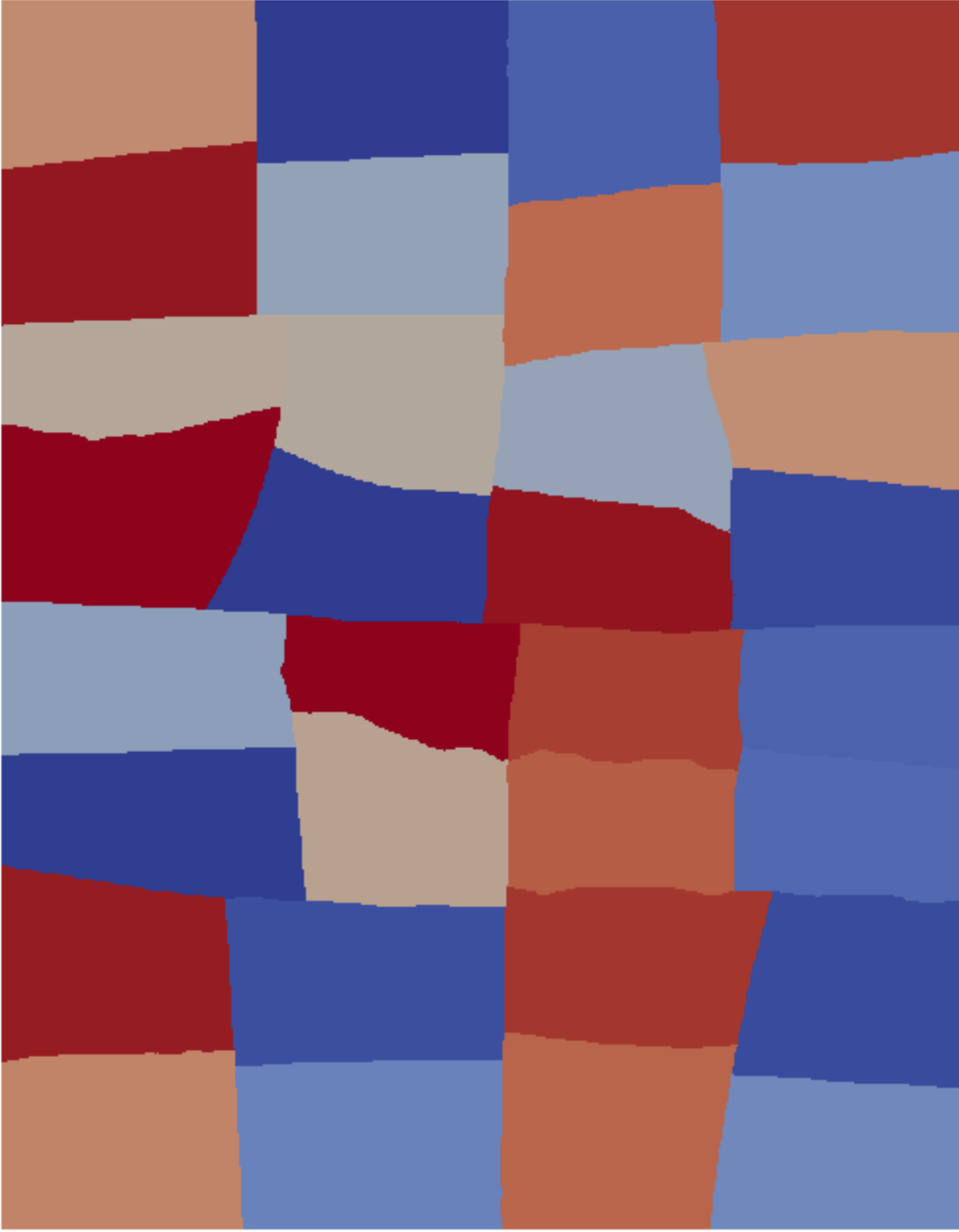}}\hfill{}\subfloat[64 parts]{\includegraphics[width=0.4\textwidth]{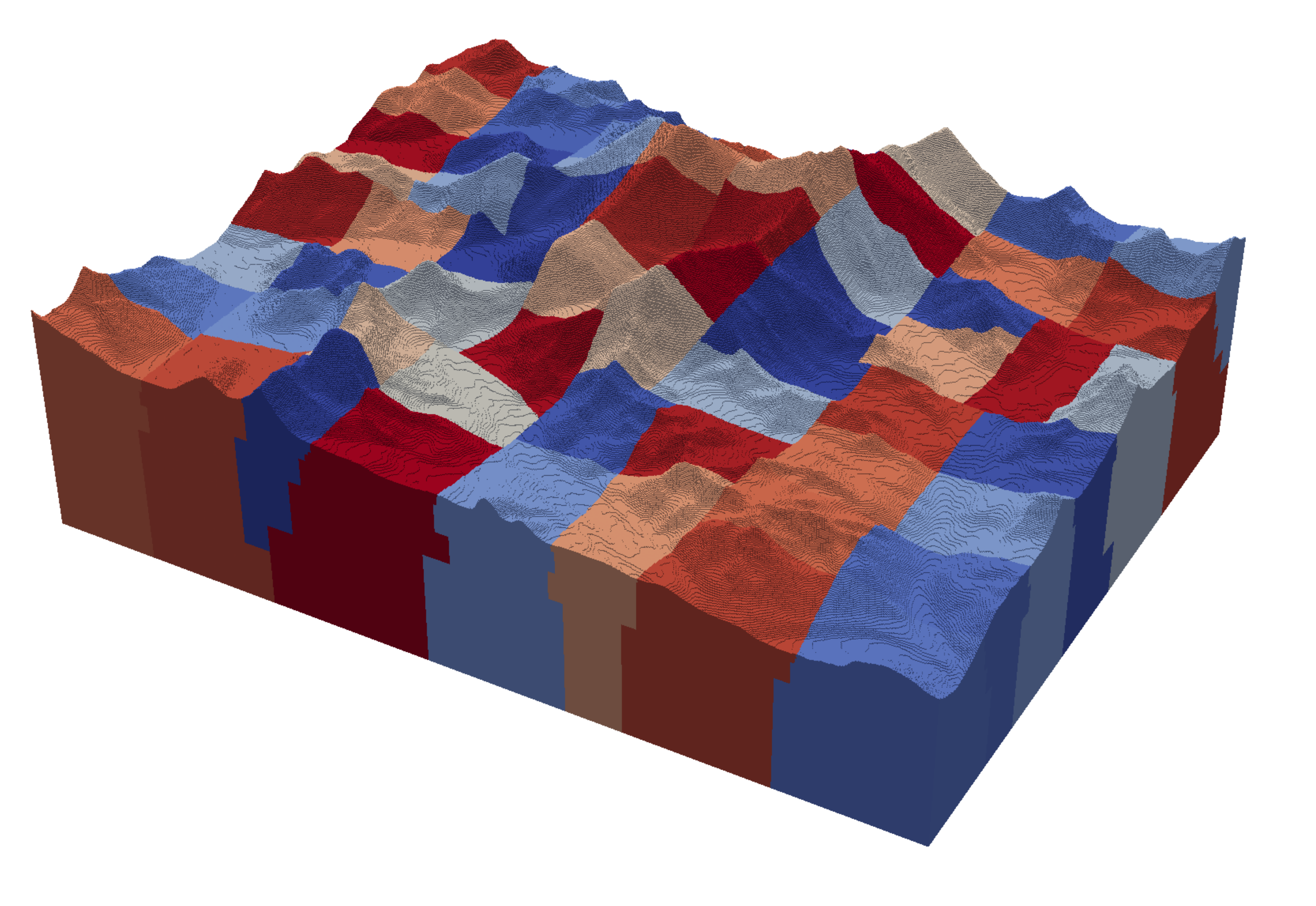}\hspace{-0.05\textwidth}\includegraphics[width=0.05\textwidth]{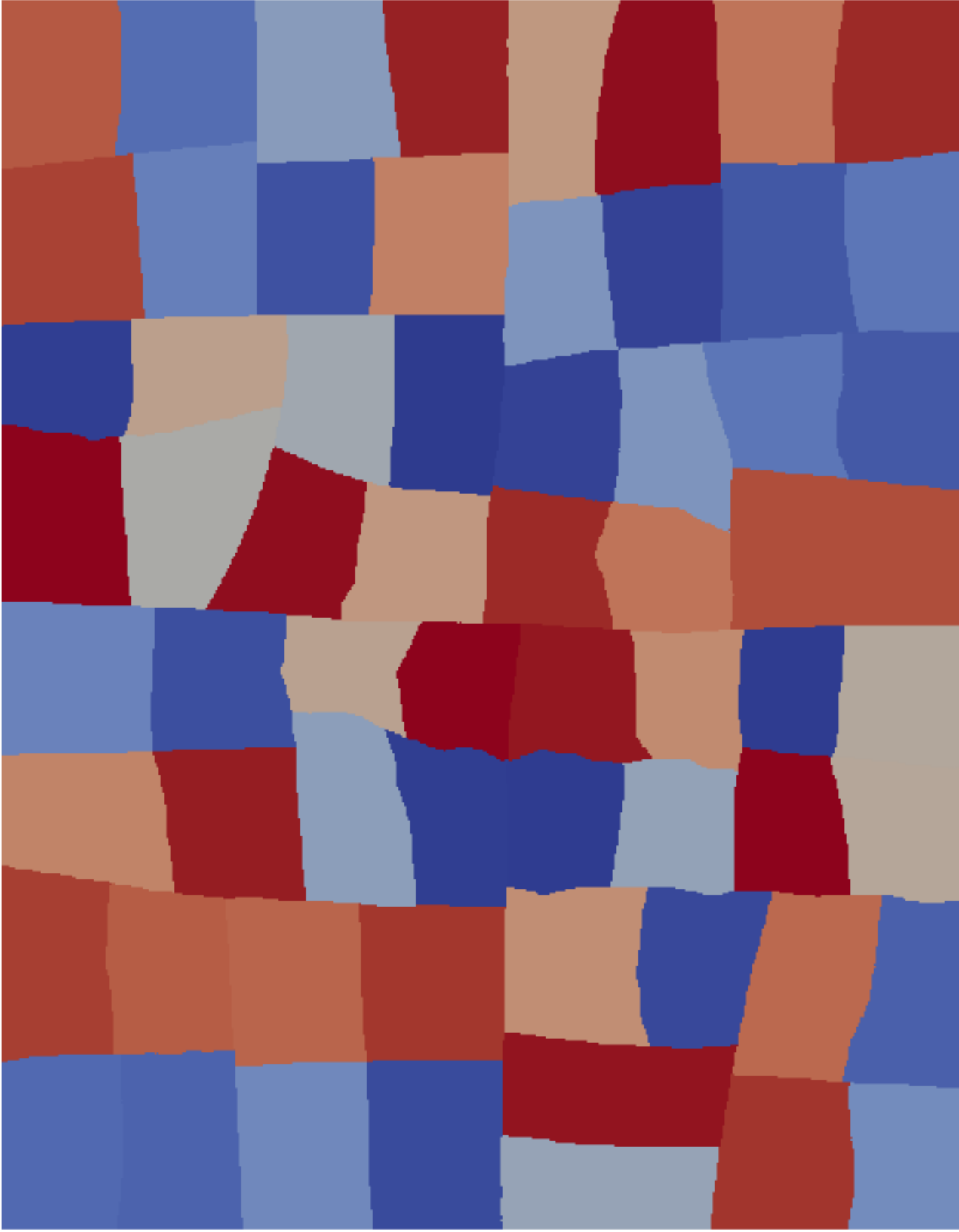}}\hfill{}

\caption{Mesh partition of the mountainous region\label{fig:Everest_part}}
\end{figure}

\clearpage{}

\section{Mesh partition of the sandwich panel\label{sec:Panel_part}}

\begin{figure}[tb]
\hfill{}\subfloat[128 parts]{\includegraphics[width=0.4\textwidth]{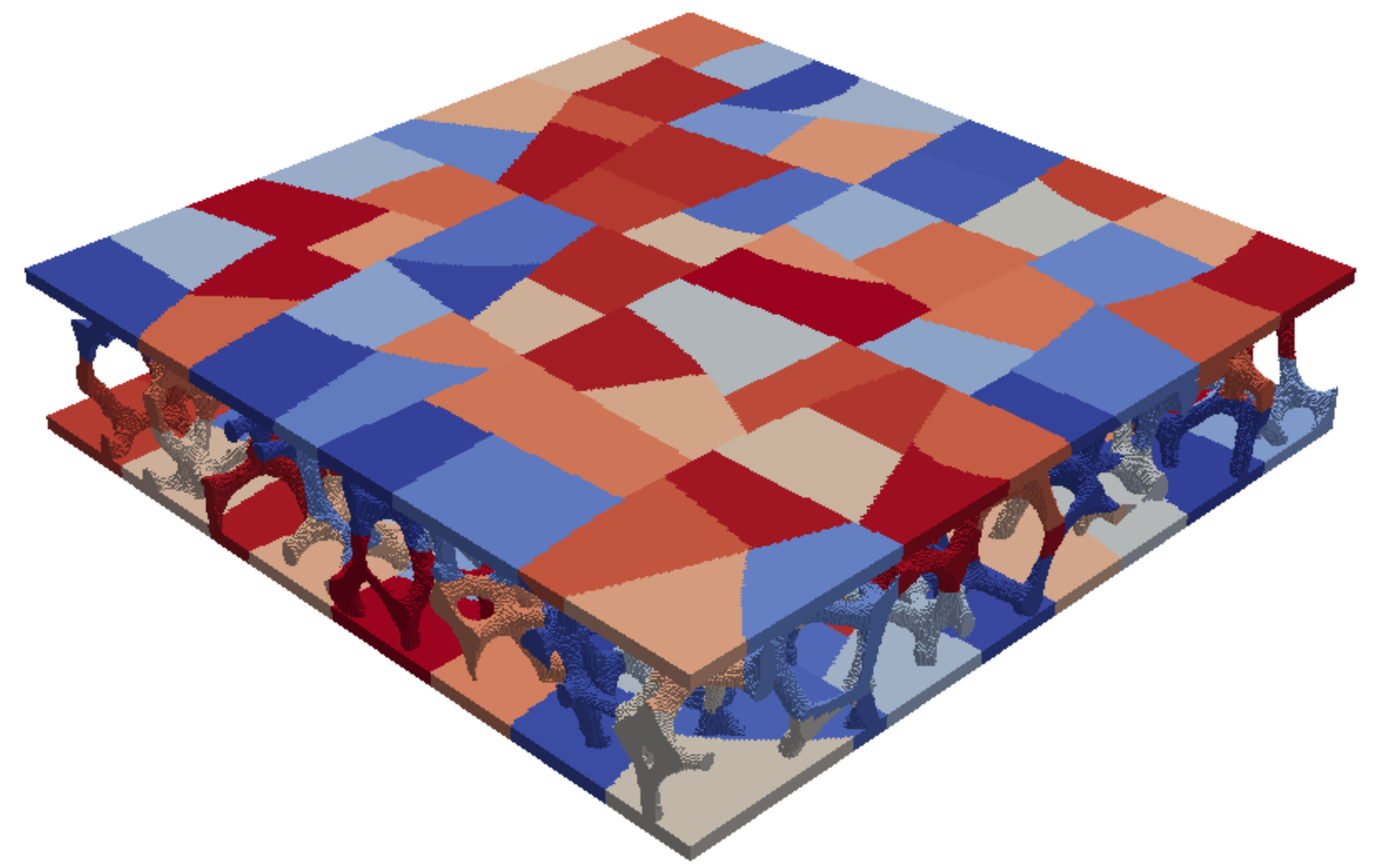}}\hfill{}\subfloat[256 parts]{\includegraphics[width=0.4\textwidth]{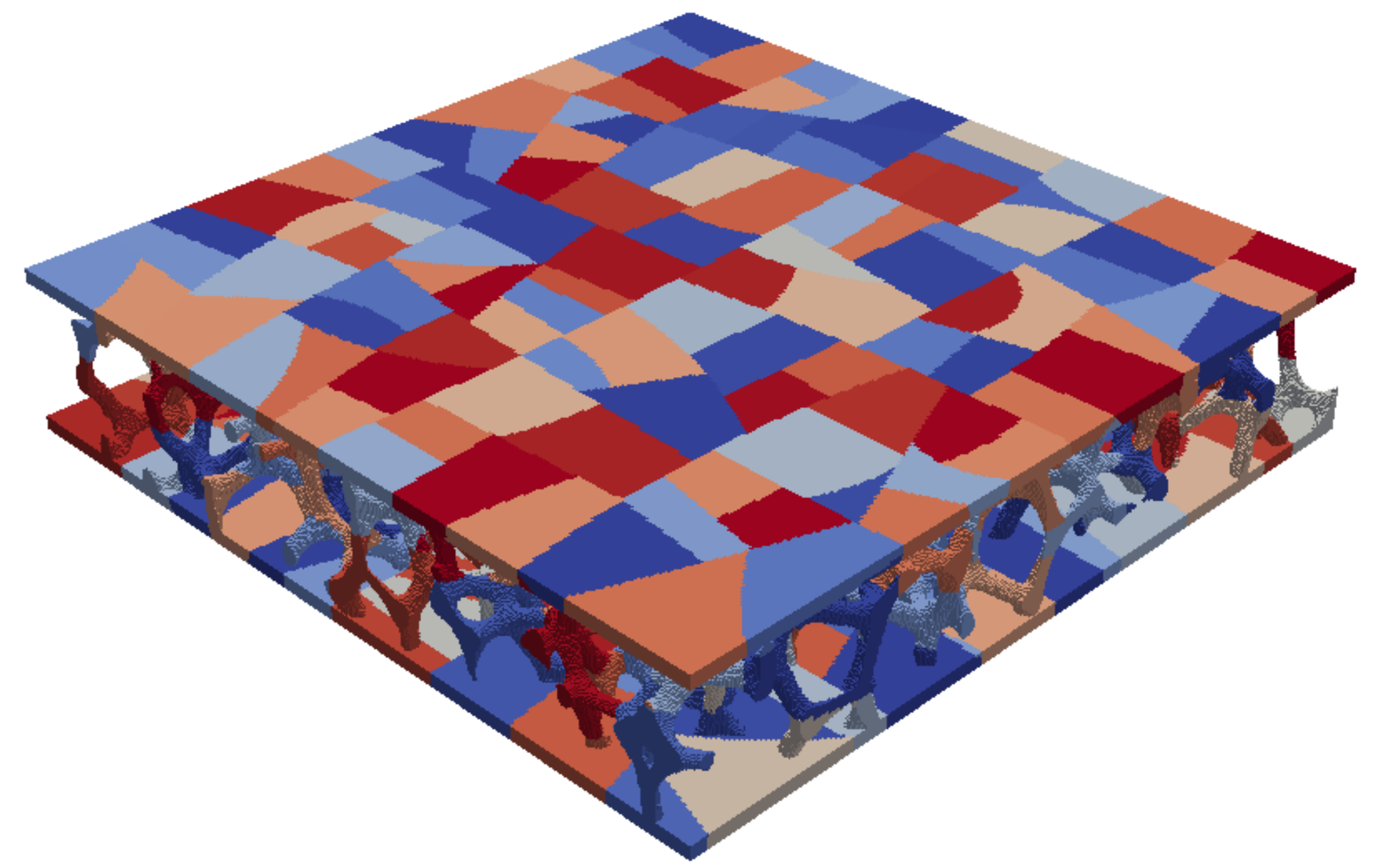}}\hfill{}

\hfill{}\subfloat[512 parts]{\includegraphics[width=0.4\textwidth]{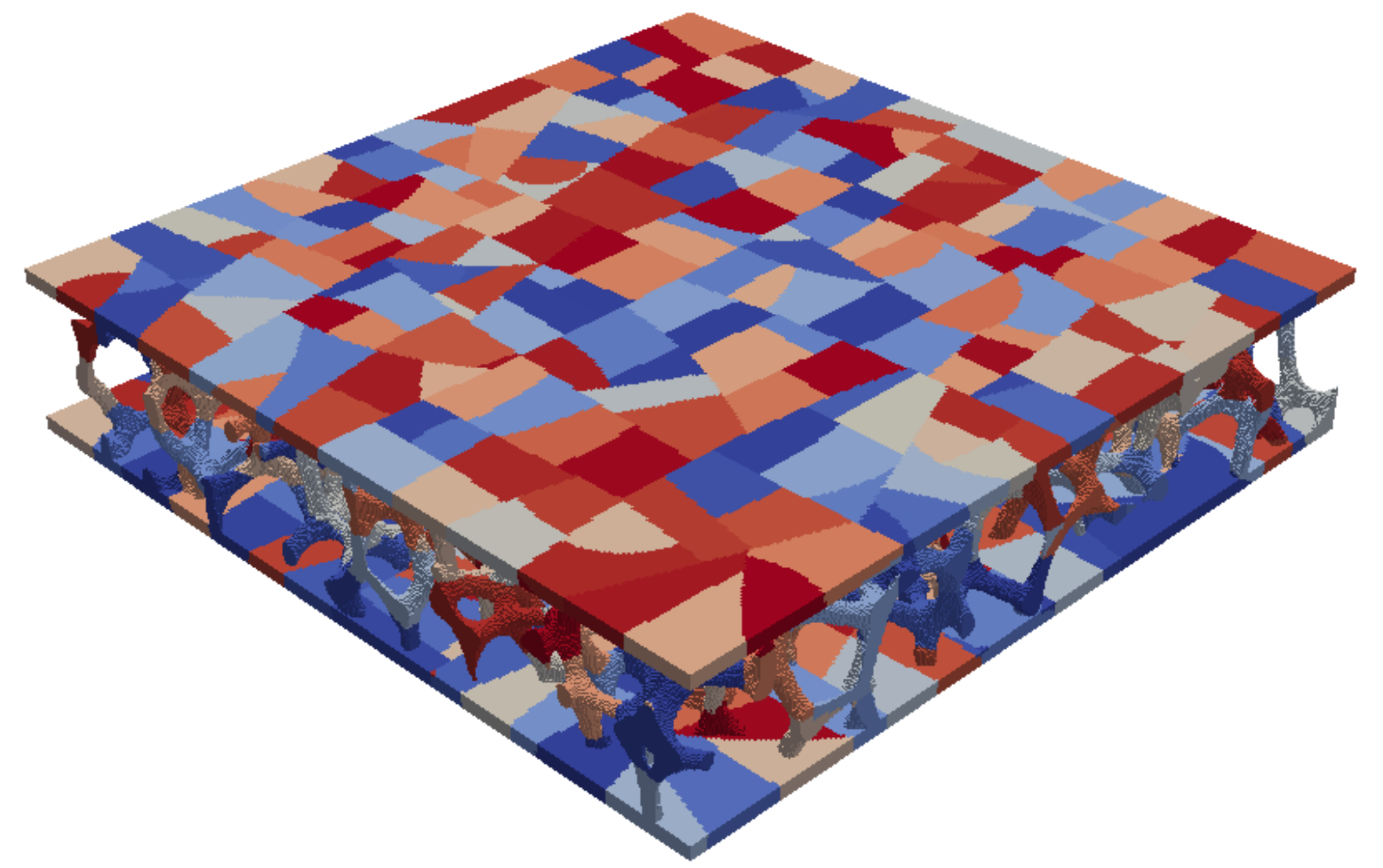}}\hfill{}\subfloat[1024 parts]{\includegraphics[width=0.4\textwidth]{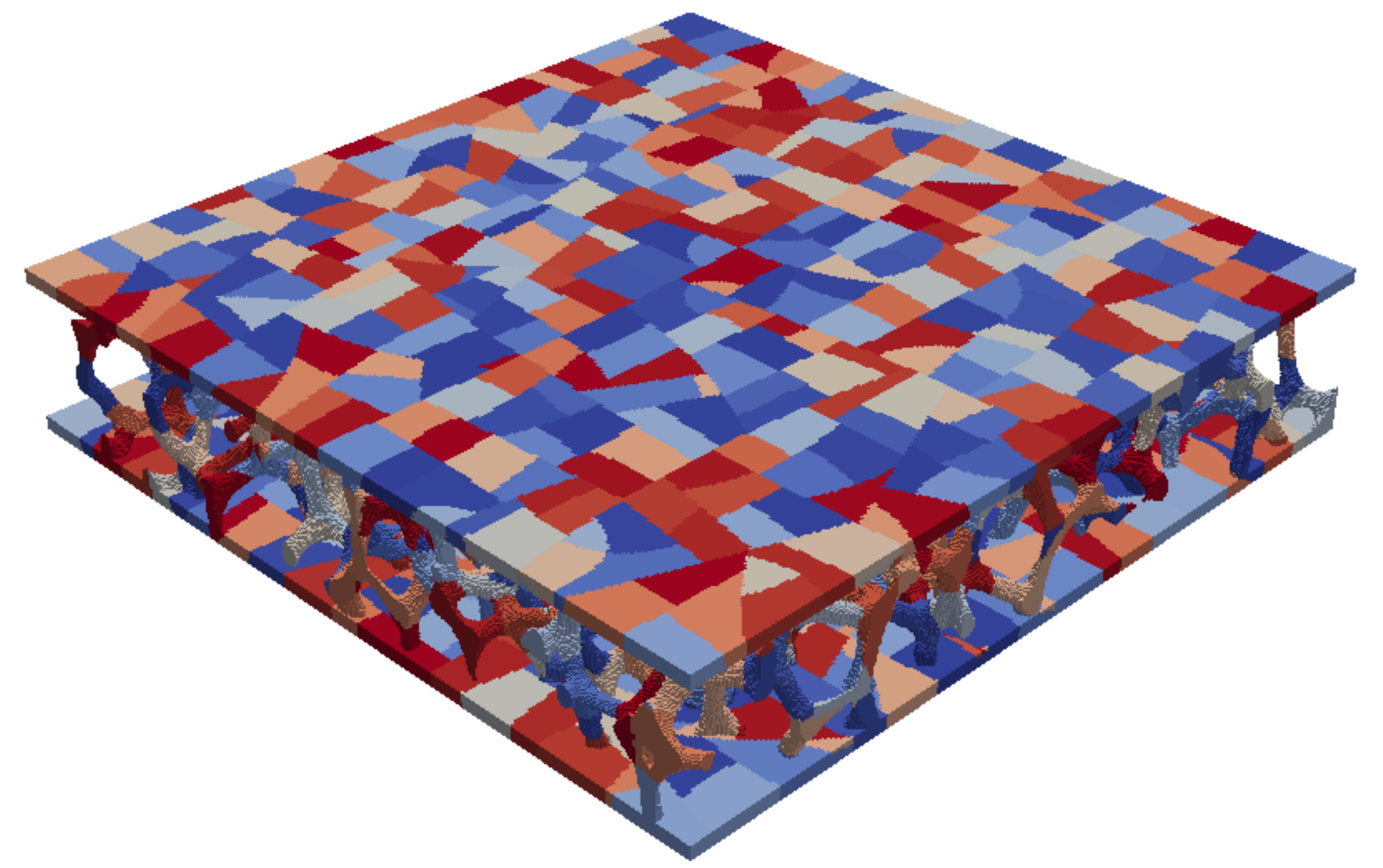}}\hfill{}

\caption{Mesh partition of the $1\times1$ sandwich panel\label{fig:Panel_part}}
\end{figure}

\end{document}